\documentclass[11pt]{ucthesis}

\usepackage[dvips]{epsfig}
\usepackage{bibsetup}

\newcommand{\ie}{{\it i.e.}}

\def\lsim{\mathrel{\rlap{\lower4pt\hbox{\hskip1pt$\sim$}}
    \raise1pt\hbox{$<$}}}         
\def\gsim{\mathrel{\rlap{\lower4pt\hbox{\hskip1pt$\sim$}}
    \raise1pt\hbox{$>$}}}         

\def\vereq#1#2{\lower3pt\vbox{\baselineskip1.5pt \lineskip1.5pt
\ialign{$#1\hfill##\hfil$\crcr#2\crcr\sim\crcr}}}

\def\asec{\hbox{$^{\prime\prime}$}}

\def\micro{$\mu$}
\def\err{$\pm$}

\def\sima{$\sim$}

\def\approxa{$\approx$}
\def\mina{\hbox{$^\prime$}}
\def\ella{$\ell$}
\def\dega{$^\circ$}
\def\sqra{$^2$}

\def\mins{\hbox{$^\prime$}\ }
\def\secs{\hbox{$^\prime$}\hbox{$^\prime$}\ }
\def\ells{$\ell$\ }
\def\degs{$^\circ$\ }

\def\ditto{\tt "}

\def\maxipol{{\sc maxipol}}
\def\maxima{{\sc maxima}}
\def\maximai{{\sc maxima-i}}
\def\maximaii{{\sc maxima-ii}}

\def\boomerang{{\sc boomerang}}
\def\cbi{{\sc cbi}}
\def\max{{\sc max}}
\def\cat{{\sc cat}}
\def\dasi{{\sc dasi}}
\def\vsa{{\sc vsa}}
\def\cobe{{\sc cobe}}
\def\dmr{{\sc dmr}}
\def\pique{{\sc pique}}
\def\polar{{\sc polar}}
\def\dirbe{{\sc dirbe}}
\def\firas{{\sc firas}}
\def\iras{{\sc iras}}
\def\issa{{\sc issa}}
\def\madcap{{\sc madcap}}

\begin{document}

\title{MAXIMA: Observations of CMB Anisotropy}
\author{Bahman Rabii}
\degreeyear{2002}
\degreesemester{Fall}
\degree{Doctor of Philosophy}
\chair{Professor George F. Smoot}
\othermembers{Professor Adrian T. Lee\\
Professor Chung-Pei Ma}
\numberofmembers{3}
\prevdegrees{S.B. (Massachusetts Institute of Technology) 1996\\
S.B. (Massachusetts Institute of Technology) 1996\\
M.A. (University of California, Berkeley) 1999}
\field{Physics}
\campus{Berkeley}

\maketitle
\approvalpage
\copyrightpage

\begin{abstract}

This document describes the Millimeter Anisotropy eXperiment IMaging
Array (\maxima), a balloon-borne experiment measuring the temperature
anisotropy of the Cosmic Microwave Background (CMB) on angular scales of
10\mins to 5\dega .  \maxima\ data are used to discriminate between
cosmological models and to determine cosmological
parameters.

\maxima\ maps the CMB using 16 bolometric detectors observing in
spectral bands centered at 150~GHz, 230~GHz, and 410~GHz, with 10\mins
resolution at all frequencies.  The combined receiver sensitivity to
CMB anisotropy is \sima 40~\micro K~$\sqrt{sec}$, the best reported
by any CMB experiment.  Systematic errors are rejected by using four
uncorrelated spatial modulations, multiple independent CMB
observations, heavily baffled optics, and strong spectral
discrimination.  Observation patterns are well cross-linked and
optimized for the extraction of cosmological information.  Pointing is
reconstructed to an accuracy of 1\mina . Absolute calibration
uncertainty of 3-4\% is the best achieved by any sub-orbital CMB
experiment.

Two \maxima\ flights were launched from the National Scientific
Balloon Facility in Palestine Texas in 1998 and 1999.  During a total
of 8.5 hours of CMB observations, 300~deg$^2$ of the sky were mapped,
with \sima 50~deg$^2$ overlap between the two flights.  The observed
region was selected for low foreground emission and post-flight data
analysis confirms that foreground contamination is negligible.

Cosmological results are presented from the 1998 flight, \maximai, in
which 122~deg$^2$ of sky were mapped over 3 hours.  A maximum
likelihood map with 3\mins pixelization is obtained from the three
most sensitive and best tested detectors.  The angular power spectrum
derived from this map shows a narrow peak near $\ell = 200$, and is
consistent with inflationary Big Bang models.  Within these models,
cosmological parameters are estimated, including total density
$\Omega_{tot}=0.9_{-0.16}^{+0.18}$, baryon density $\Omega_b h^2 =
0.033 \pm 0.013$, and power spectrum normalization $C_{10} =
690_{-125}^{+200} \mu K^2$.  In combination with recent supernova
observations, we obtain additional constrains on the matter density
$\Omega_{m}=0.32{+0.14\atop-0.11}$ and the dark energy density
$\Omega_{\Lambda} = 0.65{+0.15\atop -0.16}$.  All parameter estimates
are presented at 95\% confidence.

The final chapter is a discussion CMB polarization anisotropy,
including an overview of \maxipol, the polarization sensitive
follow-up to \maxima.  Measurements of CMB polarization are an
essential complement to those of temperature anisotropy.

\abstractsignature
\end{abstract}

\begin{frontmatter}

\begin{dedication}
\addcontentsline{toc}{chapter}{Dedication}
\null\vfil
\begin{center}
\large To Andrew P. Kitchen.
\end{center}

\end{dedication}

\chapter*{Preface}
\addcontentsline{toc}{chapter}{Preface}

	\maxima\ is one of the first experiments to map the Cosmic
Microwave Background on sub-degree angular scales.  My work on the
project began in the spring of 1997.  I have been involved with
preparation and operations for both flights, data analysis, and design
and modification of hardware.

	This document is intended as a general overview of the
\maxima\ experiment, including goals, hardware, data reduction and
analysis, and results.  Emphasis is placed on experimental techniques.
Chapters~\ref{chap:point} and~\ref{chap:calib} are treatments of areas
in which I have been particularly active: pointing and responsivity
calibration.  Later chapters present data analysis, results, and
systematic error tests.  The final chapter deals with future work on
CMB polarization.

	Another \maxima\ PhD dissertation, \cite{CDWThesis}, is in
preparation.  Though some of the general information overlaps with
that in this document, \cite{CDWThesis} includes a detailed look at
the detector system and optics, that are only summarized here, and
treats pointing, observations, and calibration in less detail.  In
this sense, the two dissertations are complementary.

\hfill --- Bahman Rabii\\
\noindent
\emph{Berkeley}\\
\emph{September, 2002}
\clearpage

\renewenvironment{acknowledgements}
{\begin{alwayssingle}
\chapter*{\acknowledgename}
\addcontentsline{toc}{chapter}{\acknowledgename}
}
{\end{alwayssingle}}

\tableofcontents
\listoffigures
\listoftables
\begin{acknowledgements}

	Working on \maxima\ has been an incredibly rewarding and
sometimes painful experience.  My deepest thanks go to the entirety of
the collaboration, especially to my advisor George Smoot and to
Celeste Winant, Adrian Lee, Shaul Hanany, and Paul Richards.  I was
also lucky to work with a great data analysis team and am especially
grateful for the tireless efforts of Radek Stompor.

	Thanks also to the staff at NSBF who never failed to go beyond
the call of duty, to Barth Netterfield, Enzo Pascale, Paolo
DeBernardis, and Andrea Boscaleri for the pointing system, and to
Patrick Bonnefil and Diane Linklater for the company in Palestine.

	I want to thank Baba, Azi, Faramarz, and Shahriar for their
years of support and guidance, and Uncle Sohrab, my image of a
scientist.

	Mad props to Toastizens past and present, the LeConte crew,
the DNA-ers, the Warpheads, the Floreys, and team Sensomatic.  You all
made the last six years more fun than you can shake a stick at.  Ted and
Xina, it was more than worth moving out here to get to know you.
Quincy, Jordan, Dave - I am still lucky to have friends like you.  My
most heartfelt thanks to Anca.

\newpage

\noindent {\bf Added in Proof:}

\bigskip

\noindent I have committed a great error by failing to credit the following
individuals:

\bigskip

\noindent Brad Johnson (University of Minnesota) for his efforts
during the \maximaii\ field campaign.

\bigskip

\noindent Jeffrey Collins, and Huan Tran (University of California),
and Brad Johnson, Tomotake Matsumura, and Tom Renbarger (University of
Minnesota) for their work on \maxipol .

\end{acknowledgements}
\end{frontmatter}

\chapter{Motivation and Background}\label{chap:motiv}

The first section of this chapter is a brief review of the Cosmic
Microwave Background and its use as a probe of cosmology, especially
inflationary Big Bang models.  Theoretical details have been
thoroughly explored in published literature (e.g. \cite{HuDodelson},
\cite{Kosowsky}) and are not repeated here.  Section~\ref{motiv:past}
presents the history of CMB observation.  Section~\ref{motiv:obser}
deals with technical considerations of measuring the temperature
anisotropy.

\section{The Cosmic Microwave Background}\label{motiv:cmb}

The Cosmic Microwave Background (CMB) is nearly uniform blackbody
radiation at a temperature of 2.725\err 0.002~K.  It is believed to be
of cosmological origin: the heavily redshifted emission from a hot,
optically thick period in the early universe.  The existence of low
temperature background radiation in a Big Bang universe was proposed
in 1948 by George Gamow and further explored in 1950 by his colleagues
Ralph Alpher and Robert Herman.  The CMB was detected by Arno
Penzias and Robert Wilson in 1964.  Its existence remains one of the
strongest pieces of evidence for the Big Bang.  Further measurements
of the CMB -~its spectrum, temperature anisotropy, and polarization~-
provide information about the structure and evolution of the universe
(\cite{RPP}).  \maxima\ and other experiments of the late 1990's have
measured the temperature anisotropy on sub-degree angular scales to
test cosmological models and obtain estimates of cosmological
parameters.

\subsection{Fundamental Implications of the CMB}

By the early 1990's, two extremely powerful statements could be made
of the CMB.  First, its spectrum is that of an astoundingly precise
blackbody.  Second, its temperature anisotropy is extremely small.
These two facts have become cornerstones of modern cosmology.

\begin{figure}[ht]
\centerline{\epsfig{width=4.0in,angle=0,file=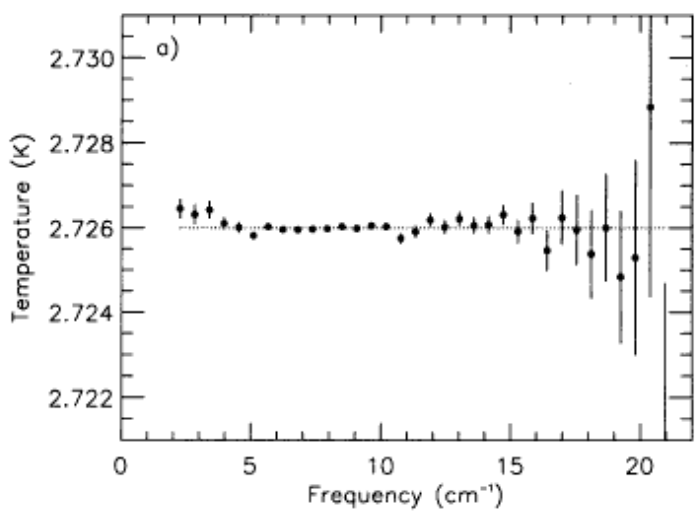}}
\caption[Measured CMB Spectrum]{The spectrum of the CMB as measured by
the \cobe\ \firas\ instrument.  Data are plotted in units of blackbody
equivalent temperature.  The vertical zero is suppressed to make the
error flags visible.  This plot is taken from \cite{MatherSpectrum};
subsequent improvements in calibration changed the best fit temperature
from 2.726~K to 2.725~K (\cite{MatherSpectrum2}).}
\label{fig:firas}
\end{figure}

The CMB spectrum, best measured by the \firas\ instrument of the the
\cobe\ satellite (\cite{MatherSpectrum2}), shows no statistically
significant deviations from a thermal spectrum and has a mean Compton
y-parameter of $<$10$^{-5}$.  Subsequent experiments have measured
spectral distortions near galaxy clusters due to the Sunyaev-Zeldovich
effect.  This spectrum has only been explained by Big Bang cosmology.
The universe began in a hot, dense, expanding state.  At \sima 300,000 years after the Big Bang, the universe consisted mostly of baryons, photons, and dark matter, with the baryons and photons in
thermal equilibrium.  As it expanded, adiabatic cooling allowed
electrons and protons to combine, breaking the thermal equilibrium and
suddenly increasing the mean free path of photons to greater than the
present horizon size of the universe.  This transition from thermal
equilibrium to photon free streaming, referred to as recombination or
last scattering, occurred at a redshift of \sima 1100.  Since the time
of last scattering, CMB photons have cooled with the cosmological
redshift to their present temperature of 2.725~K.  Because Doppler
shifts introduce no spectral distortions, the CMB spectrum today is
nearly identical to that of the thermalized universe just before last
scattering.

While the almost perfect spectrum of the CMB answers fundamental
questions about the evolution of the universe, the homogeneity of the
CMB temperature has introduced new mysteries.  The largest temperature
variation of the CMB is a dipole of about $10^{-3}$~K caused by the
peculiar velocity of the earth relative to the CMB rest frame.  Apart
from this dipole, temperature anisotropies are only about one part in
$10^5$.  These small fluctuations are the seeds needed for
gravitational condensation to produce the structures observed in the
modern universe.  But how were these fluctuations generated?  And why
is the temperature so homogeneous on large scales?  Answers to both of
these questions are provided by the concept of inflation (\cite{InflationBook}, \cite{GBellido}).

Large scale homogeneity suggests that the universe at the time of last
scattering was at a constant temperature over what is now the
observable universe.  This region was not in causal contact at the
time.  According to the dynamics of simplistic Big Bang models, it would 
never have been in causal contact and could not have come to thermal equilibrium.  This ``horizon
problem'' is the most compelling evidence that such
models are inadequate.  
Inflation proposes that a period of rapid acceleration
increased the scale factor of the universe by \sima 50 orders of
magnitude during the first $10^{-32}$ seconds.  Regions were causally
connected (and in thermal equilibrium) before they were separated by
inflation, and would remain at the same temperature at the time of
last scattering and beyond.

Inflation is also a
solution to two further mysteries, the ``flatness problem'' and the
``defect problem''.  The flatness problem is one of coincidence; the
present density of space is very close to the critical value required
for a spatially flat universe.  However, the growth of the universe
through radiation and matter dominated phases causes divergence from
flatness by more than 50 orders of magnitude.  The present situation
is difficult to explain without a mechanism to have forced
extraordinary flatness in the early universe.  The strong,
accelerating expansion of inflation provides such a mechanism.
Inflation can also explain the lack of observed topological defects,
such as magnetic monopoles, resulting from phase transitions in the
very early universe.  It occurred after these transitions and spread
the defects apart, reducing their number density to roughly one per
present horizon volume.

	Inflation provides a mechanism for generating the primordial
fluctuations needed to seed later structure formation (\cite{Liddle}).  This is an
important prediction and has been the key to observational exploration
of inflationary models.  The signature of inflationary structure
generation in CMB anisotropy is discussed in the following section.

\subsection{CMB Temperature Anisotropy}

\begin{figure}[ht]
\centerline{\epsfig{width=4.5in,angle=0,file=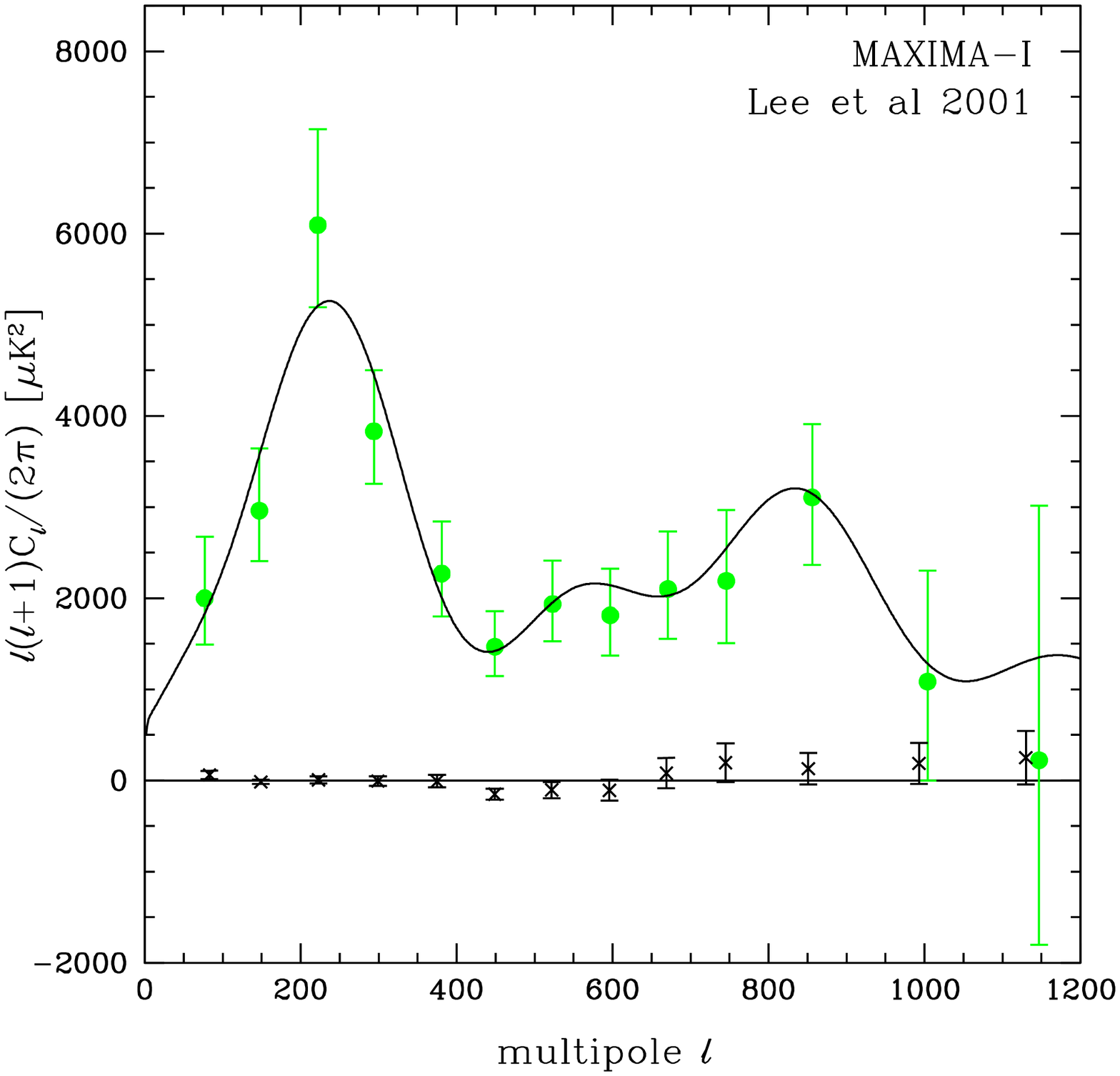}}
\caption[CMB Anisotropy Angular Power Spectrum]{ An example of a CMB
power spectrum, including measured data points and a model curve
calculated for an inflationary Big Bang cosmology with a dark energy
component.  Data shown are the results of the \maximai\ experiment from
\cite{LeeResults}.}
\label{fig:leecmbps}
\end{figure}

The study of CMB temperature anisotropy is useful in two ways.  First,
it can distinguish between cosmological models, especially between
those dominated by primordial density fluctuations and those dominated
by primordial stress fluctuations.  Second, it can be used to
determine cosmological parameters, especially the total density of
the universe.

Temperature anisotropy is often quantified with an angular power
spectrum, as in Figure~\ref{fig:leecmbps}, with CMB power in units of
$\ell(\ell + 1)C_{\ell}/(2\pi)$ (or equivalently $\Delta {T_l}^2$) on
the y-axis and \ells on the x-axis.  Here, \ells is the angular
multipole number, inversely proportional to the angular scale, and
$C_{\ell} \equiv \langle a_{lm} \rangle_m$ is the mean spherical
harmonic coefficient at a given \ella .

\subsubsection{Adiabatic vs Isocurvature Fluctuations}

CMB temperature anisotropy is the result of fluctuations in the
baryon-photon fluid at the time of recombination.  Generically, there
are two orthogonal types of fluctuation: adiabatic (density) and
isocurvature (stress) fluctuations.  In adiabatic fluctuations, all
species (baryons, photons, CDM, etc.) have fixed ratios, but overall
density varies spatially.  Adiabatic fluctuations directly seed the
gravitational growth of structure.  For isocurvature fluctuations, the
overall energy density is uniform, but there are variations in the
number densities of various species.  Isocurvature fluctuations
causally relax into density fluctuations, indirectly seeding
gravitation growth.  Density variations grow through gravitational
collapse into the structures of the present universe.

The angular power spectrum of the CMB is very different for primarily
adiabatic and primarily isocurvature models.  In adiabatic models, the
power spectrum shows the familiar pattern of several narrow peaks (see
the following section and Figure~\ref{fig:leecmbps}).  In isocurvature
models, these peaks are replaced by at most one broad hump.

Only inflationary models produce significant adiabatic fluctuations on
all scales (\cite{Liddle}), while models with significant numbers of
topological defects provide the stresses needed for isocurvature
models.  For this reason, the concepts of inflation and adiabatic
fluctuations have often been paired together as competing with defects
and isocurvature fluctuations.  Current data
(e.g. Chapter~\ref{chap:results}) are consistent with adiabatic
fluctuations, though the possibility remains of a hybrid universe with
subdominant isocurvature fluctuations.  This possibility is best
explored through the study of CMB polarization
(Chapter~\ref{chap:future}).

\subsubsection{Adiabatic CMB Anisotropy}

In standard inflationary cosmologies, the universe prior to
recombination consists primarily of photons, baryons, and
collisionless dark matter.  Photons and baryons are tightly coupled,
while the dark matter is not.  The initial spatial spectrum of density
fluctuations is close to the scale-free Harrison-Zeldovich spectrum,

\begin{equation}
| {\delta}_k  | ^2 
\equiv
\left| \frac{\delta \rho}{\rho}  \right| ^2 
= A k
,\end{equation}
\noindent where $\rho$ is the average density, k is a Fourier
wavenumber, $\delta \rho$ is the density fluctuation on that scale,
and A is a constant.  Under linear gravitational collapse
$| {\delta}_k | ^2 $\ grows with the square of the scale
factor.  This is the state before recombination, outside the sound
horizon.  As the modes fall within the sound horizon, photon pressure
counters the effects of gravity, causing harmonic acoustic
oscillations of the photon/baryon fluid.  In contrast, the dark matter
is coupled to the photons only by gravity via perturbations in the spacetime metric and collapses monotonicly.  After
recombination, baryons are decoupled from photons and quickly couple
to the larger density variations of the dark matter and purely
gravitational collapse resumes.

Density fluctuations at the time of recombination are
imprinted on the CMB; the phase of a given mode is determined by the
time between the start of its acoustic oscillations and recombination.
This phase sets the observed CMB power at a given scale.  The smooth
variation of phase with scale leads to evenly spaced ``acoustic
peaks'' in the angular power spectrum.

Density fluctuations cause CMB temperature fluctuations through three
mechanisms.  First, density variations in the photon/baryon fluid
cause adiabatic heating and cooling; denser regions emit hotter
photons.  This is the dominant mechanism at \ella $>$100 and is the
source of the acoustic peaks.  Second, the photons in potential wells
are gravitationally redshifted as they climb out; the mechanism causes
denser regions emit cooler photons.  This is referred to as the
Sachs-Wolfe effect and is the dominant mechanism at large angular
scales, \ie~those that did not fall within the sound horizon and
oscillate before recombination.  Third, the local velocity of the
photon/baryon fluid at the time of recombination imparts a Doppler
shift to the CMB photons.  Velocity extrema are 90\degs out of phase
with density extrema, causing another set of peaks between the
acoustic peaks.  The Doppler shifts are always sub-dominant and these
out of phase peaks do not appear distinctly in the power spectrum.

Two other effects are also considered primary anisotropies: the
integrated Sachs-Wolfe effect, and photon diffusion damping.

The integrated Sachs-Wolfe (ISW) effect is the net gravitational
redshift as photons pass in and out of potential wells after last
scattering.  In a static universe, the infall blueshift and the exit
redshift would cancel exactly, but if potential wells grow or decay as
photons pass through them, there is a net redshift or blueshift.  The
ISW effect may occur after recombination if the universe is not fully
matter dominated (``Early'' ISW) or, in an open or dark energy
universe after matter domination has ended (``Late'' ISW).  The ISW
effect contributes to CMB anisotropy at large and moderate scales
(\ells up to 300).

Photon diffusion damping causes an exponential decay in the power
spectrum at small angular scales.  Because recombination is not
instantaneous, the CMB does not provide a snapshot of the an
arbitrarily thin surface in the early universe.  The finite thickness
of the surface emitting CMB photons suppresses CMB structure at \ella
\approxa 1000 and higher.

Effects which add or modify CMB anisotropies after recombination
(other than ISW) are often called secondary anisotropies.  Secondary
anisotropies, such as a diffuse Sunyaev-Zeldovich effect or
reionization, are most significant at small angular scales
(\ella$>$2000) and are most likely to be observed by high resolution
interferometric experiments.

\subsection{Cosmological Parameters from Temperature Anisotropy}

\begin{figure}[ht]
\centerline{\epsfig{width=4.2in,angle=0,file=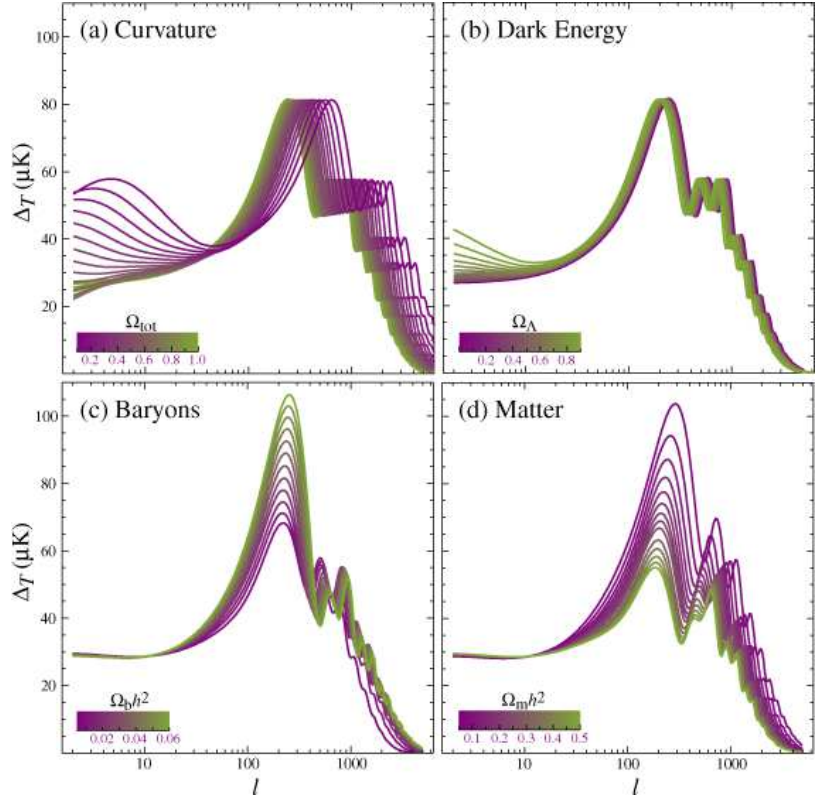}}
\caption[Parameter Sensitivity of CMB Anisotropy]{The dependence of
the anisotropy power spectrum on cosmological parameters.
Inflationary cosmologies are assumed.  {\bf{Top Left:}} The power
spectrum is most sensitive to total density, or equivalently to
spatial curvature.  Lower density moves the acoustic peaks to smaller
scales (higher \ella ), and also increases large scale power from the
ISW effect.  {\bf{Top Right:}} For a given curvature, there is some
sensitivity to the dark energy density.  The shifting of the acoustic
peaks is degenerate with the stronger effect of curvature.
{\bf{Bottom Left:}} Baryon density affects relative peak heights
because of its influence on the zero point of the acoustic
oscillations.  {\bf{Bottom Right:}} Matter density affects the total
power in the acoustic peaks as well as shifting peaks and influencing
their relative heights.  Figures by \cite{HuDodelson}.}
\label{fig:paramsens}
\end{figure}

Sub-degree scale measurements of CMB temperature anisotropy can be
used to find cosmological parameters.  In particular, the CMB power
spectrum is sensitive to $\Omega_{tot}$, the total density of the universe;
$\Omega_b$, the baryon density; $n_s$ the primordial spectrum of
density fluctuations; and $\tau_c$, the optical depth to reionization.
There is a lesser degree of sensitivity to other parameters such as
$\Lambda$, the vacuum energy density; and $\Omega_m$ the matter
density.  The effects of some of these parameters on the CMB power
spectrum are illustrated in Figure~\ref{fig:paramsens}.

The strongest sensitivity is to the total density $\Omega_{tot}$.  For
$\Omega_{tot} = 1$ (flat space), the power spectrum first peaks at
$\ell \simeq 220$.  For lower density (positive curvature) the first
peak is at higher \ella , while for higher density (negative
curvature) the first peak is at lower \ella .  Other peaks are shifted
proportionately.  This effect is relatively insensitive to the physics
at the time of recombination.  Because the redshift to last scattering
and the size of the sound horizon are well estimated, the physical
scale corresponding to the first peak can be calculated.  The apparent
angular scale depends primarily upon the curvature of light rays and
is a direct measurement of the curvature of space.

Other parameters reflect earlier physics.  The change in relative peak
amplitudes with baryon density, for example, is a consequence of the
gravitational attraction of baryons shifting the zero-point of the
acoustic oscillations before recombinations.

The impact of various parameter changes on CMB power spectra has been
widely discussed in the literature, and can be calculated using
publicly available numerical tools (e.g. CMBFAST \cite{CMBFAST}).

\section{Observational History}\label{motiv:past}

In this section, we outline the history of CMB observations over the
last forty years.  A more detailed account of observations up to the
early 1990's can be found in \cite{Partridge}.

The CMB was first observed by Penzias and Wilson at Bell Labs in 1964
as an unknown `excess noise' with a blackbody equivalent temperature
of 3.5\err 1.0~K in a radio telescope observing at a wavelength of
7.35~cm.  At the same time, Robert Dicke at Princeton was promoting
the construction of a specialized telescope to detect the CMB.  It was
Dicke who first argued that the excess background observed by Penzias
and Wilson was a relic of the Big Bang (\cite{PandW}, \cite{Dicke}).
Various theories were proposed in the late 1960's to explain the
measured signal without requiring a Big Bang, though none could
account for an isotropic background with a purely thermal spectrum.

\subsection{Spectral Measurements}

Measurements of CMB intensity in other spectral bands quickly
followed.  A Princeton experiment already in progress measured a
blackbody equivalent temperature of 3.0\err 0.5~K at 3.2~cm
(\cite{RollW}).  In total, over a dozen consistent measurements of the
CMB temperature were published in the late 1960's over a range of
wavelengths from 0.33~cm to 73.5~cm (90~GHz to 0.41~GHz).
Observational efforts were somewhat reduced in the 1970's and rejoined
in the 1980's yielding increasingly convincing evidence for a precise
blackbody spectrum (\cite{smoot87}, \cite{Sironi91}).  The limitation
of these measurements, made with radio telescopes, was their inability
to measure the high frequency Wien region of the CMB spectrum.

Higher frequency measurements were not successful until the end of the
1970's, when bolometric receivers on balloon-borne and rocket-borne platforms
were used to overcome the problem of atmospheric emission.  The first
experiment to confirm the expected reduction of power in the Wien tail
of the CMB spectrum was conducted by Paul Richards and Dave Woody of
UC Berkeley.  The experiment, a balloon-borne Michelson interferometer
with a bolometric detector, constrained CMB power over a range of 75~GHz to
720~GHz (4~mm to 0.4~mm) giving clear evidence of a peak in the frequency
spectrum (\cite{WoodyR}).  These measurements were further refined
over the next decade by several groups (\cite{Gush81},
\cite{Peterson85}, \cite{Matsumoto88}).

In the early 1990's, the \firas\ instrument on the \cobe\ satellite
provided a definitive measurement of the spectrum over the range of
5~GHz to 500~GHz (1.0~cm to 0.1~mm).  These data reliably disprove any
significant overall distortion of the CMB from a purely thermal
spectrum (\cite{MatherSpectrum2}).

\subsection{Anisotropy Measurements}

\begin{figure}[ht]
\centerline{\epsfig{width=2.5in,angle=0,file=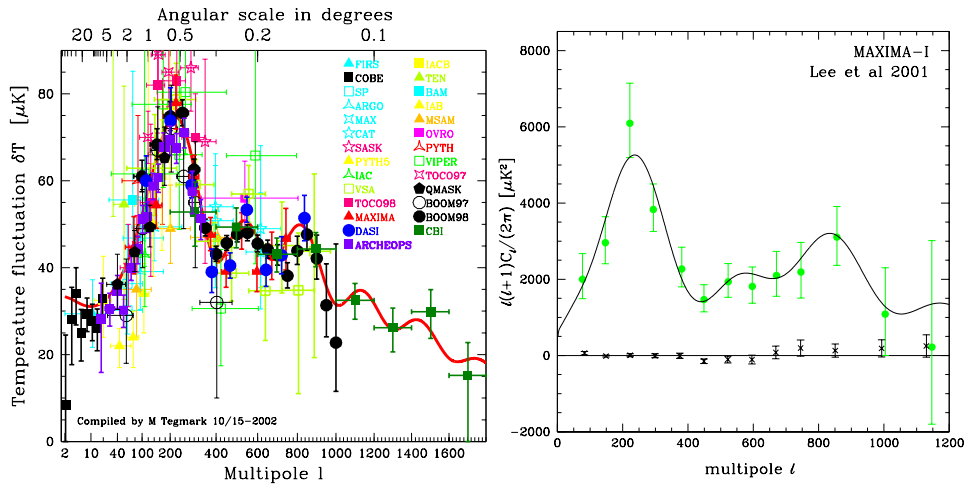}
\epsfig{width=2.5in,angle=0,file=figures/color/max3arcmtop.eps}}
\caption[Current Measurements of CMB Anisotropy]{{\bf{Left:}} A
composite of CMB anisotropy measurements from over two dozen data sets
including \maxima, \boomerang, \dasi, \cbi, \vsa, and Archeops (Max Tegmark 10/02).
{\bf{Right:}} The measurement from the \maximai\ alone, along with the
best fit model curve.  Points scattered around zero are difference
data from subtracted maps of independent detectors.}
\label{fig:curbest}
\end{figure}

The original Penzias and Wilson measurement sets an upper limit of
\sima 20\% on CMB anisotropy.  This was quickly improved to \sima
$10^{-3}$ by the use of differential instruments sensitive only to the
variations in the background (\cite{PartridgeW}).  Continuing
improvements led to the detection of the CMB dipole and eventually the
intrinsic anisotropy.

In the late 1970's, the dipole was measured using radiometers on
balloons and high altitude aircraft (e.g. \cite{DipoleSGM}).  Over the
next decade, several groups obtained increasingly precise
measurements; the results from the mid-1980's are quite similar to the
current best value, 3.358\err 0.023~mK from the \dmr\ instrument of the
\cobe\ satellite (\cite{dmr}, \cite{Dipole}).

Smaller scale anisotropy was not detected until much later.  By the
mid-1970's an upper limit of \sima $10^{-3}$ was established for
anisotropy on angular scales as small as 1\mina , primarily by ground
based observations.  The next generation of experiments included high
altitude observations and early bolometric receivers. By the
mid-1980's, the upper limit on small scale anisotropy was well below
$10^{-4}$.

In the early 1990's, two balloon experiments including \max\ (the
predecessor to \maxima) gave statistical detections of CMB anisotropy
at the $10^{-5}$ level ({\cite{MAX2}, \cite{Meyer91}, \cite{Alsop}).
Shortly thereafter, the \cobe\ \dmr\ provided a completely unambiguous
detection of anisotropy at angular scales of 7\degs and higher (\cite{Gorski}).

From the early 1990's to the present, anisotropy measurements have
pushed to increasingly small angular scales and increasingly large
angular dynamic range.  Data from the ground-based experiments Saskatoon 
(\cite{Sask}) and \cat\ (\cite{CAT}) together provided the first evidence of an acoustic peak near $\ell
=220$ in the angular power spectrum.  At present, the balloon
experiments \maxima\ (\cite{HananyResults}, \cite{LeeResults}),
\boomerang\ (\cite{BOOM}), and Archeops (\cite{Archeops}) and ground-based interferometers \dasi\
(\cite{DASI}), \cbi\ (\cite{CBI}), and \vsa\ (\cite{VSA}) have provided
consistent measurements of the CMB power spectrum, including high
signal-to-noise measurements of the first acoustic peak.  While no
single experiment has so far measured higher acoustic peaks with high
signal-to-noise, the combined data strongly suggest the existence of
at least two more peaks.

Strong upper limits on polarization anisotropy have been obtained with
the \pique\ (\cite{PIQUE}) and \polar\ (\cite{POLAR}) experiments and,
very recently, the \dasi\ experiment is believed to have made a
detection (\cite{DASIpol}).  Study of polarization anisotropy is an
active field of CMB research, and is the aim of \maxipol, the follow-up
to \maxima\ (Chapter~\ref{chap:future}).

\section{Technical Considerations}\label{motiv:obser}

Experimental efforts to tap the enormous potential of the CMB have
yielded great results since the early 1990's.  The recent rapid
progress of the field owes both to improved detector technologies and
to a strong commitment by the observational community.

The small size of CMB anisotropy compared to astronomical foregrounds,
side-lobe sources, atmospheric emission, and the background loading of
the CMB itself presents a serious challenge.  Time domain noise
correlations are problematic for data analysis and require a carefully
planned scan strategy.  A further complication has been the relatively
late development of detector technologies in the optimum frequency
range of 20~GHz to 300~GHz.

\subsection{Optical Signals}

CMB anisotropy is much smaller than optical backgrounds and parasitic
signals.  These unwanted signals are of three general types: constant
loading, variable but non-sky stationary parasitics, and sky
stationary parasitics (foregrounds).  Stable optical backgrounds,
referred to as loading, come from the CMB, the atmosphere, and the
telescope.  For a properly optimized detector system, loading
contributes purely statistical photon counting noise
(\S \ref{bolos:response}).

Unstable but non-sky stationary signals come from side-lobe sources such
as the Sun, Moon, and Earth and from variations in atmospheric or
instrumental loading.  Small contributions of this type can be
acceptable given a number of observations of each sky region; over
repeated observations, their effects will tend to cancel out.  Large
signals, and those that are not purely uncorrelated with CMB, are a
major problem (\S \ref{sys:concerns}).

Sky stationary parasitic signals, referred to as foregrounds, are of
astronomical origin (e.g. Galactic dust and radio point sources).
Foreground contamination can only be controlled by observing regions
of the sky in which it is small and/or well understood (\ie~sky
selection), by observing at optical frequencies with lower foreground
sensitivity, and by spectral discrimination (\S \ref{sys:foreg}).

\subsection{Detector Technologies}

Despite rapid advances made in the past decade, detectors in the
20~GHz to 300~GHz range are not yet fully developed.  Continuing
improvements in detector technology have enabled tremendous progress
in CMB cosmology and will continue to do so in the future.
Presently, two technologies are widely used by CMB experiments:
bolometers, which are used in \maxima, and high electron mobility
transistors (HEMTS).

Bolometers are total power square law detectors, best suited to
observations at optical frequencies of 90~GHz and higher.  Some
bolometers, as in \maxima, are coupled to incident light via thin
metal films, for large optical bandwidth and polarization independent
sensitivity.  Antenna coupled bolometers are being developed for faster response
times and polarization discrimination.  The chief advantage of
bolometers is their high sensitivity; single \maxima\ detectors have
achieved noise equivalent temperatures of less than 100~\micro
K~$\sqrt{sec}$, which is comparable to the photon noise limit.  In
practice, bolometric experiments have also benefited from the relative
small effect of extragalactic point sources at their operating frequencies.

Bolometer technology is in a period of rapid development.  In the
early 1990's bolometers were often hand made, suffered significant
performance variations between devices, and were rarely used in
arrays.  \maxima-era bolometers are made by a combination of
photolithography and manual construction, and are typically operated
in small arrays.  Recent work has focused on the creation of arrays of
hundreds of well matched bolometers with little or no manual
construction.  The success of these efforts is critical for future
challenges such as the measurement of CMB polarization anisotropy.

The main disadvantage of bolometers is their operational complexity.
Bolometers require extremely low operating temperatures (100~mK for
\maxima), attainable only with sophisticated cryogenic systems and
they are relatively sensitive to microphonic noise.  In addition,
since bolometers are total power sensors, they are insensitive to the
phase of incident radiation and cannot be used for interferometry.
Finally, the atmosphere is relatively emissive at bolometric
frequencies, making ground-based observations difficult.

HEMTs are coherent amplifiers best suited to observations at
frequencies below 90~GHz.  They inherently preserve phase and
polarization information.  Modern HEMT based experiments usually take
advantage this phase sensitivity for interferometry.  HEMTs operate
optimally at temperatures of tens of Kelvins, which can be readily
achieved and maintained with simple cryogenic systems.  At the lower
optical frequencies of HEMT-based systems, atmospheric emission is
relatively weak, and most HEMT systems are ground-based.

The main disadvantage of HEMT based experiments is lower
sensitivity caused by their quantum noise limit at CMB frequencies and their relatively small optical bandwidth.  A
secondary disadvantage is the strong influence of radio point sources
at their lower operating frequencies.

\subsection{Noise Correlations}

A common feature of all CMB experiments is low time stream
signal-to-noise combined with noise correlations.  Time domain noise
correlations are caused by detector time constants and by drifts in
detector temperatures or amplifier gains.  Correlations also result
from the instrumental high pass filtering often used by CMB
experiments to suppress backgrounds.  Because of correlations, CMB
temperature at a given location on the sky cannot be accurately
determined by simply averaging the detector output at that position.

The effects of temporal correlations are minimized by an observation
pattern that revisits of each position on the sky over several
uncorrelated time scales and along a variety of spatial trajectories
(cross-linking) as discussed in Section~\ref{point:scans}.

Even with an excellent scan strategy, noise correlations must be
analyzed explicitly (\S \ref{data:noise} to
\S \ref{data:ps}); this is the reason for the computational
intensity of CMB data analysis.

\chapter{The MAXIMA Experiment}\label{chap:exper}

		This chapter describes the \maxima\ experiment.  A
brief introduction (\S \ref{exper:expover}) and an outline of the
scientific goals and approach (\S \ref{exper:goals}) are followed by
discussions of the telescope and optics (\S \ref{exper:teles}), the
detectors (\S \ref{exper:bolos}), and the cryogenic receiver (\S
\ref{exper:receiv}).

	Pointing is discussed in detail in Chapter~\ref{chap:point}.
More detailed discussions of optics and detectors can be found in
\cite{CDWThesis}.  \cite{LeeRome} also describes the instrument.

\section{Introduction}\label{exper:expover}

	\maxima\ is a balloon-borne telescope designed to measure the
anisotropy of the CMB over a wide range of angular scales ($\ell =$ 35
to 1500).  Over the course of two flights, in 1998 and 1999, high
resolution observations have been made of 300~deg$^2$ of the sky.
Results have been released (\cite{LeeResults}, \cite{HananyResults})
and cosmological implications have been explored by members of the
\maxima\ collaboration (\cite{StomporResults}, \cite{BalbiResults}) as
well as third parties.

	The experiment is based at the University of California,
Berkeley, and includes collaborators from the University of Minnesota
and worldwide (Appendix~\ref{chap:collab}).

\section{Goals and Design}\label{exper:goals}

	The primary scientific objectives of \maxima\ are to
distinguish between models of cosmological structure formation, and to
measure parameters within these models.  The $\ell$-space coverage and
resolution of the experiment are well suited for measurement of the
first three acoustic peaks of inflationary models.  Measurements in
this region are a powerful tool for testing the general predictions of
inflation and for parameter estimation (Chapter~\ref{chap:motiv},
\cite{HuDodelson}).

	In addition, \maxima\ data have been a realistic test of
analysis methods and tools (\cite{StomporData}).  Treatments have been
developed for problems such as mild beam asymmetry (\cite{WuBeam}),
foreground discrimination, scan synchronous noise, and detection of
spatial non-gaussianity (\cite{WuGauss}).

	\maxima\ has been used to test new technologies.  In
particular, \maxima\ is the first CMB experiment to have used 100-mK
spiderweb bolometers, similar to those planned for the Planck
Surveyor.  The combination of these detectors
(\S \ref{exper:bolos}, \cite{BockBolo}) and an adiabatic
demagnetization cooling system has provided instrumental sensitivity
of \sima 40~\micro K~$\sqrt{sec}$, the best reported by any CMB
experiment.

\subsubsection{Experimental Concept}

	\maxima\ is a bolometric instrument, making CMB observations
at relatively high optical frequencies (150~GHz and higher).  In order
to avoid atmospheric emission, observations are made from an altitude
of \sima 40~km during multiple balloon flights.  The relatively short
duration of the balloon-borne observations is offset by the use of a
16-element array of single color photometers with extremely sensitive
detectors.  Good angular resolution (10\mina ) and large sky coverage
make the experiment sensitive over a wide range of angular scales.
The compact and well cross-linked scan pattern is optimized for
extracting the angular power spectrum.  The use of three spectral
bands allows discrimination between the CMB and foreground sources.
\maxima\ benefits from exceptionally precise pointing reconstruction
(1\mina ) and accurate calibration (4\%).  The instrument is designed
to consistently survive balloon flights, and has been successfully
recovered after a test flight and two science flights.

	A bolometric receiver, including cooled optics, and an
external primary mirror are mounted on a two axis attitude controlled
frame.  Data are collected during balloon flights lasting one night.
There have been two such flights providing 3~hours and 8~hours
respectively at altitudes of 36~km to 39~km.  Each flight consists of
a pair of cross-linked CMB observations covering areas of 122~deg$^2$
and 225~deg$^2$ with a signal-to-noise of \sima 5 at the 10\mins beam
size.  Each flight also includes calibrations from the CMB dipole and
from a planet.  More details about the flights are available in
Chapter~\ref{chap:flights}.

	The detector system is a high sensitivity array of spiderweb
bolometers cooled to 100 mK by means of adiabatic demagnetization (\S
\ref{receiv:cryo}).  The array consists of 16 single color pixels,
each projected onto the sky with a 10\mins FWHM beam size.  Eight
detectors observe at 150~GHz, four at 230~GHz, and four at 410~GHz.
Optical bandwidths are 45~GHz, 65~GHz, and 35~GHz respectively.  The
\maxima\ array provides a combined sensitivity of \sima 40~\micro
K~$\sqrt{sec}$ (See Appendix~\ref{chap:NET} for single detector
sensitivities).  Overall detector responsivity calibration
(Chapter~\ref{chap:calib}) is obtained from the CMB dipole with an
accuracy about 4\%.

	The telescope is an off-axis Gregorian system with a 1.3-m
diameter primary mirror providing a 10\mins beam size (FWHM) for all
detectors (\S \ref{exper:teles}).  Due to fluctuating atmospheric emission at
our observing wavelengths, the telescope is mounted on a high-altitude
balloon-borne frame.

	The primary attitude sensor is a boresighted CCD star camera.
Pointing reconstruction is accurate to \sima 1\mins for CMB
observation.  Requirements on the selection of observing regions, scan
pattern, cross-linking, and scan speed are presented in
Section~\ref{point:requirements}.  A key element of the \maxima\
observation strategy is the active modulation of the primary mirror,
which moves the telescope beams \err 2\degs at 0.45 Hz.

\begin{figure}[ht]
\centerline{\epsfig{width=3.0in,angle=0,file=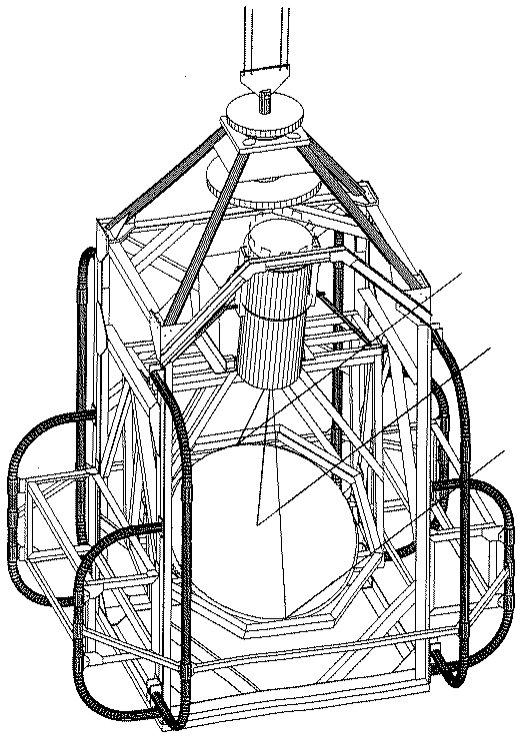}}
\caption[\maxima\ Telescope]{A mechanical drawing of the \maxima\
telescope from an elevated front/side perspective.  Rays representing
the telescope beam are shown reflecting from the primary mirror into
the cryogenic receiver.  Electronics housed in the rectangular boxes
on the sides of the instrument include the pointing system, data
multiplexers and digitizers, and telemetry and command interfaces.
Near the top of the telescope are motors controlling azimuthal
orientation.  The inner assembly consisting of the primary mirror and
the receiver is tilted relative to the outer frame to aim the
telescope in elevation.}

\label{fig:gon_wframe}
\end{figure}

\subsubsection{MAXIMA Data Sets}

	Data were collected in August 1998 (\maximai) and June 1999
(\maximaii).  The \maximai\ data have been analyzed to produce a
40,000-pixel map of 122~deg$^2$ of the cosmic microwave background and
this map has been used to estimate the angular power spectrum.

	The lowest multipole bin (largest angular scales) measured by
\maxima\ spans $\ell$=36 to $\ell$=110.  Two factors limit our ability
to measure anisotropy at the largest scales: sky coverage and low
frequency noise.  The highest multipole bin (smallest angular scales)
measured spans $\ell$=1086 to $\ell$=1235.  Limiting factors in our
high $\ell$ measurements are: beam size and characterization,
scan speed and detector time constants, pointing reconstruction,
and integration time per pixel.

\section{Telescope and Optical System}\label{exper:teles}

\subsection{Optics}\label{teles:optics}

\begin{figure}[hp]
\centerline{\epsfig{width=5.5in,angle=270,file=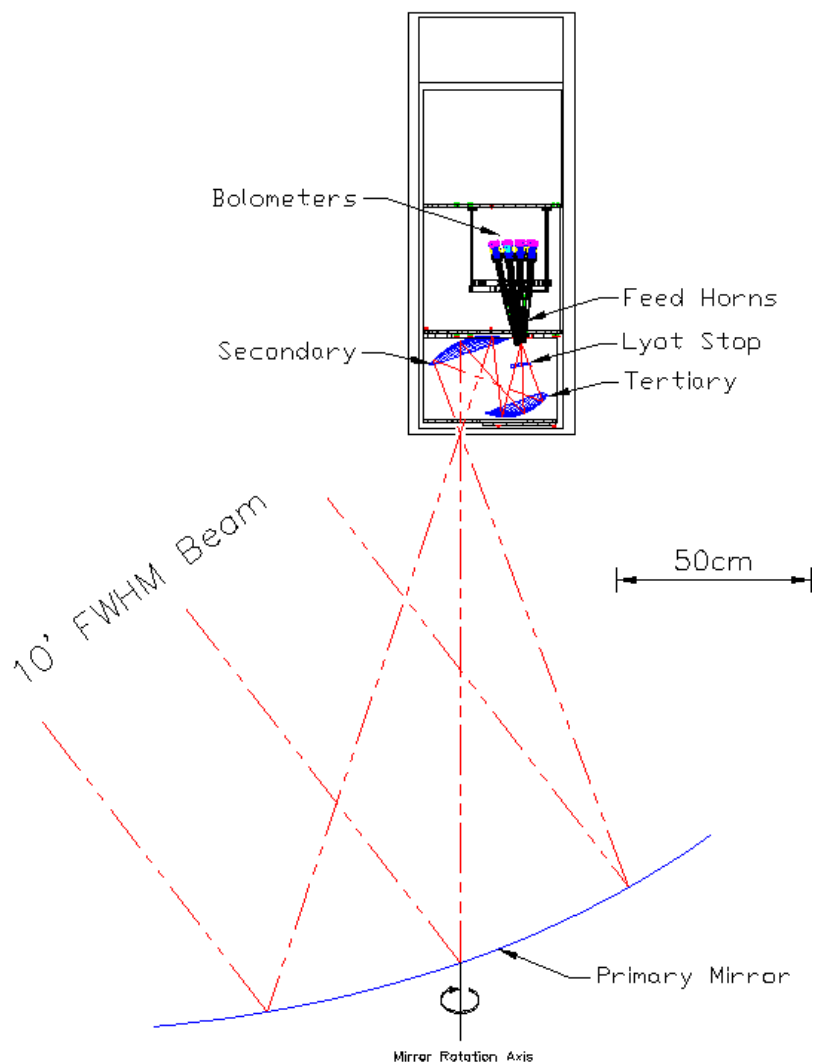}}
\caption[Optical System]{The optical system of the \maxima\ telescope.
The telescope is a fast (f/1) Gregorian system, with a prime focus at
the window of the cryogenic receiver.  The primary mirror, a 1.3-m
diameter underilluminated paraboloid at ambient temperature, is modulated
about the indicated axis.  The secondary and tertiary reimaging
mirrors correct aberrations from the primary.  A 4-K Lyot stop defines
the illumination of the primary.  An array of feed horns channels
light to the bolometers, which are held in resonant cavities.  Optical
filters are located at the prime focus, the Lyot stop, and after the
feed horns.  All detector channels have a beam size of 10\mins FWHM.}
\label{fig:optics}
\end{figure}

\begin{figure}[ht]
\centerline{\epsfig{width=2.5in,angle=0,file=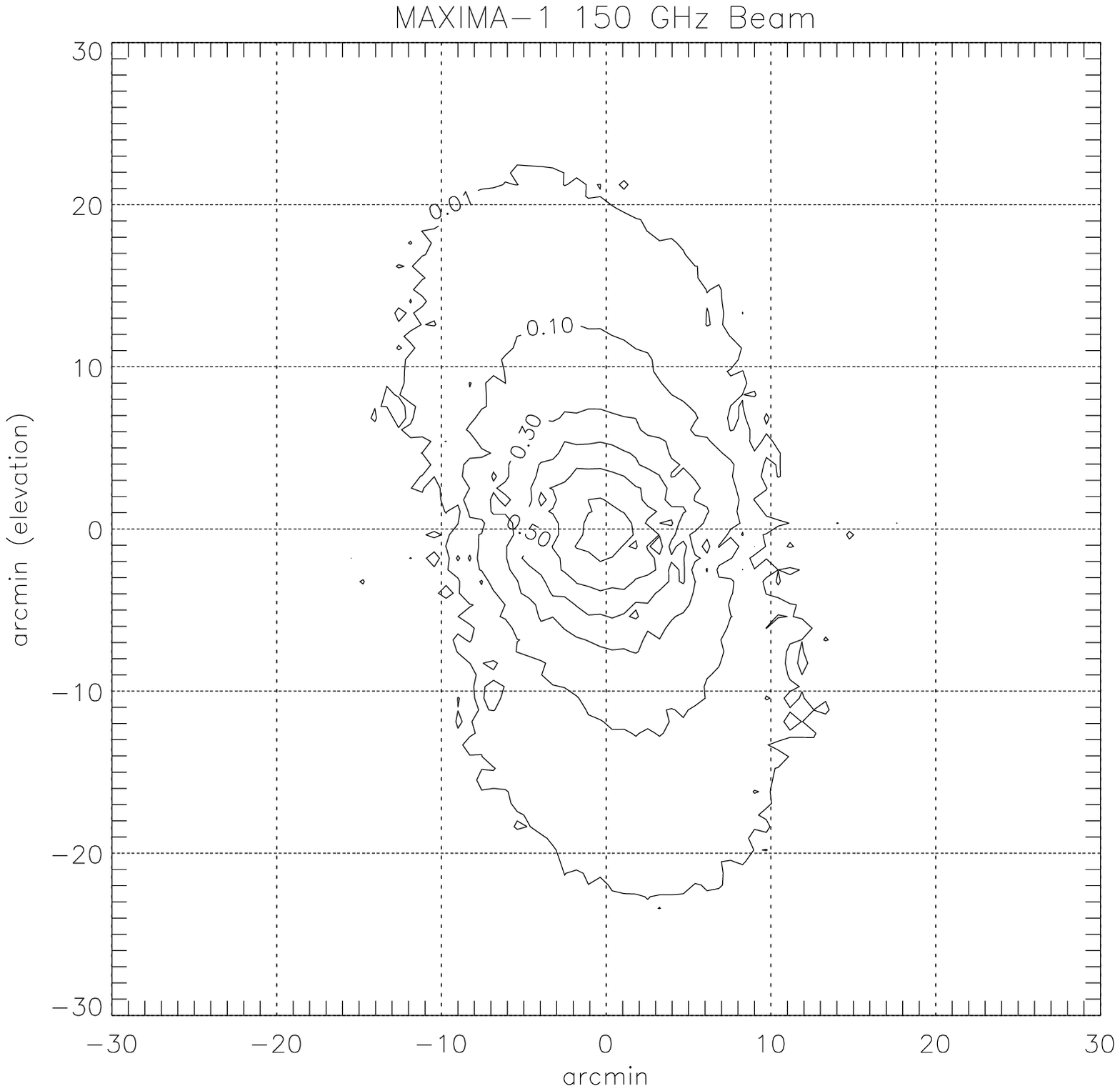}
\epsfig{width=2.5in,angle=0,file=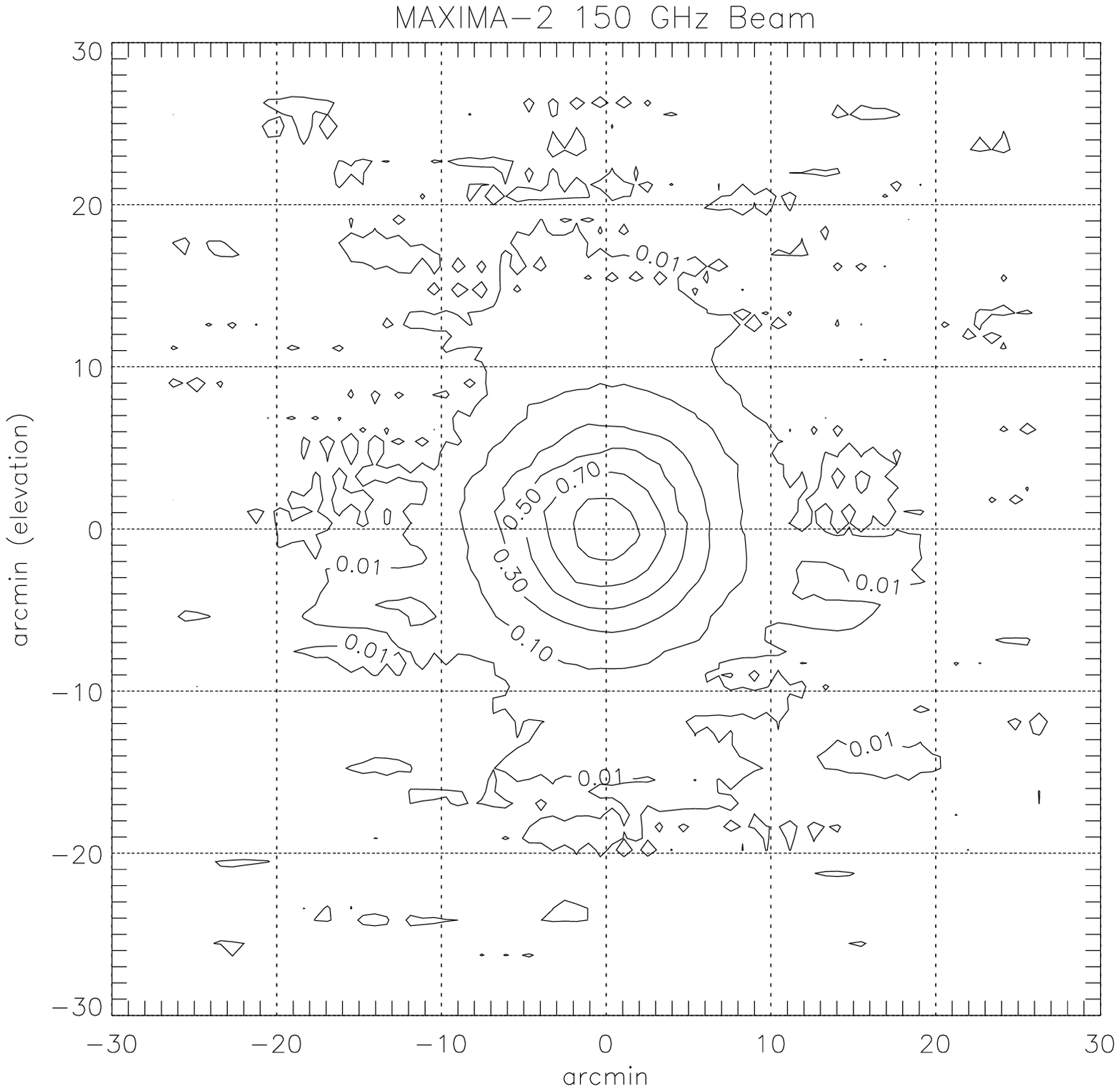}}
\caption[Beam Maps]{Beam maps of the same 150-GHz detector in \maximai\
(left) and \maximaii\ (right).  This is flight data from observations of
Jupiter (\maximai) and Mars (\maximaii).  Contours shown are 90\%,
70\%, 50\%, 30\%, 10\%, and 1\% of the maximum response.  The 1\%
contour in the \maximaii\ beam map is noisy because of the relatively
low intensity of Mars.  Beams were more symmetric and gaussian in
\maximaii, due to better telescope focusing.
\cite{WuBeam} shows that the beams in both cases can be approximated
as symmetrical.}
\label{fig:beams}
\end{figure}

		The \maxima\ telescope is a fast (f/1), off-axis
Gregorian system consisting of a 300-K primary mirror and two cold
reimaging mirrors.  The primary mirror, produced by Dornier
Satellitensysteme, is an underilluminated 1.3-m off-axis paraboloid
constructed from a lightweight honeycomb material.  The mirror is
actively modulated during observations.  The prime focus is located
near the window of the cryogenic receiver.  Two cold mirrors inside
the receiver are off-axis ellipsoids with corrections to compensate
for aberrations from the primary.  A cold lyot stop between the tertiary
mirror and the focal plane helps to define the beams and strongly
suppress telescope side-lobes.

		The cold optics (21-cm secondary mirror, 18-cm
tertiary mirror, and lyot stop) are contained within a heavily
baffled, liquid He-cooled optics box.  The optics box maintains a
temperature of approximately 3~K during flight.  All non-optical
surfaces inside the optics box are coated in a combination of Stycast
epoxy\footnote{A filled epoxy produced by Emerson and Cuming.}, carbon
black powder, and glass beads.  The resulting material absorbs stray
far infrared radiation with high efficiency (\cite{BockThesis}),
further reducing the potential for side-lobe response.

\begin{figure}[ht]
\centerline{\epsfig{width=2.5in,angle=0,file=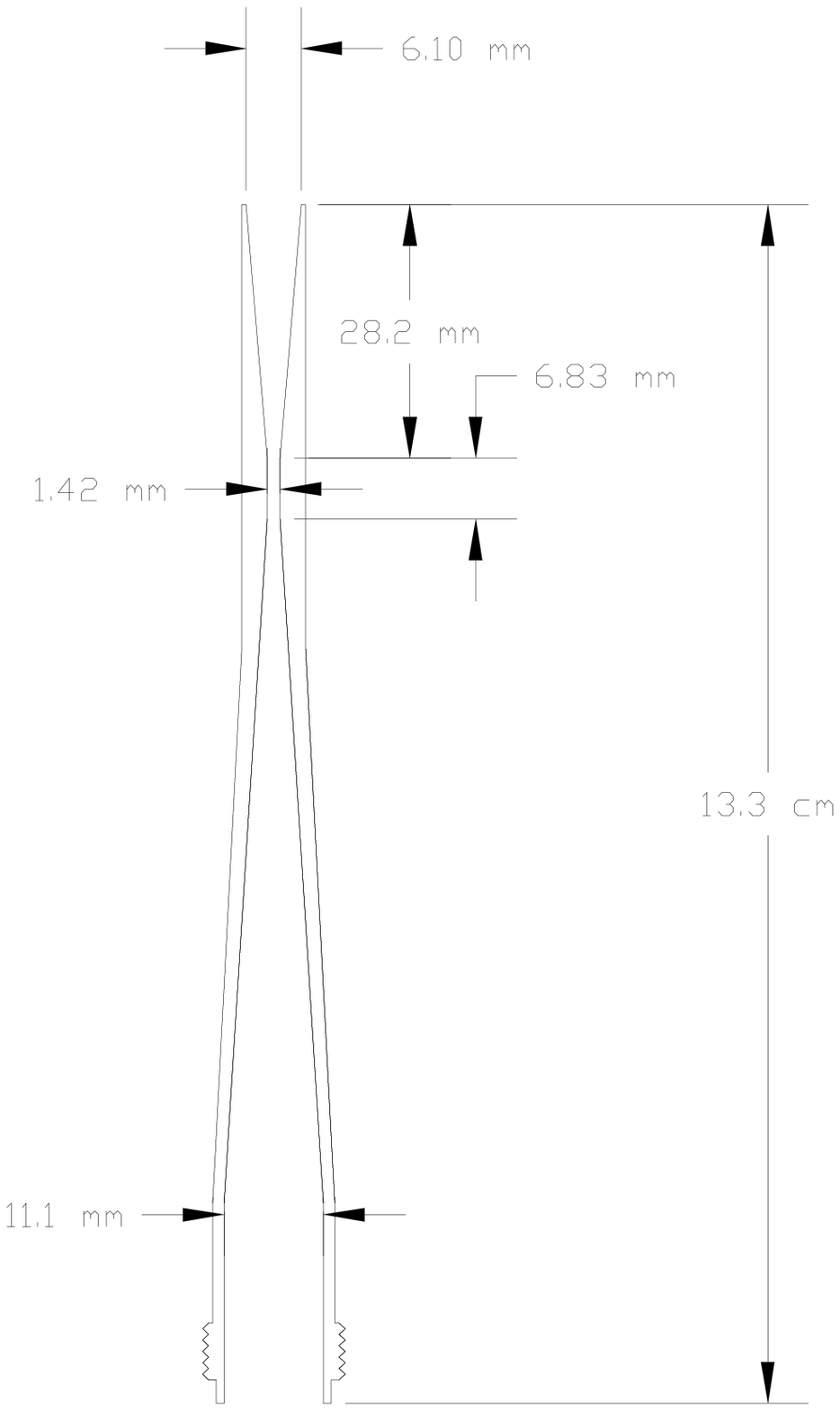}
\epsfig{width=2.5in,angle=0,file=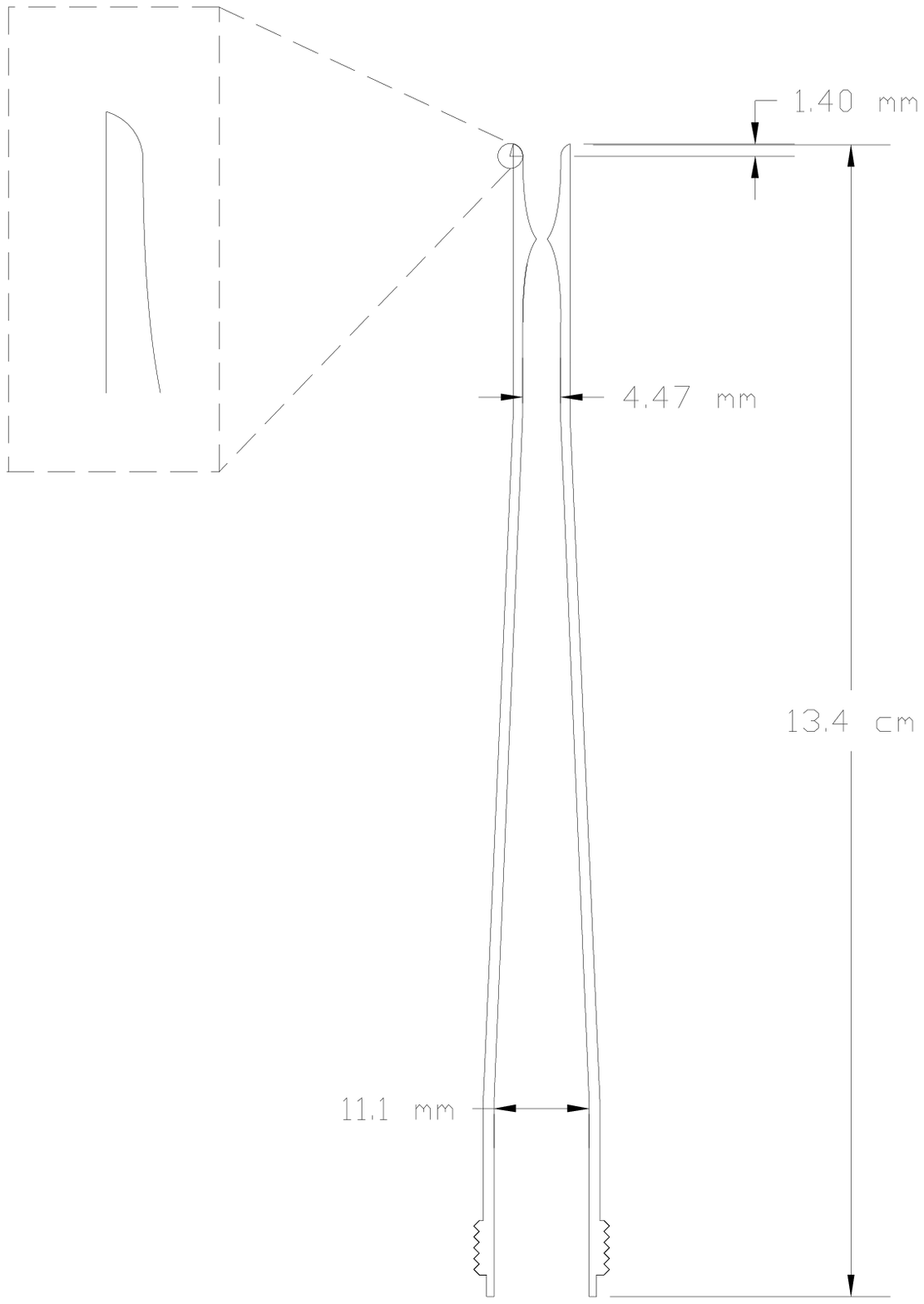}}
\caption[Bolometer Feedhorns]{Bolometer Feedhorns.  The narrow
openings (top) are at the focal plane.  The bases are screwed into a
liquid $^4$He-cooled plate.  Radiation leaving the horns passes into 100-mK
assemblies holding band-defining optical filters and the bolometers.
{\bf{Left:}} A straight-walled, singled-moded 150-GHz feedhorn.  A
conical section (4.7\degs flare) leads from the focal plane to a
1.4-mm diameter straight waveguide.  A second conical section
(3.7\degs flare) leads to the exit of the feedhorn.  {\bf{Right:}} A
multi-moded feedhorn for the 230-GHz and 410-GHz channels.  The entrance
of the horn consists of back-to-back Winston cones.  These are coupled
to a conical section (2.4\degs flare) leading to the exit of the
horn.}
\label{fig:horns}
\end{figure}

	The focal plane array consists of the entrances of 16 copper
feedhorns.  The feedhorns for the 150-GHz detectors are single-moded
and consist of back-to-back straight cones connected by a length of
cylindrical waveguide (Figure~\ref{fig:horns}, left).  The feedhorns
for the 230-GHz and 410-GHz detectors are multi-moded and consist of
back-to-back Winston horns coupled to a straight cone
(Figure~\ref{fig:horns}, right).  Both types of feedhorn end in a
straight cylinder, screwed into a liquid $^4$He-cooled plate.  Light
exiting a horn passes through a 0.5-mm gap before entering a 100-mK
cylindrical waveguide holding optical filters and terminating in the
bolometer cavities.

		A neutral density filter (NDF) can be
inserted into the optical path between the secondary and tertiary
filters.  The NDF has a transmittance of 1\% and is used to simulate
in-flight optical loading during ground tests.

		The main lobes of all telescope beams were measured in
flight by observing the planets Jupiter and Mars.  Planets are
effectively point sources for our beam size and are detected with
signal-to-noise ratios of 100 to 1000 (Chapter~\ref{chap:calib}).

		The 1.3-m primary mirror is continuously modulated
about the optic axis (Figure~\ref{fig:optics}) in a rounded triangle
wave pattern with an amplitude of 4\degs at a frequency of 0.45~Hz.
Section~\ref{chopper} describes the primary modulation and its role in
the \maxima\ scan pattern.

\subsubsection{Far Side Lobes}

	Spurious signals from bright sources outside the main lobe of
our telescope beams are a potential source of systematic errors.  The
internal baffling of the cold optics and the beam-defining lyot stop
strongly suppresses side-lobe response.  In addition, the outside of
the telescope is heavily baffled.  This baffling is particularly
effective at low elevation angles, blocking ground emission.  Side
lobe measurements are discussed in Section~\ref{sys:sidelobe}.

\subsection{Spectral Bands}\label{bands}

	Bolometers are broad-band detectors.  Optical filters are used
to define spectral bands, which are chosen for atmospheric
transparency and foreground discrimination.  Band-defining mesh
filters for each channel are located in light pipes before the
bolometer cavities.  These are cooled by the ADR to 100~mK.  Each of
the 230-GHz and 410-GHz detectors uses two filters for band
definition: a capacitive lowpass filter and an inductive highpass
filter.  The 150-GHz detectors use only a lowpass filter; the lower
edge of this band is defined by the size of the feedhorn.  Sample
spectra from each band are shown in Figure~\ref{fig:spectra}.

\begin{figure}[ht]
\centerline{\epsfig{width=3.0in,angle=90,file=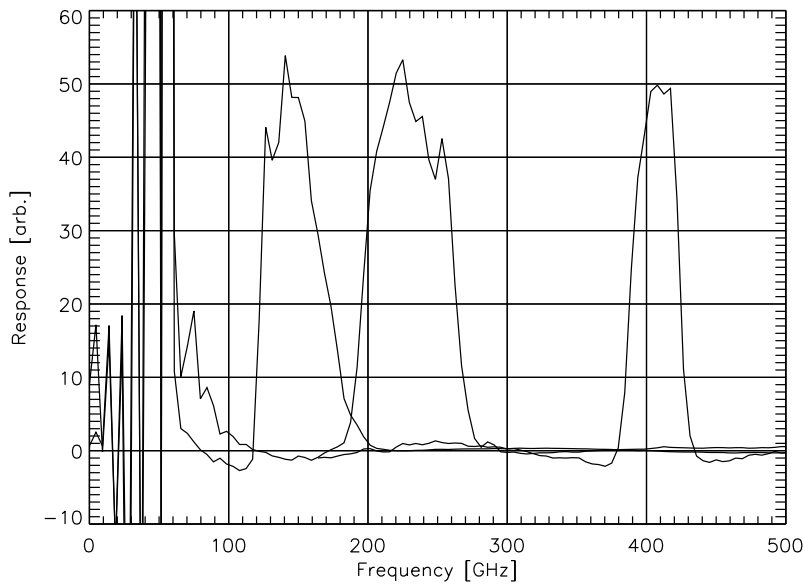}}
\caption[Spectral Bands]{The measured spectral response of the
\maxima\ detectors.  The peaks at 150~GHz, 230~GHz, and 410~GHz are
measurements of sample detectors in each of our three observing
frequencies.  The three spectra are overlaid for comparison.  The
spectrum of the CMB peaks within the 150-GHz band.  The measurement
noise at low frequency does not represent real spectral leakage.}
\label{fig:spectra}
\end{figure}

	Three lowpass filters are placed before the focal plane.
These serve to reduce the optical load on the 100-mK stage and to
block high frequency resonant leaks in the band-defining filters.  Two
of the lowpass filters are located near the prime focus, just inside
the cryostat window.  The first (closer to the sky) is a reflecting
capacitive mesh, with a cutoff at 18~cm$^{-1}$ (540~GHz).  The second
is an absorptive alkali halide filter with a cutoff at 55~cm$^{-1}$
(1650~GHz).  A third lowpass filter, located at the lyot stop, is a
capacitive mesh with a cutoff at 16~cm$^{-1}$ (480~GHz).  The first
reflective filter is cooled to 77~K; the other two filters are cooled
to 4~K.

		The optical filters used in \maxima\ have been
constructed at QMW and Cardiff.

\subsubsection{Metal Mesh Filters}

	Metallic mesh interference filters provide excellent out of
band rejection and in band transmission at far infrared wavelengths.
All of the band-defining filters and two of the initial lowpass
filters are of this type.  The theory of these filters has been widely
studied (\cite{Ulrich}, \cite{Irwin}).  Mesh filters of the kind used
in \maxima\ are described in detail in \cite{LeeThesis}.

	The filters consist of dielectric substrates (1.5-\micro\ taut
Mylar) with thin metallic mesh coatings (0.2-\micro\ copper).  The
metal is etched in a repetitive pattern by photolithography.  The
periodicity of the mesh is smaller than the radiation wavelength.  The
mesh thickness is negligible.  The spaces between the metal mesh can
be treated as transmission lines; different mesh spacings and
geometeries correspond to different equivalent oscillatory circuits as
found by \cite{Ulrich}.  Each filter consists of several layers of
substrate and mesh, separated by spacing rings, pressed together, and
glued.

	Induced currents in the mesh give rise to reflected and
transmitted waves.  Absorption caused by ohmic losses in the mesh or
by dielectric effects in the thin mylar are of order $10^{-3}$ or
less.

\subsection{Multicolor Array}

\begin{figure}[ht]
\centerline{\epsfig{width=3.0in,angle=270,file=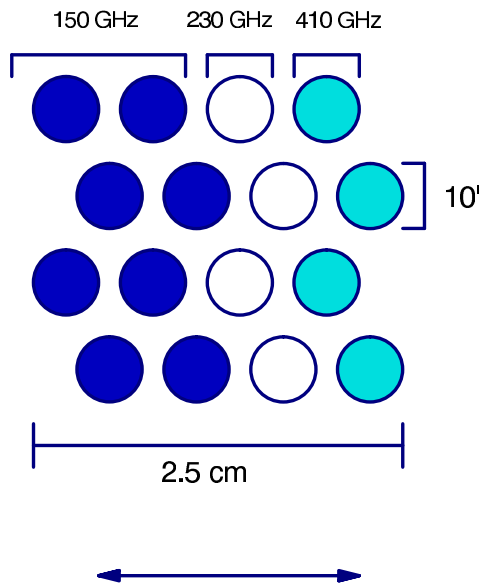}}
\caption[Focal Plane Array]{The layout of the \maxima\ focal plane.
The arrows indicate the scan direction (azimuth modulation at constant
elevation).  All 16 channels project onto the sky with a 10\mins FWHM
beam-size.}
\label{fig:focal}
\end{figure}

		The \maxima\ focal plane is located at the exit of the
optics box and consists of the entrances to the 16 single color
feedhorns.  Each feedhorn channels light through band-defining optical
filters to a bolometer.  The focal plane is laid out in four rows of
four pixels; each row is projected onto a constant elevation on the
sky, and consists of two detectors at 150~GHz, one at 230~GHz, one at
410~GHz (Figure~\ref{fig:focal}).

\section{Bolometers}\label{exper:bolos}

		\maxima\ uses an array of high sensitivity bolometers
fabricated at JPL (\cite{BockBolo}).  Here we present a very brief
review of the important concepts in bolometric detection.

		Bolometers are incoherent square law detectors most
often used in submillimeter and far infrared applications.  Their main
advantages are high optical bandwidth (30\% in \maxima) and low noise.
The primary disadvantages are their sensitivity to microphonic noise
and their low operating temperature; the latter is the reason for the
use of an adiabatic demagnetization refrigerator in \maxima\
(\S \ref{cryo:friges}).  Bolometers are the best detectors for
observations of the CMB at frequencies of 90~GHz and higher.

\begin{figure}[ht]
\centerline{\epsfig{width=3.25in,angle=270,file=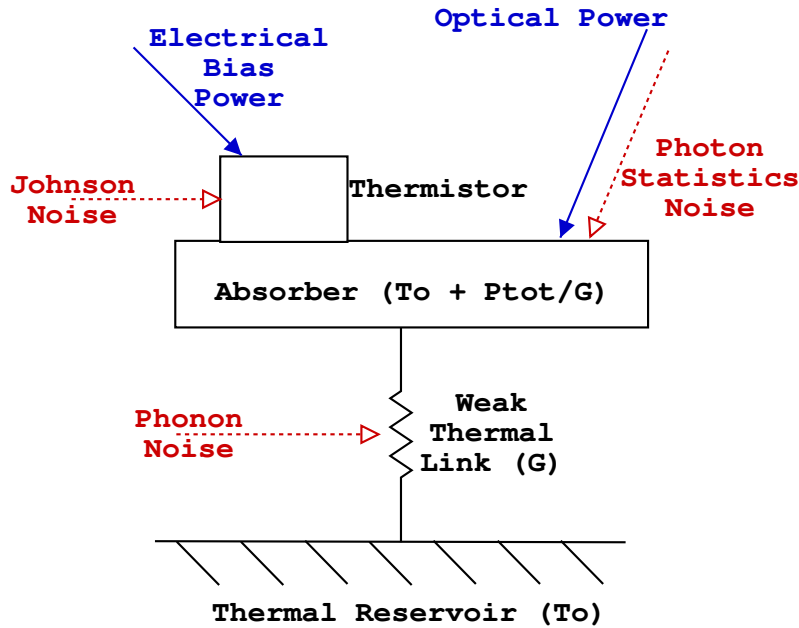}}
\caption[Bolometer Power and Noise]{A schematic of bolometer power and
noise inputs.  An absorber coupled bolometer consists of an absorber
and an electrically biased thermistor, and a weak thermal link, $G$,
to a thermal reservoir at a fixed temperature, $T_o$.  Power inputs
are shown in blue (solid arrows).  $P_{tot}$ is the total optical and
electrical power.  Noise sources are shown in red (dotted arrows).
Amplifier noise is not introduced in the bolometer itself and is not
represented here.}
\label{fig:bolofig}
\end{figure}

	In general, bolometers consist of an optical coupling, such as
a metal film or antenna, and an electrically biased detector element,
such as a semiconductor thermistor or a superconductor near a
transition edge.  Photons absorbed by the coupling deposit energy in
the detector element causing an electrical signal.  The bolometer is
weakly coupled to a thermal reservoir, $T_o$ by a weak thermal
conductance, $G$.  The bolometer temperature, $T_{bolo}$, is

\begin{equation}
T_{bolo} = T_o + \Delta T = T_o + \frac{P_{opt} + P_{bias}}{G}
,\end{equation}

\noindent where $P_{opt}$ and $P_{bias}$ are the optical and
electrical power inputs of the bolometer (Figure~\ref{fig:bolofig}).
In the simplest case, $P_{bias}$ and $G$ are nearly constant, and
$\Delta T_{bolo}$ varies linearly with $P_{opt}$.

	For a \maxima\ bolometer, the optical coupling is a metallic
absorber, impedance matched to free space, in a resonant (1/4
wavelength long) cavity.  The absorber is a layer of gold coated onto
a spiderweb shaped substrate of silicon-nitride.  The detector element is a
Neutron Transmutation Doped (NTD) germanium thermistor produced at
LBNL (\cite{NTD}).  The NTD is mounted at the center of the spiderweb
and is biased with a constant electrical current.  The thermal
reservoir is the 100-mK ADR.  The bolometer operates at \sima 130~mK.

	\maxima\ is the first experiment to use
microlithographed spiderweb absorbers at 100~mK.  This design provides fast
response time, a small cross-section to cosmic rays, and relatively
low microphonic response.

\subsection{Responsivity and Noise}\label{bolos:response}

	The theory of bolometer noise and responsivity optimization
is developed elsewhere (\cite{BoloReview}, \cite{SabrinaBolo}).
Spiderweb bolometers of the type used in \maxima\ are described in
\cite{BockBolo}.  \maxima\ bolometer optimization is found in
\cite{CDWThesis}.

	Overall sensitivity to CMB temperature fluctuations is
characterized by noise equivalent temperature (NET), with units
$[CMB$~$Temp.]$\ $[Time^{0.5}]$, given by,

\begin{equation}
NET = \frac{NEP}{S_{CMB}}
.\end{equation}

\noindent $S_{CMB}$ is responsivity to the CMB, defined by the change
in bolometer voltage per change in CMB temperature ($V\cdot T^{-1}$),
and $NEP$ is the noise equivalent power in the detector output voltage
($V\cdot sec^{0.5}$).  NET describes sensitivity according to,

\begin{equation}
\Delta T_{cmb} = \frac{NET}{\sqrt{t_{obs}}}
,\end{equation}

\noindent where $\Delta T_{cmb}$ is the noise in the measured temperature of
an observed region and $t_{obs}$ is the time spent observing that
region.  The NET generally varies with signal frequency, but single values
are often quoted, representing averages over a band of signal frequencies used in
the experiment.  The NET is often presented in units of
$[CMB$~$Temp.]$\ $[Freq.^{-0.5}]$, representing the noise density per
unit bandwidth.  In this form, the quantity is larger by a factor $\sqrt{2}$,
because 1 second of integration samples a bandwidth of 0.5~Hz.

	The white noise NET for a well optimized bolometer is improved
by reducing the bolometer operating temperature and reducing the
optical background loading.  The overall NET is a quadrature sum of
NETs from different noise sources: heat sink thermal fluctuation noise, Photon
counting noise, thermistor Johnson noise, and amplifier noise.

	Thermal fluctuation NET is given by $\sqrt{4kT^2G}$, where $T$ is the
bolometer temperature and $G$ is the thermal conductivity of the heat
sink.  Reduced thermal reservoir temperature ($T_o$) allows a properly
optimized bolometer to operate at lower temperature.
The dependence of thermal fluctuation noise NET on $T_o$ varies
with the type of thermal link between the detector and the thermal
reservoir.  $T_o^{1.5}$ holds for metals, while
$T_o^{2.5}$ holds for insulators and superconductors.  In
practice, the thermal link consists of multiple materials with
different scalings.  Reductions in optical loading allow roughly
linear reductions in $G$, which is directly proportional to thermal fluctuation
NET.

	Photon noise is the quantum statistical fluctuation in photon
flux.  Its importance is minimized by increasing the number of `signal' photons
and decreasing the total number of photons absorbed by the detector.
For a given signal size (\ie~optical band and optical
efficiency), photon NET decreases as at least the square root of
optical background loading.  Because the loading can not be reduced
below the CMB flux, photon NET represents the fundamental limit of
bolometer sensitivity in a given optical band.

	Johnson noise NET is given by $\sqrt{4kTR/|S|^2}$, where $T$ is
the bolometer temperature, $R$ is the electrical resistance of the bolometer,
and $S$ is the optical responsivity.
Electrical resistance is matched to the input impedance of the amplifier.
The Johnson noise NET is reduced by
decreasing temperature and increasing responsivity.
Assuming properly
optimized bolometers, responsivity increases as $P_{tot}^{-n}$, where
$P_{tot}$ is the total optical and electrical load and $n$ is at least 1.

	Amplifier noise is the quadrature sum of voltage noise and current noise
through the thermistor.  The noise level (as opposed to NET) is not strongly affected by
temperature or optical load in a well optimized system.  Amplifier NET
is most strongly affected by responsivity, and is minimized by reduced
optical background loading and the use of high quality amplifiers.

	Appendix~\ref{chap:NET} presents tabulated values of white
noise NETs achieved in the \maxima\ flights.

	Low freqeuncy noise originates from variations in the bolometer
electrical resistance and thermal reservoir temperature.  The former
term is relatively small for modern, well fabricated bolometers. In
\maxima, the thermal reservoir fluctuations are substantial, due to
the very low thermal mass of the 100-mK stage and the mechanical
coupling of telescope modulation to the refrigerators.  These
fluctuations dominate the bolometer noise at frequencies below \sima
0.5~Hz.  An additional contribution of low frequency noise from the
amplifiers can be avoided by AC-biasing the detectors, as in \maxima.

\subsection{Biasing and Readout}\label{bolos:bias}

		The \maxima\ bolometers are AC biased using 10-nA
(100-mV) RMS sine waves at \sima 300~Hz.  AC biasing provides strong
rejection of low frequency electronic noise, particularly in the
cryogenic preamplifiers.  The exact bias frequency is chosen before
flight to minimize narrow band microphonic pickup.

		All bias signals are phase locked and have fractional
amplitude variations of less than $10^{-6}$.  The AC signals from the
detectors are bandpass filtered and demodulated using a lock-in
amplifier referenced to the bias generator.  The demodulated signal is
processed using a 15-mHz highpass filter to remove 1/f noise not
rejected by the AC bias (primarily caused by detector temperature
drifts).  In addition, a four-pole butterworth lowpass filter with a
19-Hz cutoff is used to eliminate high frequency noise.  The phase
shifts caused by these filters are measured before flight and are
removed in data analysis.

		AC Bias generators and lock-in readout electronics are
mounted to the outside of the receiver at ambient temperature.

\subsection{Cryogenic Preamplifiers and Microphonics}\label{receiv:jfet}

		NTD Bolometers are high impedance devices (\sima
5~MOhm for \maxima) with correspondingly high sensitivity to
microphonic noise in wiring between the detector and the first
amplifier.  To minimize this sensitivity, preamplifiers are placed
within the cryogenic receiver, very close to the detectors.  These
differential amplifiers each consist of a matched pair of cryogenic
JFETs\footnote{Infrared Laboratories TIA JFETS} operating at 150~K
inside a sealed cavity within the liquid $^4$He-cooled portion of the
cryostat.  Between the detectors and the preamplifiers are \sima 8
inches of wiring and \sima 4 inches of circuit board traces.  The
wiring is potted in epoxy or varnished to a rigid surface over most of
its length.  The stiffness of the wiring and traces minimizes
microphonic response near and below the bolometer bias frequency
(\sima 300~Hz).

		The JFET amplifiers typically contribute
non-negligible, but subdominant white noise (\sima 5~$nV/\sqrt{Hz}$).
Though the amplifiers are the dominant source of 1/f noise in the
bolometer signal (1/f knee \sima 10~Hz), this low frequency noise is
rejected by the AC biasing of the detector. The performance of the
JFETs has been acceptable, but with little margin.  For slightly lower
detector noise, better JFETs would have been needed.

\subsection{Time Constants}\label{bolos:tcs}

		Detector time constants limit telescope scan speed and
act as a lowpass filter on the data.  \maxima\ detector time constants
vary from 1~msec to 10~msec (typically 6~msec to 8~msec).  The slowest
detectors, at 10~msec, are fast enough for the combination of
4\dega/sec scanning speed and 10\mins FWHM beam size, using the
$\frac{FWHM}{2\tau}$ criterion of \cite{Hanany_TC}.

		The filtering effects of detector time constants, like
those of electronic filters, must be deconvolved during data analysis.
Time constants are measured before flight, to an accuracy \err
0.5~msec.  Flight data from planet observations are used to refine
this to \err 0.2~msec (\cite{CDWThesis}).

\section{Receiver}\label{exper:receiv}

\begin{figure}[ht]
\centerline{\epsfig{width=4.5in,angle=0,file=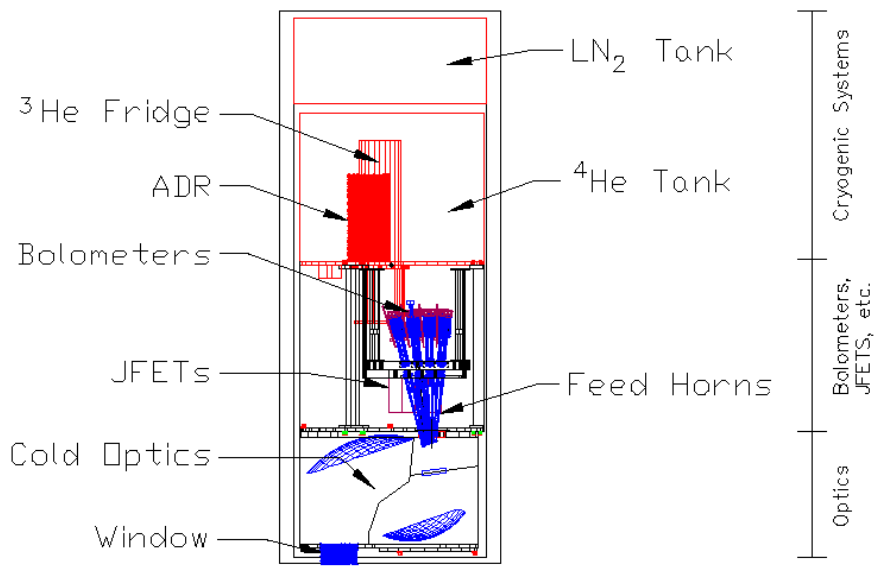}}
\caption[\maxima\ Receiver]{A mechanical drawing cutaway of the
\maxima\ cryogenic receiver.  The bottom section of the receiver
contains heavily baffled, liquid $^4$He-cooled optics and the internal
relative calibration source.  The middle holds the bolometers and
bolometer feedhorns, cryogenic JFET preamplifiers, and thermal
switches.  The top section of the receiver contains the cryogenic
systems, with the low temperature refrigerators (liquid $^3$He and
adiabatic demagnetization) surrounded by the open-cycle liquid $^4$He
tank.}
\label{fig:receiver}
\end{figure}

		The \maxima\ cryostat (Figure~\ref{fig:receiver}) houses
the secondary optics, the bolometer array and preamplifiers, an
optical calibration source, and the cryogenic coolers.  There are also
a number of diagnostic devices including `dark' detector channels not
exposed to the CMB and a variety of internal temperature monitors.

\subsection{Cryogenics}\label{receiv:cryo}

	The \maxima\ receiver makes use of four cooling systems:
open-cycle LN$_2$ and liquid $^4$He tanks, a closed-cycle liquid
$^3$He refrigerator, and an adiabatic demagnetization refrigerator.

\begin{table}
\begin{center}
\begin{tabular}{ccccc}
Cooler & LN$_2$ & $^4$He & $^3$He & ADR 
\\ \hline\hline 
\\Temperature (Kelvins) & 50 & 2-3 & 0.35 & 0.1
\\Hold Time (Hours) & 24 & $>$30 & $>$36 & 12
\\Thermal Cycle & Open & Open & Closed & Closed
\\ \hline
\end{tabular}
\caption[\maxima\ cryogenic systems]{Temperatures, cooling durations,
and type of thermal cycle for the \maxima\ cooling systems.  Numbers are
quoted for flight conditions (high altitude, nighttime).  All
cryogenic systems have ample capacity for a \maxima\ balloon flight.}
\label{tab:holdtimes}
\end{center}
\end{table}

	A 13-liter LN$_2$ tank cools an outer layer of radiation
shielding to 77~K.  This temperature drops to \sima 50~K when the
LN$_2$ tank is exposed to vacuum, as in flight.  The LN$_2$
temperature radiation shields are covered with thin, low emissivity
aluminum foil.

	Inside the LN$_2$-cooled space is a 21-liter, open-cycle
liquid $^4$He tank and an additional layer of shielding at liquid
$^4$He temperature.  The outer shell of the cold optics box serves as
part of this radiation shielding.  The outside of these shields is low
emissivity aluminum, while the inner surfaces are coated with a
blackening mixture.  The blackened interior absorbs high temperature
radiation that leaks past the shields.

		Within the liquid $^4$He temperature space are the
optics, the detectors, the JFET preamplifiers, the sub Kelvin coolers
and a variety of thermometers.  The optics and most electrical
components are thermally linked to the coldplate, a 1.0-cm thick
copper plate forming the bottom of the liquid $^4$He tank.

		Various locations in the liquid $^4$He space range in
temperature from 4~K to 6~K, depending on thermal load and proximity to
the helium tank.  When the liquid $^4$He is exposed to vacuum, for
testing or in flight, these temperatures drop to 2~K to 3~K.  This
causes a significant drop in the background loading of the bolometers.

		Inset into the liquid $^4$He tank are the adiabatic
demagnetization refrigerator (ADR) and the liquid $^3$He refrigerator.
All wiring entering the receiver passes through the liquid $^4$He and
LN$_2$ tanks and is made of low thermal conductivity stainless steel
leads, ending in cold radio frequency filters (\S \ref{receiv:rfi}).
		
\subsubsection{Low Temperature Refrigerators}\label{cryo:friges}

	Two cooling stages beyond liquid $^4$He are used to reach
sub-Kelvin temperatures.  The first is a closed-cycle $^3$He
refrigerator using 40~liters (stp) of $^3$He which provides a
temperature of \sima 350~mK under flight conditions.  The other is an
adiabatic demagnetization refrigerator (ADR) (\cite{SaltPill}) which
provides a temperature of 100~mK.  The \maxima\ ADR consists of
40~grams of ferric ammonium alum (FAA), a high permeability
ferromagnetic salt, inside a 2.5-Tesla electromagnetic coil.

	Mechanical supports for the ADR stage are constructed from
thin-walled, low-conductively Vespel\footnote{A polymer resin produced
by DuPont.} tubes and taut kevlar string.  Liquid $^3$He-cooled copper
straps are used to intercept all supports for the 100-mK stage,
reducing the conductive thermal load from the 4-K stage by more than a
factor of ten.

	An electrically controlled superconducting solenoid is used as
a heat switch, magnetically closing gold plated flanges to create a
thermal link between the liquid $^3$He refrigerator and the ADR.  It
is used to provide the isotemperature phase of the ADR thermal cycle.

	The heat load on the ADR - hundreds of nanowatts - causes a
temperature increase of \sima 3~mK per hour.  This is slow enough that
continuous temperature control is not needed.  Periodic corrections
are made via ground-based commanding of the ADR electromagnet current.
During the \maximai\ flight the ADR field was adjusted twice after the
initial cool down (\sima 2 hours between adjustments) with temperature
drifts of $<$5\% between adjustments.  During the \maximaii\ flight, a
problem with the ADR magnet control electronics made commanding
impossible.  Over \sima 6 hours of \maximaii, the ADR temperature
drifted by \sima 20\%.  During both flights, the effects of
temperature drift on bolometer responsivity were monitored using the
internal calibration source.

\subsection{Internal Relative Calibrator}\label{receiv:stim}

		Bolometer responsivity varies over the duration of a
flight, primarily because of variations in the temperature of the
nominally 100-mK ADR.  These variations are monitored using a stable
internal calibration source (a stimulator) consisting of a thin
nickel-chromium layer (2~mm $\times$ 2~mm) backed with a sapphire
substrate.  The metal layer is impedance matched to radiate
efficiently into free space when heated.  When a heating current \sima
1~mA is applied, the metal warms to \sima 50~K with a time constant of
\sima 1~sec.  The heating current is maintained for 10 seconds, and is
applied every 20 minutes during flight.  The stimulator is mounted
inside the cold optics box, and fitted with a light pipe to illuminate
the focal plane array from just outside the optical path.  The
illumination of the array is not uniform, with detectors closer to the
calibrator receiving about twice the flux of the more distant
detectors.

	The source is extremely stable with negligible resistance
fluctuations and $<$1\% current fluctuations.  The on-state
calibrator temperature is further stablized by a weak, temperature
dependent thermal link to the liquid helium stage.

	The absolute flux is not well measured, so the stimulator is
used purely for monitoring of responsivity variations over time.
Absolute calibration is obtained from celestial sources (the CMB
dipole and planets).  The use of stimulator data in detector
calibration is discussed in Section~\ref{calib:relative}.

\subsection{RFI protection}\label{receiv:rfi}

		During flight, several radio transmitters are used for
telemetry.  Each radiates 15 to 40 Watts at frequencies of 1.5~GHz and
higher.  Extensive filtering is required to prevent pickup from these
sources in the detectors.

		The receiver itself consists of three metallic shells
which serve as partial Faraday cages.  External cabling is also fully
enclosed in a metallic shell.  All cables between the receiver and the
readout electronics and all cables exiting the readout electronics
pass through commercial RF filtered connectors (Amphenol FPT02 Series;
60-dB attenuation at 1.0~GHz).  Within the receiver, all wiring is
potted in 27~cm of Eccosorb CR-124, a metal-filled
epoxy\footnote{Commercially available from Emerson and Cuming.} that
acts as a radio frequency lowpass filter (30-dB attenuation at
1.0~GHz).

		Radio frequencies are filtered directly at the
detectors using an inductive/capacitive (LC) lowpass filter with a
1.0~GHz cutoff.  These filters are fully described in
\cite{CDWThesis}.

\chapter{Observations}\label{chap:flights}

\begin{figure}[ht]
\centerline{\epsfig{width=4.5in,angle=0,file=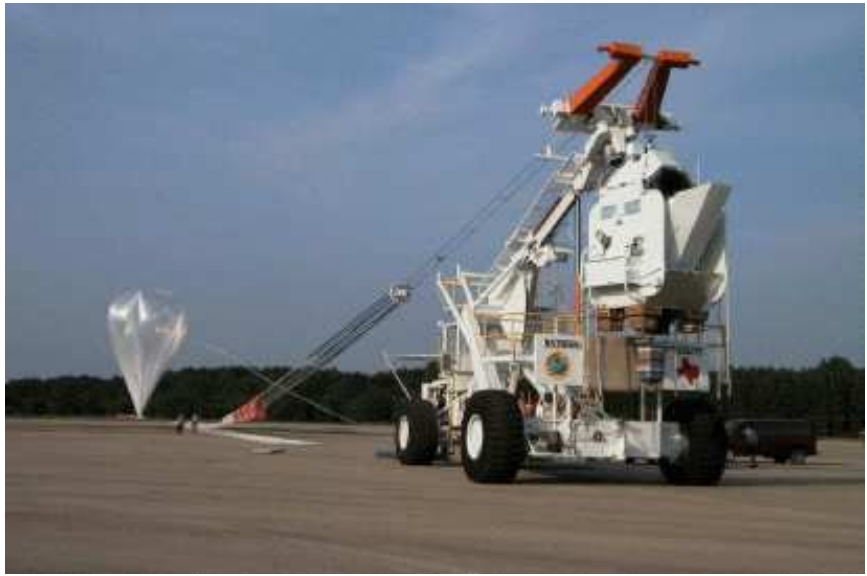}}
\caption[Launch Photo]{The \maxima\ telescope on ``Tiny Tim'', the
launch vehicle, shortly before launch on June 16, 1999 }
\label{fig:launchpic}
\end{figure}

		This chapter describes the two \maxima\ flights,
including flight conditions, scan durations, flight trajectories, and
the positions of the Sun and Moon relative to the scans.

\section{MAXIMA-I}

		\maxima\ was first flown on August 2, 1998.  Field
work began in late April and \maxima\ was flight ready on June 10.
Due to unusual weather patterns (El Ni\~{n}o), there were no launch opportunities
for nearly two months.

		The payload was launched on August 8 at 00:58 UT
(19:58 local time) from the National Scientific Balloon Facility
(NSBF) in Palestine, TX (latitude 31.8\dega N, longitude 95.7\dega W).
The maximum float altitude of 37.5~km was reached at 4:35 UT.  The
telescope traveled 189~km west and less than 1~km south before
reaching maximum altitude.  At float, the telescope drifted 405~km
west and less than 1~km north.  Descent began 3.8~hours later at 8:22
UT.  Summer flights from the NSBF in Palestine are limited to a range
of approximately 600~km.  The \maximai\ flight was relatively short
due to fast high altitude winds.

		Four observations were conducted during the flight.
First, the CMB dipole was observed in order to calibrate the
responsivity of the detectors.  The dipole observation was started
before reaching float altitude and lasted from 03:37 UT to 04:11 UT.
Next, two overlapping, cross-linked scans of CMB anisotropy were
conducted over a 122-deg$^2$ region in the vicinity of the Draconis
constellation.  These scans occurred from 04:21 UT to 05:59 UT and
06:02 UT to 07:24 UT.  Finally, observations were made of Jupiter to
characterize the telescope beams and to calibrate the 410-GHz
detectors, which are insensitive to the CMB dipole.  Jupiter was
observed from 07:30 UT until 08:04 UT.

\begin{table}
\begin{center}
\begin{tabular}{ccccccc}
Flight & Hours at          & 1st CMB     & 2nd CMB     & CMB Dipole  & Planet    & Daytime
\\     & Maximum          & Scan        & Scan        & Scan        & Scan   & Test Data
\\     & Altitude 	  & (hours) 	& (hours) 	& (hours) 	& (hours) 	& (hours)
\\ \hline\hline 
\\\maximai & 3.78 & 1.63 & 1.37 & 0.57 & 0.57 & 0.00 
\\\maximaii & 7.78 & 2.42 & 2.15 & 0.62 & 0.63 & 1.65 
\\ \hline
\end{tabular}
\caption[Flight Statistics]{Scan and flight durations
for both \maxima\ flights.  Some calibration data were collected before
the telescope reached maximum altitude.  In the case of \maximai, this
results in a total scan time greater than the time at maximum
altitude.}
\label{tab:flight_stats}
\end{center}
\end{table}

	The sun was at least 20\degs below the horizon for all
observations.  The Moon was below 20\degs elevation during CMB
observations, and below the horizon for over an hour of the second
observation.  While above the horizon, it was below 30\degs elevation
and was at least 70\degs from the scan region.  The relative position
of the Moon differed by \sima 20\degs azimuth and \sima 10\degs
elevation between the two CMB scans.  During the dipole observation,
the Moon was at \sima 30\degs elevation, \sima 20\degs below the scan.
The Moon was below the horizon during the Jupiter scan.  The Moon was
68\% full during the flight.

\section{MAXIMA-II}

		The second flight occurred in June 1999.  Field work
began in late April, and the instrument was prepared for flight in the
first week of June.  Footage of the launch and flight preparations can
be found in \cite{Nova}.

		The weather in 1999 was much more favorable for a
timely launch.  The launch occurred on June 17 at 00:07 UT (19:07 June
16, local time).  The telescope traveled 42~km east and 9~km south
before reversing direction and reaching maximum altitude 97~km west
and 1~km south of the launch position.  The maximum float altitude of
38.0~km was reached at 04:34 UT.  At float, the telescope drifted
490~km west and 42~km north.  Descent began 7.8~hours later at 12:21
UT.  The relatively slow high-altitude winds of early summer allowed
us a considerably longer flight than \maximai.

	As with \maximai, two CMB observations and two calibration
scans were conducted.  The first was an observation of Mars from 03:14
UT to 03:52 UT.  Approximately one hour was spent on maintenance
tasks.  Two overlapping, cross-linked CMB scans were conducted from
05:04 UT to 07:29 UT and 07:31 UT to 09:40 UT.  The observed region
has an area of 225~deg$^2$ and overlaps the \maximai\ region by
50~deg$^2$.  A calibration scan of the CMB Dipole was conducted from
09:42 UT to 10:19 UT.  Further data were recorded from 10:20 UT to
11:59 UT as a test of the daytime performance of the instrument.

	The sun rose to -20\degs elevation at 09:24 UT and to 0\degs
elevation at 11:17 UT.  Data collected after 10:20 UT have been used
only as test data for future daytime balloon flights.  The Moon was
17\% full during the flight and was below the horizon during the
dipole observation and both CMB observations.  During the Mars
observation, the Moon was 75\degs from the scan.

\chapter{Pointing and Attitude Reconstruction}\label{chap:point}

\begin{figure}[ht]
\centerline{\epsfig{width=3.0in,angle=90,file=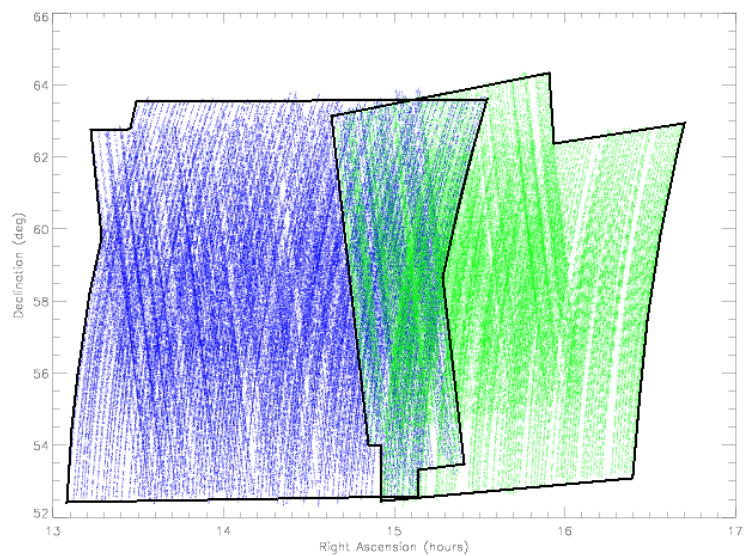}}
\caption[\maxima\ Scans]{The full reconstructed pointing for a single
detector in both \maxima\ flights.  \maximai\ is represented in green
(light gray) on the right, and \maximaii\ is in blue (dark gray) on the
left.  The scan region for each flight is boxed, and the \sima
50-deg$^2$ overlap region can be seen at Right Ascension \sima 15~hours.}
\label{fig:colorscans}
\end{figure}

		In this chapter we discuss the pointing and attitude
reconstruction of the \maxima\ telescope.  There are three main
pointing related issues.  First is scan strategy; the size and shape
of the observation pattern are chosen for high angular power spectrum
sensitivity and to minimize the effects of noise correlations and
potential systematic effects (\S \ref{point:scans}).  Second is
pointing control; we have constructed a flexible pointing system,
capable of realizing a variety of scan strategies (\S
\ref{point:hardware}).  Third is post-flight pointing reconstruction;
the orientation of the telescope is determined to 1\mins - roughly
10$\%$ of the FWHM of the telescope beams (\S \ref{point:data}).
Section~\ref{point:requirements} lists specific requirements in all
three areas.

		A number of \maxima\ collaborators were deeply
involved in pointing and attitude reconstruction, including Paolo
DeBernardis (planning; scan strategy; flight hardware), Andrea
Boscalleri (flight hardware), Barth Netterfield (flight software),
Enzo Pascale (flight software), and Amedeo Balbi (pointing
reconstruction).  In addition, George Smoot, Amedeo Balbi, Andrew
Jaffe, and Shaul Hanany developed flight planning software.

\section{Requirements}\label{point:requirements}

		In this section we summarize the various requirements
for the pointing of the \maxima\ telescope.  The implementation and
strategies used to meet these requirements are detailed later in this
chapter.

\subsection{Sky Selection}

		The absence of Galactic dust was the strongest
requirement for the observation region.  Scan regions in \maximai\
were also chosen to avoid known bright point sources.  The expected
signal from uncatalogued point sources is very small
(\cite{GJS_point}), and was not a major concern.  Dust, point sources,
and other foregrounds are discussed in Chapter~\ref{chap:systematic}.

		The scan regions for the two flights were chosen to
have a modest (\sima 50-deg$^2$) overlap, both as a consistency check
and to facilitate the combination of the data sets.

		Within these constraints, we selected scan regions
offset by roughly 40\degs azimuth from the north celestial pole.  This
was useful for technical reasons discussed in
Section~\ref{hardware:ccd}.

\subsection{Scan Speed}

		Scan speed determines the relationship between time
domain data and the spatial map of the CMB.  In particular, the
modulation with the shortest characteristic time scale sets the
minimum revisitation time for an observed spot on the sky.  The
corresponding temporal frequency is roughly the minimum at which the
data carry significant CMB information.\footnote{While not an exact
relationship, this is a useful rule of thumb.  In general, the exact
mapping between temporal and spatial frequencies is complex and best
explored through simulations.}

		The \maxima\ detectors have a low frequency knee (1/f
or steeper) in their noise profiles between 0.1~Hz and 1.0~Hz.  The
fastest modulation should be in this range or higher.  The fastest
\maxima\ modulation (0.45~Hz) is described in
Sections~\ref{point:scans} and~\ref{chopper}.

\subsection{Depth of Integration}

		An experiment with fixed integration time and partial
sky coverage makes a trade-off between the integration time available
for each pixel and the number of pixels observed.  Equivalently, the
trade-off is between noise and sample variance in power spectrum
estimation.  The optimal integration for measurement at a given
angular scale is one that yields a signal-to-noise of \sima 1 at that
scale (\cite{TegmarkGuide}).  A higher SNR with fewer pixels degrades
the power spectrum estimation fairly slowly with decreasing pixel
count.  The opposite case (low SNR, many pixels) causes a much faster
degradation in power spectrum estimation as a function of pixel count.

		This trade-off is complicated by the fact that we
are simultaneously measuring over a large number of angular scales.
Optimal integration at large scales leads to too much sky coverage at
small scales, while optimal integration at small scales leads to too
little sky coverage at large scales.  Because of the asymmetry in the
optimization, and inherent difficulty of measurements at small scales
(CMB anisotropy power drops dramatically at high \ella ), we have
optimized our integration time based on beam-sized (10\mina ) pixels.
We deem this approach optimal for an experiment measuring a large
range of angular scales.  A final consideration is systematic error
testing, which benefits from higher SNR.

		Achieved integration time and signal-to-noise per
pixel are summarized in Section~\ref{point:scans}.

\subsection{Scan Pattern}\label{requirements:spattern}

		In the case of an ideal CMB experiment, with purely
uncorrelated receiver noise, any modulation pattern yielding a
constant integration time/pixel would be acceptable.  In reality, low
frequency detector noise leads to significant noise correlations.
Given arbitrary pointing control, the effects of these correlations
are best eliminated by a random scan pattern in which pixels in the
scan region are measured with uniform probability at each detector
sampling.  This is obviously unrealistic.  In practice, good results
are obtained by the use of interlocking observations, with the scan
directions tilted to provide cross-linking
(\cite{TegmarkGuide}).\footnote{An example of cross-linking is shown
Figure~\ref{fig:crosslink}.}

		The overall shape of the scan region is also
significant.  A more compact scan region (as opposed to a highly
elongated region with a large aspect ratio or exotic shape) is
desirable, especially for measurement of large spatial modes (low
\ella ).  Such a region contains more independent modes, particularly
large modes, than an elongated region of the same area.  In addition,
edge effects (\ie~imperfections in the scan pattern near the edges) are
minimized for compact scan regions.

		Fast elevation modulations should be avoided.  The
column depth of the atmosphere varies roughly as the cosecant of the
elevation and is a significant source of background loading on the
bolometers.  A modulation of this background loading would cause a
modulation synchronous signal in the detectors.  In addition, rapid
changes in elevation angle disturb liquid cryogens in the receiver,
leading to detector instabilities.

		A good approach is repeated raster scans, tilted
relative to each other for cross-linking.  A raster-like pattern can be
achieved by a periodic modulation in azimuth alone; the rotation of
the sky effectively provides the modulation in the elevation
direction.  The \maxima\ scan pattern is of this general form.  The
details are presented in Section~\ref{point:scans}.

\subsection{Precision of Control and Reconstruction}

		The absolute position of the scan boundaries must be
controlled to \sima 0.5\degs to maintain the shape of the scan region.
Scan periods should be stable to at least 10\%, to ensure reasonable
beam overlap from one period to the next.  Elevation must be stable to
a few arcminutes during a scan period, also to ensure overlap.

		For pointing reconstruction, an accuracy of \sima
1\mins is desired.  This yields negligible pointing-based error in the
CMB power spectrum over most of our measured range.  Pointing error is
most significant at high \ells (\sima 10$\%$ at $\ell = 1000$), but
with 1\mins accuracy remains subdominant at all angular scales.

\subsection{Calibration Scan Requirements}

		In addition to CMB scans, we require observations of
both the CMB dipole, and a bright point source (a planet) for
calibration.

		The dipole is the main responsivity calibrator.  The
spatial distribution of the signal makes it difficult to observe the
dipole on time scales faster than detector drift.  We require that the
beams be scanned quickly over a large temperature contrast.  The
observation is repeated continuously for long enough to reduce
calibration uncertainty to a few percent.  The required pointing
reconstruction accuracy of the dipole scans is \sima 10\mina,
corresponding to \sima 0.1\% calibration bias.

		Planet scans are used primarily for beam measurement.
They also provide a secondary responsivity calibration.  For these
observations, we require that a telescope beam cross the planet
quickly compared to detector drift, that the beams be well sampled in
two dimensions, and that pointing reconstruction be as accurate as for
CMB observations.

		As with CMB scans, calibration scans must be conducted
at constant elevation.  Due to the higher signal-to-noise of the
calibration sources, cross-linking is not a requirement.
Section~\ref{point:scans} describes the actual observation strategy
used for both calibration sources.

\section{Scan Strategy}\label{point:scans}

\subsection{Selected Sky}

		Each \maxima\ flight has consisted of two cross-linked
observations of a single patch of the CMB sky.  The dust in these scan regions has
a predicted in-band equivalent temperature of \sima 10.0~\micro K at 150~GHz with
rms fluctuations of \sima 2.5~\micro K in units normalized to the CMB spectrum
(\cite{FORECAST}).  Tests of the spectral and
angular profiles of the observed signals, as well as cross
correlations with known dust maps, confirm the absence of significant
dust contamination in our CMB data (\S \ref{sys:foreg}).

		The \maximai\ scan region was chosen to contain no
detectable point sources.  For \maximaii, this requirement was relaxed
so that while no point source contribution is expected in the CMB
power spectrum, a few bright sources might be detectable in the
anisotropy maps, particularly at 410~GHz.

\begin{figure}[ht]
\centerline{\epsfig{width=2.5in,angle=0,file=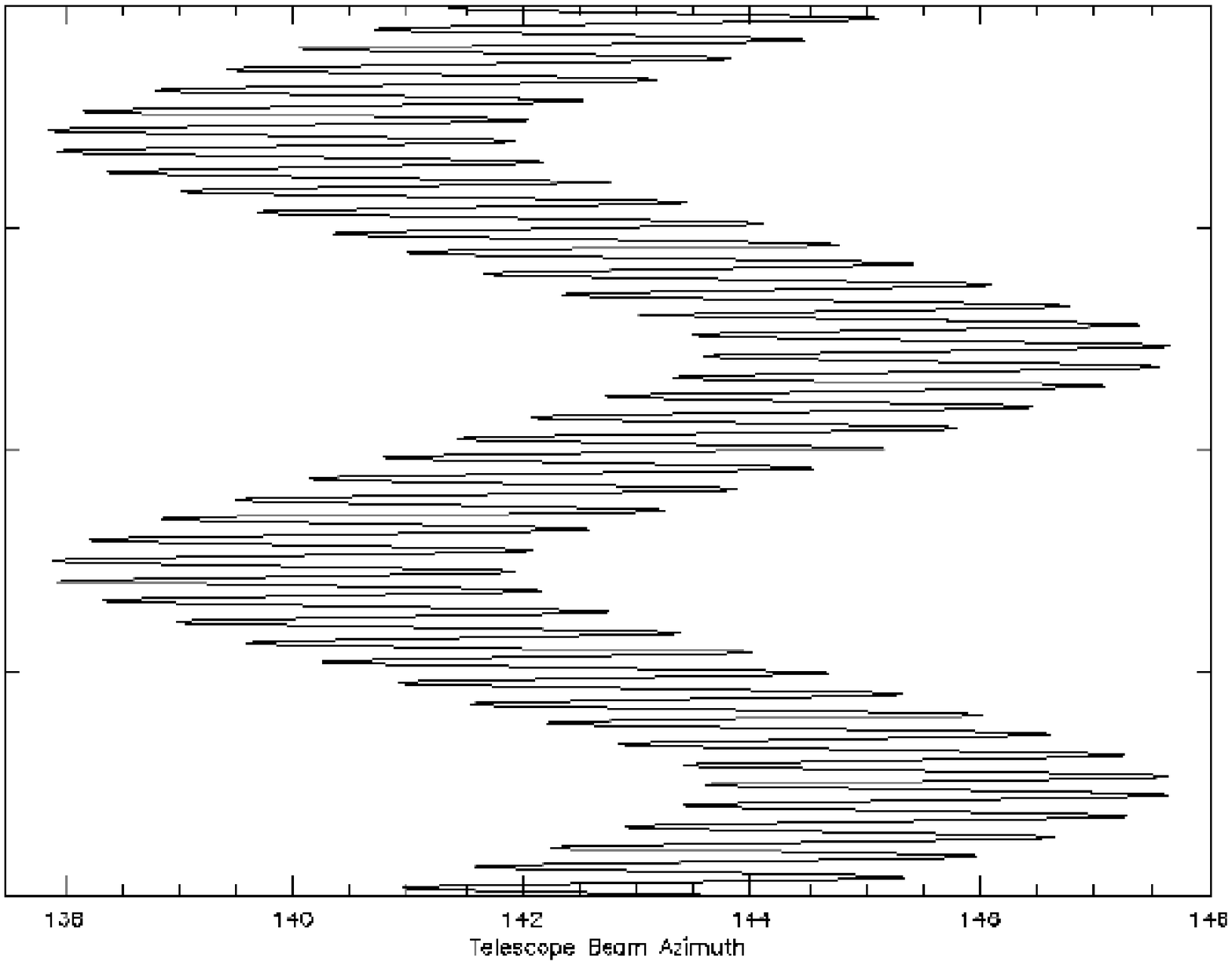}
\epsfig{width=2.in,angle=90,file=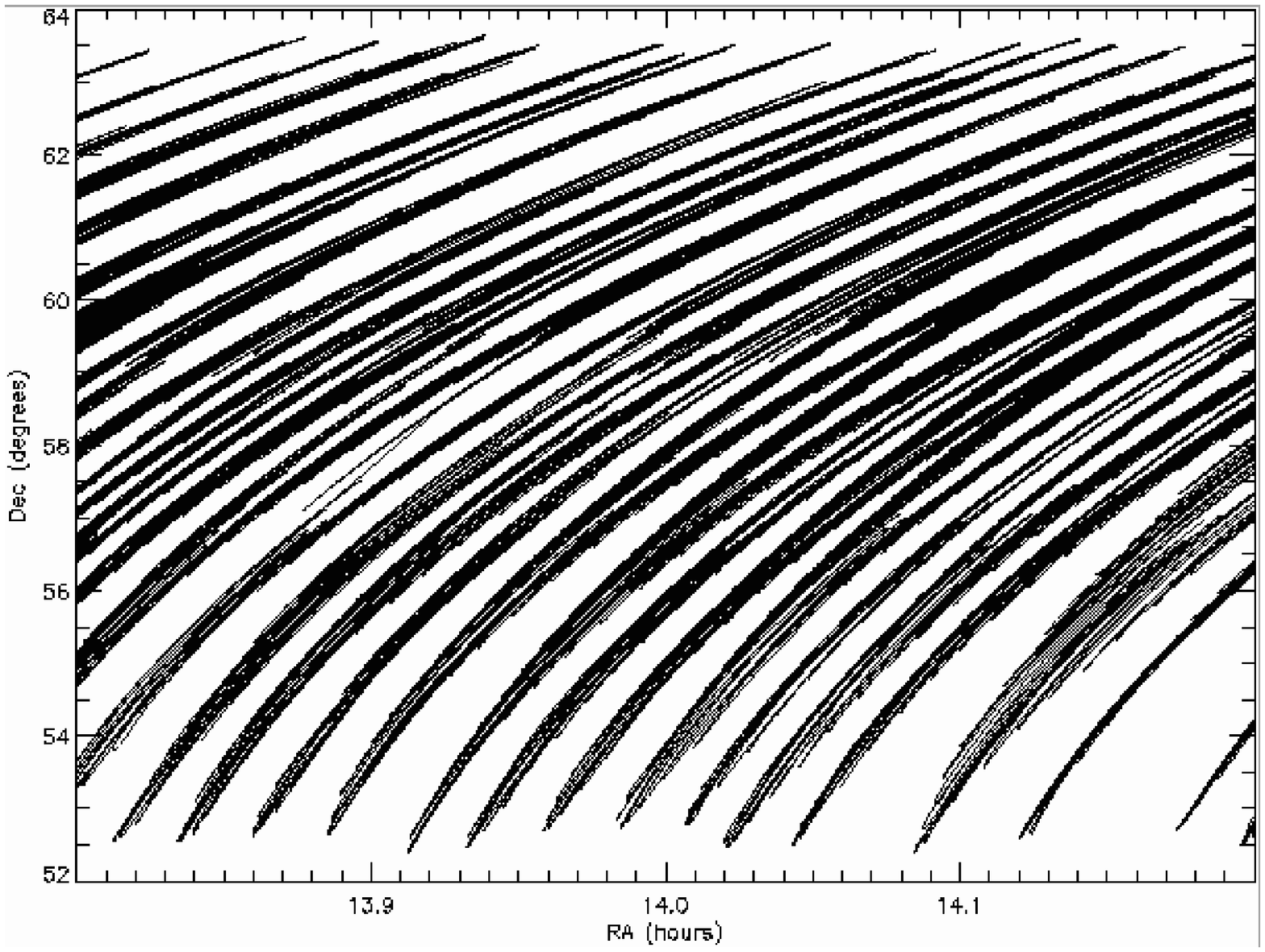}}
\caption[Azimuth Double Modulation]{{\bf{Left:}} A simulation of the
double modulation in azimuth.  The x-axis is the azimuthal position of
the telescope beams, while the y-axis is time.  The slower modulation
is caused by the motion of the entire telescope, while the faster
modulation is caused by the rotation of the primary mirror about the
optic axis.  {\bf{Right:}} The scan pattern formed in RA and
declination, combining the azimuth modulations with the rotation of
the sky (data are shown from a \maximaii\ scan).  Note that lines of
constant elevation move with the rotation of the sky, in this case
spanning the plot in a diagonal arc from the lower left to the upper
right.  The gaps seen in this scan pattern are less than half the
telescope beam size, so a continuous CMB map is obtained.}
\label{fig:doublemod}
\end{figure}

\begin{figure}[ht]
\centerline{\epsfig{width=2.in,angle=90,file=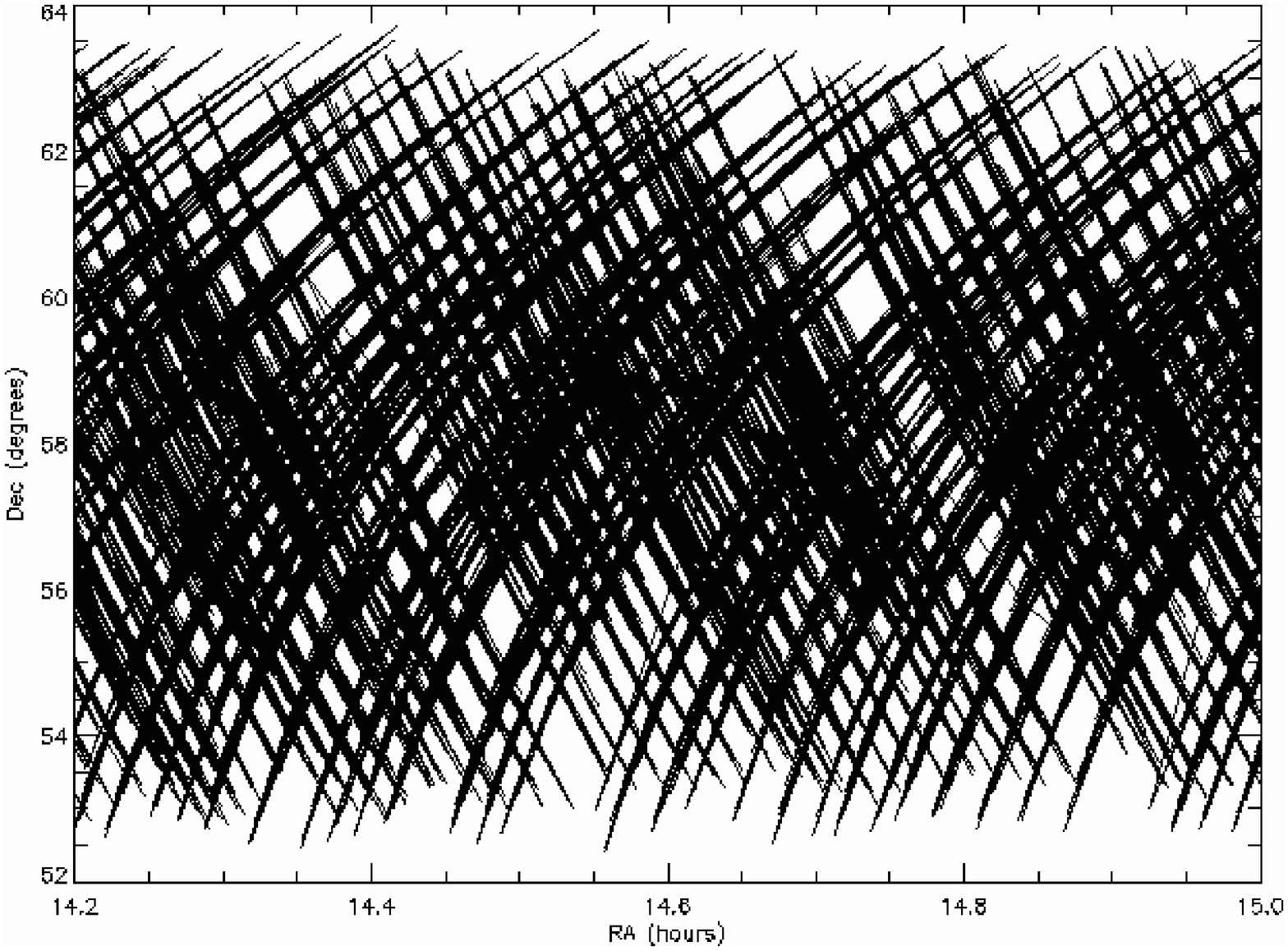}
\epsfig{width=2.in,angle=90,file=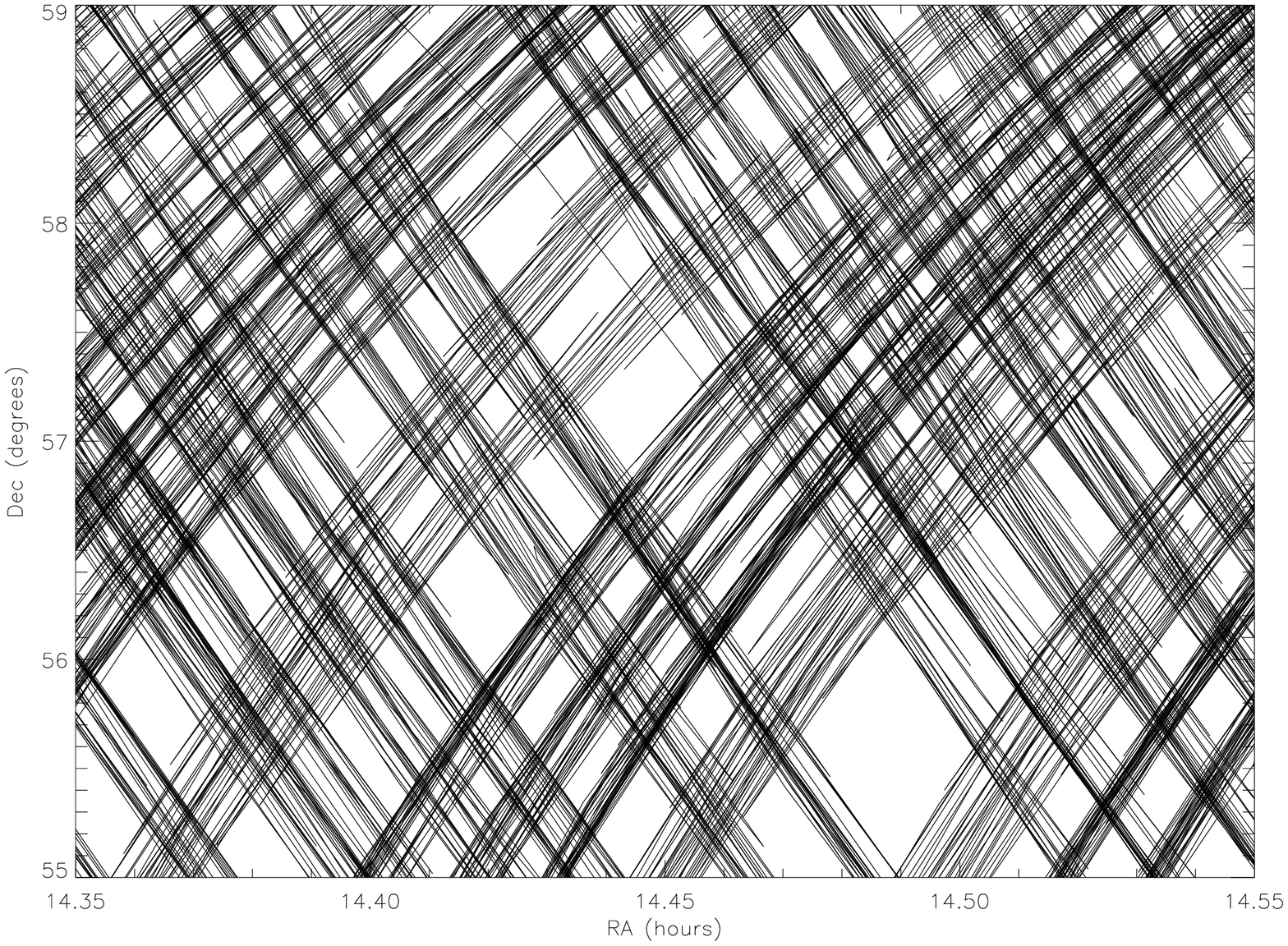}}
\caption[Cross-linking]{{\bf{Left:}} The cross-linked scan pattern as
realized in \maximaii.  The cross-linking angle averages 27\dega.
{\bf{Right:}} A blowup of the central region.}
\label{fig:crosslink}
\end{figure}

\subsection{Modulation Pattern}

		Each CMB observation is conducted at a fixed
elevation, while the telescope beams are moved actively in azimuth.
The azimuth modulation defines one dimension of the roughly
rectangular scan region.  The rotation of the sky over the duration of
the observation defines the other dimension.

		The azimuth motion consists of two independent
modulations.  The primary mirror rotation provides a relatively fast
modulation with a frequency of 0.45~Hz and an amplitude of 4.0\degs
peak-to-peak.  The motion of the entire telescope provides a slower
modulation at a frequency of 12~mHz to 25~mHz with an amplitude of
4.5\degs to 9.0\degs peak-to-peak.  The fast modulation prevents our data
from being significantly corrupted by low frequency noise.  Taken
together, the two make our data relatively robust against modulation
synchronous parasitic signals.

		The two CMB scans of each flight observe the same
region of the sky, but at different times and at different
elevation angles.  The rotation of the sky between these observations
causes the two scans to be tilted relative to each other in
sky-stationary coordinates.  This cross-linking averages 22\degs in
\maximai\ and 27\degs in \maximaii.

		The two azimuth modulations, the rotation of the sky,
and the cross-linked revisitation are uncorrelated and are on radically
different time scales, minimizing the effects of potential
scan-correlated systematics.

\subsection{Depth of Integration}

		Depth of integration is set by the total width of the
combined azimuth modulation and the rotation rate of the sky.  The
average integration time per beam-size was \sima 2.5~seconds in
\maximai\ and \sima 2.2~seconds in \maximaii.  This leads to an
average expected noise level of \sima 60~\micro K per beam size area
for our best single detector and \sima 40~\micro K for our published
combination of four \maximai\ detectors.  In practice, integration is
several times longer in the center of the observed region and shorter
near the edges.

\subsection{Calibration Scans}

		\maxima\ is calibrated in flight using both planets
and the CMB dipole.  Planet observations are conducted by pointing the
telescope directly above or below it as the planet rises or sets.  The
telescope tracks the position of the planet in azimuth, while
remaining at fixed elevation.  The primary mirror modulation moves the
beams around the planet in azimuth.  The beams are much larger than
the planet, so each pass is effectively a one-dimensional map of the
telescope response.  The planet drifts steadily through the observed
elevation giving complete two dimensional beam maps.  Each crossing is
\sima 20~bolometer samples (\sima 0.1~sec) and there are several
hundred crossings with good signal-to-noise over the course of the
observation.  The scan pattern is illustrated in
Section~\ref{calib:planet}.

		The dipole is observed by rotating the entire
telescope in azimuth at high speed (18\dega/sec; 3.3~RPM) at fixed
elevation.  The signal is detected at \sima 55~mHz.  Rotations are
conducted for about 30 minutes (\sima 100 rotation).  The primary
mirror modulation has a small amplitude (4\dega ) and has very little
impact on dipole observations; the modulator was on during the
\maximai\ dipole scan and off during the \maximaii\ dipole scan.

\section{The Attitude Control System}\label{point:hardware}

		The pointing system, or attitude control system (ACS),
serves both to control the orientation of the telescope in flight and
to acquire the data needed for post-flight pointing reconstruction.
The pointing system consists of attitude sensors, a central feedback
loop control computer, and motors.  Some of the easily interpreted
sensors are used in pointing control, while the most precise sensors,
the CCD cameras, are used after the flight for pointing
reconstruction.

\begin{figure}[ht]
\centerline{\epsfig{width=4.0in,file=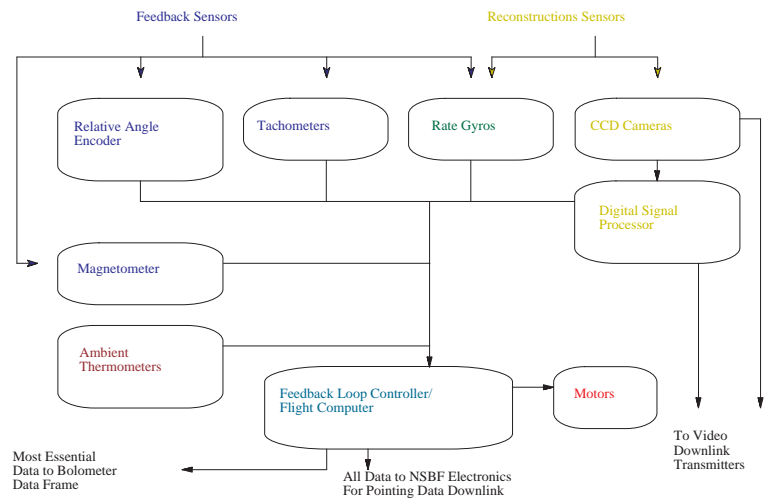}}
\caption[Pointing System Schematic]{A Schematic of the \maxima\
pointing system.  The system acquires data for for pointing control
and post-flight pointing reconstruction.  The central control computer
reads all data, commands the motors, and handles remote
communications.}
\label{fig:pointsys}
\end{figure}

\subsection{Control Electronics}

		The Feedback Loop Controller (FLC) is the computer
that controls the pointing system.  The FLC reads data from the various
sensors, applies a digital feedback algorithm, and sets the power
level for the motors.  Each of these tasks is performed once every
96~ms, synchronously with the bolometer data acquisition system.

		Signals from most sensors are sampled each cycle by
analog-to-digital converters inside the control computer.  Data from
the CCD cameras are processed by a separate computer, and passed to
the control computer every two cycles (192~ms).  GPS data (absolute
time and position) are updated once per second.  Using these data, the
computer sets power levels for the motors by generating pulse width
modulated (PWM) square-waves.

		Flexible commanding and scheduling of the pointing
system have been essential to the efficient use of limited observation
time.  Pointing normally follows one of several preprogrammed flight
schedules.  Remote (ground-based) commanding is used to switch between
schedules, to make modifications to schedules, or to take complete
manual control of the pointing.  In addition, remote commanding can be
used to modify control loop gains, to make adjustments to sensor
calibrations, and to set parameters for the CCD image processing.  The
FLC also generates a digital data frame for transmission to the
ground.  This includes sensor data and status information.

\subsection{Pointing Control}\label{point:control}

\subsubsection{Sensors}

		Feedback control is based on azimuthal rotation
velocity, as measured by a rate gyroscope.  Two other gyroscopes
measure pitch and roll velocities but are used only for post-flight
diagnostics.  The gyroscopes are obtained commercially\footnote{BAE
Systems Vibrating Structure Gyro} and have an accuracy of \sima
0.01\dega/sec.  Though they are very sensitive, the gyroscopes have
substantial low frequency drifts, primarily due to ambient temperature
fluctuations.  Drifts are calibrated once per gondola scan period, and
have little impact on pointing control.

		Absolute azimuth is measured using a two axis
magnetometer.  The magnetometer is extremely precise ($<$0.5\mina) in
differential measurement, but is highly non-linear due to the magnetic
properties of the telescope.  Pre-flight measurements are used to
calibrate the magnetometer to an absolute accuracy of \sima 1\dega.

		Absolute elevation is measured by an optical angle
encoder between the inner assembly (receiver and primary mirror) and the
outer frame of the telescope.  The accuracy of this measurement
depends on the balancing of the telescope (\sima 0.1\dega) and on
long time scale pendulum motion (\sima 0.5\dega, varying over tens
of minutes).  The differential accuracy of this elevation measurement
is \sima 1\mina.

		The CCD star cameras, described in
Section~\ref{hardware:ccd}, provide the most accurate measurement of
telescope orientation.  They are used for post-flight reconstruction
rather than pointing control.

\subsubsection{Motors}

\begin{figure}[ht]
\centerline{\epsfig{width=5.0in,angle=0,file=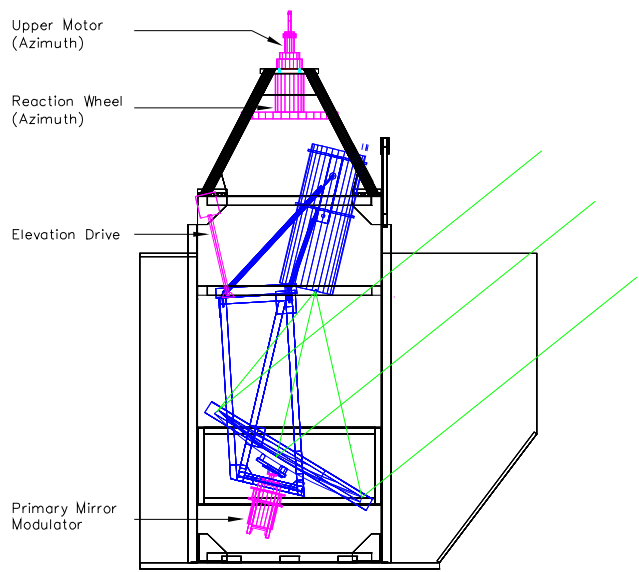}}
\caption[Pointing Motors]{A mechanical drawing showing the layout of
the telescope motors.  Two motors near the top of the telescope
control azimuthal orientation by driving against a reaction wheel and
the cables from the balloon respectively.  A linear actuator/servo-arm
tilts the inner assembly, pointing the telescope in elevation.  A
motor below the primary mirror modulates it at relatively high speed
(0.45~Hz, \err 2\degs amplitude) in azimuth.}
\label{fig:motors}
\end{figure}

		Three motors are used to point the \maxima\ telescope
and a fourth, described in Section~\ref{chopper}, is used to modulate
the primary mirror.  Two motors, located near the top of the telescope
frame, are used for pointing in azimuth.  One of these drives a
reaction wheel with a moment of inertia of 10~kg$\cdot$m$^2$ (\sima
0.5\% that of the telescope).  The other torques against suspension
cables connected to the balloon, which has a much greater moment of
inertia than the telescope.  Both motors are direct drive (ungeared)
and have a torque of \sima 35~N$\times$m with our maximum power of
12~Watts.  The light reaction wheel provides fast response, while the
other motor keeps the speed of the reaction wheel low by dumping
angular momentum into the balloon.  The rotational velocities of these
motors are monitored by tachometers.

		Elevation control is provided by a geared motor
connected to a linear actuator arm.  The arm is fixed between the
outer assembly of the telescope frame and the inner assembly of the
receiver and primary mirror.  The inner assembly is balanced about the
rotation axis, so the load on this motor is very small.

\subsubsection{The Control Loop}

\begin{table}
\begin{center}
\begin{tabular}{c|cccc}
\\CMB & \maximai & \maximai & \maximaii & \maximaii
\\Observation & Scan 1 & Scan 2 & Scan 1 & Scan 2
\\ \hline\hline
\\Max Scan Speed (deg/sec)&0.29& 0.30& 0.29& 0.26
\\RMS Velocity Error (\ditto)& 0.022 & 0.028& 0.043 & 0.057
\end{tabular}
\caption[Pointing Performance]{A summary of pointing control in both
\maxima\ flights.}
\label{tab:point_perform}
\end{center}
\end{table}

		Motor power is determined in a digital control loop.
Pointing control in azimuth and elevation are not strongly coupled and
may be considered separately.

		In azimuth, we use a feedback system based on the
rotational velocity measured by a rate gyroscope.  In CMB scans, the
target velocity is constant, except during turnarounds in the scan
direction.  During turnarounds the target velocity varies linearly
with time.  The absolute position is not used directly in the control
loop; it is instead used to trigger these turnarounds.  This is a
reasonable approach for our experiment - as long as the telescope is
scanning steadily, in the correct scan region, the exact position
at any given moment is unimportant.

		The azimuth control algorithm determines power for two
motors from three inputs, for a total of six gains.  The power to the
flywheel motor, $P_{fly}$, and the power to the upper motor, $P_{up}$,
are given by,

\begin{equation}
\left(  \begin{array}{c}P_{fly} \\P_{up}  \end{array} \right)
=
\left( \begin{array}{ccc}G_{fly,v\_az} & G_{fly,dv\_az} & G_{fly,v\_fly}\\G_{up,v\_az} & G_{up,dv\_az} & G_{up,v\_fly}\end{array} \right)
\cdot 
\left( \begin{array}{c}v\_az\_err \\\frac{d}{dt}v\_az\_err\\  v\_fly\end{array} \right)
,\end{equation}

\noindent where $v\_az\_err$ is the target rotational velocity minus
the measured rotational velocity, and $v\_fly$ is the rotation rate of
the flywheel relative to the telescope frame.  $v\_az\_err$
and$\frac{d}{dt}v\_az\_err$ are the usual error terms in a PD control
loop.  $v\_fly$ is used for two related but distinct purposes.  In
flywheel control, it serves as a correction for the back EMF generated
by the flywheel motion.  In upper motor control, it is used to
transfer the angular momentum of the flywheel to the balloon.  There
is no term for upper motor velocity, because it is essentially the
same as the rotational velocity of the entire telescope.  At our scan
speeds (up 0.3\dega/sec during CMB scans), back EMF in the upper motor
is negligible.

In practice, several of the gain terms are not needed.  Velocity based
feedback systems are not stabilized by a derivative term, so both
$G_{fly,dv\_az}$ and $G_{up,dv\_az}$ are set to zero.  In addition, it
is possible to set either of $G_{fly,v\_az}$ or $G_{up,v\_az}$ zero.
For \maximai, $G_{up,v\_az}$ was set to zero.  In this configuration,
the pointing is directly controlled by the flywheel motor, while the
upper motor acts only to reduce the angular momentum of the flywheel.
For \maximaii\ both of these gains were finite, though $G_{fly,v\_az}$
was relatively small.  In this configuration the upper motor
controlled the pointing directly, with some contribution from the
flywheel.  The pointing accuracy of both flights is summarized in
Table~\ref{tab:point_perform}.  The \maximai\ gains are somewhat
better in terms of pointing performance, though both sets of gains
were adequate for our needs.\footnote{Before the \maximaii\ flight,
there was some evidence that the receiver was more susceptible than
usual to vibrational noise pickup.  As a safety measure, greater use
was made of the slower moving upper motor.}

The elevation control formula is based on the measured angle of the
telescope inner assembly relative to the outer frame.  In this case,
the power to the elevation drive, $P_{el}$, follows a typical PD
scheme,

\begin{equation}
P_{el} 
=
\left( \begin{array}{cc}G_{el,p\_el} & G_{el,dp\_el}\end{array} \right)
\cdot 
\left( \begin{array} {c}p\_el\_err\\ \frac{d}{dt}p\_el\_err\end{array} \right)
,\end{equation}

\noindent where $p\_el\_err$ is the difference between the target and
measured elevation.  A derivative term is useful for stability, so
both gains are finite.

\subsection{Primary Mirror Modulation}\label{chopper}

	The primary mirror is continuously rotated from side to side
about the axis indicated in Figure~\ref{fig:optics}.  The motion is a
rounded triangle scan with an amplitude of \err 2\degs and a frequency
of 0.45~Hz.  This modulation superimposed on that of the entire
telescope yields the scan pattern in Figure~\ref{fig:doublemod} (left
panel).

	The mirror motion is the highest frequency modulation in the
experiment.  This sets the limit below which the data are much less
susceptible to 1/f noise; because a location on the sky is revisited
on time scales of 2.2 seconds, noise on longer time scales has
little impact on the data.  Considering sky rotation and the motion of
the telescope, the actual beam overlap between consecutive mirror
scans is \sima 99\%

	Because the mirror rotation axis is not vertical, there is
also a small modulation in elevation - a bowed pattern in which the
extremes of the mirror modulation rise slightly in elevation.  The
elevation motion could in principle lead to a scan synchronous signal
in the bolometers due to atmospheric emission.  In practice, the
elevation motion ($<$2\mina) is not large enough to generate a
detectable signal.

	The mirror is actuated by a DC motor with solid state PID
control electronics.  The motor was obtained commercially and the
control electronics were built at Berkeley.  Mirror position is
controlled to an accuracy of 1\mina .

\subsection{CCD Cameras and Image Processing}\label{hardware:ccd}

		We use two CCD star cameras as our main absolute
pointing sensors.  CCD data are not used in pointing control, but are
used for post-flight pointing reconstruction.  One camera, referred to
as the primary or boresighted camera, is mounted on the inner
telescope assembly and is boresighted with the telescope beams.  The
second camera, referred to as the secondary or offset camera, is
positioned on the outer frame so that north celestial pole star
(Polaris) is in the camera's field of view during CMB observations -
approximately 40\degs to the right of the boresight, at a fixed
elevation angle of about 31\dega .  Camera data are processed by a
devoted computer with a video digital signal processor (DSP).

		CCD cameras provide an accurate and reliable direct
measurement of the telescope orientation and have a number of
advantages over other instruments such as magnetometers, rate
gyroscopes, or differential GPS systems.  Camera data are not
susceptible to drifts caused by temperature fluctuations, as
gyroscopes and magnetometers are.  CCD cameras are more easily and
precisely calibrated than magnetometers.  A differential GPS receiver
would have many of the advantages of a star camera, but with lower
precision ($>$3\mina).  While CCD data are affected by a rotational
degeneracy, the small offset from the telescope beams to the
boresighted camera's field of view minimizes this effect.

		The boresighted camera data are used for the final
pointing reconstruction and are accurate to \sima 0.5\mina.  The
secondary camera is only accurate to \sima 15\mina, due to the offset
of the measurement from the telescope boresight.  It is used to
identify stars in the primary camera.
	
	The main disadvantage of CCD cameras is that the limited field
of view must contain a sufficiently bright star.  Camera optics allow
a trade-off between field of view and angular resolution.  The
boresighted camera has a field of view of 7.17\degs by 5.50\degs with
a pixel size of 0.84\mins by 0.69\mina.  The resolution is further
improved to \sima 0.5\mins in flight by software interpolation.  The
secondary camera has a larger field of view (14.34\degs by 11.00\dega)
and lower resolution (\sima 1.0\mina).

		The boresighted camera reliably detects stars of V
magnitude 5.0 or brighter.  Stars of V magnitude 5.0 to 6.0 are
detected intermittently, and stars dimmer than V magnitude 6.0 are
rarely detected.  This sensitivity is sufficient for all of \maximai\
and most of \maximaii.  Source brightness is not an issue for the
secondary camera, which always views Polaris (V magnitude = 2.0).

		A small area near the center of the \maximaii\ scan
region contains no bright stars.  Here we rely on heavy interpolation
using data from the rate gyroscope.

\subsubsection{Image Rate and Phase Lag}

		The DSP processes an image once every two cycles of
the FLC.  During CMB observations, the DSP alternately
processes data from each of the two cameras (1 image per 384~ms from
each camera).  During planet observations, the DSP only processes data
from the boresighted camera (1 image per 192~ms).

		The processing time of CCD data in the DSP causes a
200-ms delay between pointing data and bolometer data.  The cameras
internally sample the CCD chips at 30~Hz, asynchronous to the rest of
the system.  This causes a jitter up to 33~msec in the image processing
delay, for an overall delay of 200\err 16~ms.  At our gondola scan
speeds, 200~ms translates to 2\mins to 5\mins of gondola rotation,
which is a significant fraction of our 10\mins beam-size and must be
taken into account.  The \err 16-ms jitter translates to \sima
0.15\mins RMS pointing uncertainty.

\section{Pointing Reconstruction}\label{point:data}

		Pointing reconstruction is based primarily on data
from the CCD cameras.  First, data from the secondary CCD camera are
used to identify sources observed with the boresighted camera.  Next,
boresighted camera data are used to find the pointing solution.  Where
camera data are unavailable, a combination of rate gyroscope data and
numerical interpolation is applied.  Finally, the effect of the
primary mirror modulation is included.  These four steps, and their
contributions to the final pointing uncertainty, are described here.

\subsection{CMB Scan Reconstruction}
\begin{figure}[ht]
\centerline{\epsfig{width=4.0in,angle=0,file=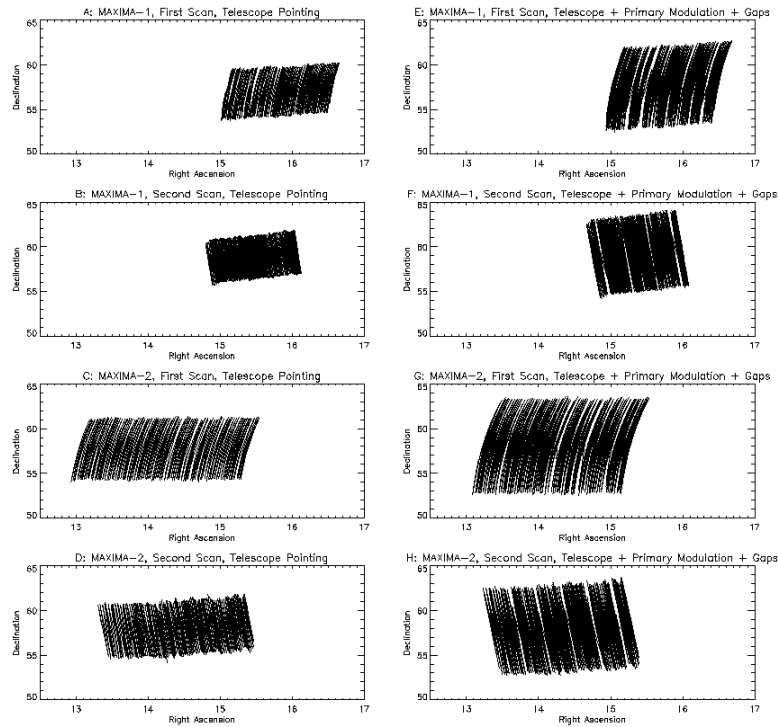}}
\caption[Pointing Reconstruction of Four CMB Scan]{Pointing
Reconstruction for the four \maxima\ CMB observations, two in each of
\maximai\ and \maximaii.  Plots A, B, C, D show the scan pattern of
the telescope, without the effect of the primary mirror modulation.
Plots E, F, G, H show the same scan patterns with the addition of the
4\dega, 0.45-Hz primary mirror modulation.  Plots E, F, G, H include
small gaps, corresponding to subdivisions in the data that coincide
with internal calibration events.}
\end{figure}

\subsubsection{Step 1: Identification of Guide Stars}

		The first stage in pointing reconstruction is the
identification of the stars observed in the boresighted CCD camera.
The camera acquires data from the two brightest stars in its field of
view at any given time.  During a CMB scan, several thousand
detections are made of 16~to~40 different stars of magnitude 6.0 or
brighter (Table~\ref{tab:guide_stars}).

		In principle, guide stars can be identified from their
relative positions, combined with data from the magnetometer and the
telescope elevation angle.  The accuracy of this method is limited by
the magnetometer calibration, the lack of simultaneous measurements of
many guide stars, and the slow pendulum motion of the telescope.
Instead, we identify stars using data from the secondary CCD camera.
The secondary camera is offset from the boresight so that a known
star, Polaris, is constantly in its field of view during CMB
observations.  This provides boresight pointing information with an
accuracy of \sima 15\mins which is sufficient to identify $>$95\% of
the stars detected in the boresighted camera.  The majority of
unidentified ``stars'' are noise in the CCD/DSP system.

\subsubsection{Step 2: Reconstruction Based on Known Guide Stars}

		Up to two stars are located from each image processed
by the boresighted CCD camera.  These data are the basis of the pointing
reconstruction.  Pointing is determined at each time sample from each
known star using a series of standard coordinate transformations.

		Often, pointing can be determined simultaneously from
two different stars in the same field of view.  In these cases the
discrepancy between the two pointing solutions is used to estimate
the overall error in the CCD measurement.  The typical discrepancy is
\sima 0.5\mins (Table~\ref{tab:pointing_error}).

\begin{table}
\begin{center}
\begin{tabular}{c|cccc}
\\CMB & \maximai & \maximai & \maximaii & \maximaii
\\Observation & Scan 1 & Scan 2 & Scan 1 & Scan 2
\\ \hline\hline
\\ Number of Stars & 16 & 16 & 39 & 30
\\\hline
\\Frames w/ 2 Stars & 6929 & 6185 & 10627 & 6077
\\Frames w/ 1 Stars & 4977 & 4879 & 7504 & 7854
\\Frames w/ 0 Stars & 3601 & 2490 & 4821 & 6339
\end{tabular}
\caption[Guide Stars]{The top row is the number of guide
stars used for pointing reconstruction of each \maxima\ CMB scan.  The
next three rows give the number of times 0, 1, or 2 guide stars were
found in a CCD camera image.}
\label{tab:guide_stars}
\end{center}
\end{table}

\subsubsection{Step 3: Interpolation}

	Data are gathered from the boresighted camera once every
384~ms during CMB scans, during which time the telescope moves \sima
8\mina.  Star data are obtained from most, but not all, CCD images.
In rare cases, no stars are detected by the camera for up to eight
seconds at a time, making accurate interpolation more difficult.  Near
the center of the \maximaii\ scan, there is a region with very few
guide stars where interpolation must be used more often over longer
time scales.

		The initial interpolation is from intermittent CCD
data (384~ms or longer between updates) to the consistent 96~ms period
of the control electronics.  In the azimuth, data from the rate
gyroscope are integrated to find position.  Because the gyroscope is
calibrated continuously from the azimuthal CMB scan, this process is
very accurate.  In the elevation direction, the rate gyroscopes are
more difficult to calibrate.  However, the motion of the telescope in
elevation is extremely slow and small, so these data are safely
interpolated numerically.

	A second interpolation from the 96-ms period to the 4.8-ms
period of the bolometer data is purely numerical.  On these shorter
time scales, the telescope beams move much less ($<$2\mins during CMB
scans) and there is no danger of introducing significant pointing
error.

		As a test of accuracy, the azimuth data are
reinterpolated numerically and compared to the gyroscope-based
interpolation.  The difference between the two is used to estimate the
pointing error introduced by interpolation.  Though the RMS
discrepancy between these two methods is very small, the distribution
has extreme outliers corresponding to regions of the sky with few
bright stars.  Interpolated regions with a difference of greater than
3.3\mins are not used in data analysis.

\subsubsection{Step 4: Primary Mirror Modulation}

		In the final stage in pointing reconstruction we add
the effect of the primary mirror modulation.  The angle of the primary
mirror is measured to several arcseconds by a linear variable
differential transformer (LVDT).  The LVDT data are calibrated both
before flight and during flight using data from the planet
observation.

		The motion of the primary mirror moves the telescope
beams primarily in azimuth.  However, there is a small motion in
elevation which is much more difficult to calibrate and is a source of
pointing uncertainty.  The elevation motion depends upon the zero
position of the mirror.  This is measured to \sima $1^o$, which leads
to a conservative pointing uncertainty of about 0.8\mins RMS.

		The absolute boresight offset of each detector, as
measured from planet observations, is included in the pointing
reconstruction at this stage.

\subsection{Pointing Uncertainty}

		The 0.8\mins uncertainty of the primary mirror
modulation is the largest error term in the \maxima\ pointing
solution.  Though purely systematic, the scan pattern and cross-linking
tend to blur out the effect.  In addition, other sources of pointing
uncertainty further randomize the total error.  The overall pointing
error is approximated as a 1\mins gaussian blur.  Simulations show
that the effect of such a pointing error on the angular power spectrum
is a 10\% reduction at \ella =1000 and that this reduction
scales roughly as \ella\sqra.

		Pointing uncertainty is a subdominant source of CMB
power spectrum error at all values of \ella .  While it is possible to
compensate the power spectrum for the reduction caused by pointing
error, we have not done so because it is relatively small, and because
our model of the pointing error as gaussian is not exact.

\begin{table}
\begin{center}
\begin{tabular}{c|ccc|cc}
\\			& \multicolumn{3}{c|}{\bf{Random Errors}} & \multicolumn{2}{|c}{\bf{Systematic Errors}}
\\	 		& CCD 		& Camera 	& Interpolation	& Detector	 	& Primary
\\	 		& Camera 	& Timing	&		& Offset		& Modulation
\\ \hline\hline
\\\maximai 		& 0.46\mina & 0.15\mina & $<$0.001\mina & 0.25\mina & 0.81\mina
\\\maximaii 		& 0.58\mina & \ditto & (see caption) & \ditto & \ditto
\end{tabular}
\caption[Pointing Reconstruction Uncertainties]{Sources of error in
pointing reconstruction.  ``CCD Camera'' is the error in determining
the coordinates of the guide star in the CCD image.  ``Camera
Timing'' is the effect of the timing uncertainty of image acquisition.
``Interpolation'' is the error caused by interpolation over periods
without camera data.  This is negligible for all of \maximai\ and most
of \maximaii .  However, for \sima 20\% of the \maximaii\ scan region
there are very few stars, increasing the use of interpolation and
raising the rms error to \sima 1\mina .  ``Detector offset'' is the
zero-position uncertainty of the telescope beams in azimuth and
elevation.  ``Primary Modulation'' is the uncertainty from the
rotation of the primary mirror.}
\label{tab:pointing_error}
\end{center}
\end{table}

\subsection{Planet Scan Reconstruction}

		Reconstruction of planet scans is very similar to that
of CMB scans with several simplifications.  Because there is always a
known bright source, the planet itself, in the telescope boresight,
source identification is not an issue.  Data are acquired every
192~ms, and a source is found in every image, allowing easy and
accurate interpolation.

		The addition of primary mirror modulation and channel
specific boresight offsets is handled somewhat differently than in CMB
scans.  The calibration of the primary mirror modulator and the offset
of each beam position are taken as parameters.  These
parameters, as well as bolometer time constants, are fit to the
bolometer data using the pointing solution and the known planet
position, calibrating both the primary mirror motion and the
detectors' spatial offsets (\cite{CDWThesis}).

\subsection{CMB Dipole Scan Reconstruction}\label{pdata:dipole}

		During CMB dipole scans, the telescope is fully
rotated in azimuth at 20\dega/sec (18 second period).  At this speed
the CCD cameras provide no meaningful information.  Data from the rate
gyroscopes and the magnetometer, as well as CCD data before and after
the dipole observation, are used to reconstruct pointing.

		These data yield a pointing reconstruction roughly one
order of magnitude less accurate than that of the CMB and planet
observations ($<$10\mins error).  This is sufficient for the dipole
measurement, causing an error of \sima 0.1\% in the absolute
calibration.  The pointing is confirmed using the strong, localized
signal of the Galactic plane in the bolometers.

\chapter{Calibration}\label{chap:calib}

		In this chapter we discuss the calibration of detector
responsivity.  Section~\ref{calib:intro} defines responsivity and the
\maxima\ calibration strategy.  Sections~\ref{calib:dipole}
and~\ref{calib:planet} cover the absolute calibrations from the CMB
dipole and planets.  Section~\ref{calib:relative} deals with the time
dependence of the calibration.  Section~\ref{calib:final} discusses
the use of all these data to determine the overall calibration.
Section~\ref{calib:preflight} describes pre-flight responsivity
testing.

		See Appendix~\ref{chap:calparms} for a complete
summary of calibration parameters and uncertainties for both \maxima\
flights.

\section{Definition and Overview}\label{calib:intro}

		Detector data are recorded in units of bolometer
voltage.  The ratio $\frac{\Delta V_{detector}}{\Delta T_{cmb}}$ is referred
to as the responsivity.  Measurement of this value using known signals
is referred to as responsivity calibration.

		Responsivity is strictly a function of both the
angular scale and the temporal frequency of the observed signal.  The
finite resolution of the experiment leads to an effective reduction of
responsivity for features on small angular scales.  This reduction is
referred to as the beam window function or $B_{\ell}$.  $B_{\ell}$ is
unity at small \ella , as for the dipole calibration, and drops off at
high \ella.  For the planet calibration, the ``beam dilution'' factor
(\S \ref{planet:data}) accounts for the beam function.  For CMB
data analysis, the explicit $B_{\ell}$ is applied to the angular power
spectrum (\S \ref{data:ps}).

		Similarly, temporal filters (e.g. bolometer time
constants, electronic filters) make responsivity a function of
frequency.  These filters are deconvolved from the data in the early
stages of analysis; the responsivity of the deconvolved data is not a
function of frequency.

		Absolute responsivity calibration uses two known
sources during each \maxima\ flight: the CMB dipole and a planet.
Measurements of the CMB dipole give the best absolute calibration for
the 150-GHz and 230-GHz detectors.  Observations of planets (Jupiter
in \maximai\ and Mars in \maximaii) are used to calibrate the 410-GHz
detectors, and are used as a consistency check for the dipole
calibration.  Observations of planets are also used to measure the
size and shape of the telescope beams.  This is not discussed here,
but may be found in \cite{CDWThesis}.

		Responsivity depends on bolometer properties, optical
loading, electrical bias power, and the temperature of the bolometer
thermal reservoir.  The thermal reservoir temperature, controlled by
the adiabatic demagnetization refrigerator, varies significantly
causing responsivity fluctuations.  An internal millimeter wave source
(stimulator) is used to periodically measure responsivity changes.

		The \maximai\ data have a calibration error of 4$\%$,
while the \maximaii\ data have a calibration error of 3$\%$.  These
are the most accurate calibrations achieved by any sub-orbital CMB
experiment.

\section{CMB Dipole}\label{calib:dipole}

	The CMB dipole is the Doppler shift in the observed CMB
temperature resulting from the motion of the Earth relative to the CMB
rest frame.  The dipole amplitude of 3.358\err 0.023~mK has been
measured by the \cobe\ \dmr\ (\cite{Dipole}).  The dipole is the main
calibrator for the 150-GHz and 230-GHz detectors.  It has two main
advantages over point source calibrators.  The first is the optical
spectrum: because the dipole is a small Doppler variation on the
2.725~K of the CMB, it has exactly the same spectral profile as the CMB
anisotropy.  Uncertainties in the detectors' spectral response do not
affect the calibration.  The second advantage of dipole calibration is
the angular scale of the signal.  Because the dipole is \sima 1000
times larger than our telescope beams, uncertainties in the beam
window function do not affect the calibration.

	The 410-GHz detectors, used to confirm the absence of Galactic
dust and atmospheric foregrounds in our data, are deliberately much
less responsive to the CMB spectrum.  Dipole calibrations for these
channels have very low signal-to-noise.  Planet observations are used
to calibrate the 410-GHz detectors.

\subsection{Dipole Observations}

\begin{table}
\begin{center}
\begin{tabular}{c|cc|cc}

\\ & Dipole & Dipole & Observation & Observed
\\ & Amplitude & Elevation & Elevation & Amplitude
\\ \hline\hline
\\\maximai\ & 3.195 mK & 20\dega & 50\dega & 2.04 mK
\\\maximaii\ & 3.010 mK & 48\dega & 32\dega & 1.15 mK
\end{tabular}

\caption[Dipole Scan Parameters]{The CMB dipole signals measured in
each \maxima\ flight.  ``Dipole Amplitude'' includes the effect of the
Earth's motion around the Sun.  ``Dipole Elevation'' gives the angle
from the dipole direction to the horizon during the observation.  For
our observing pattern (azimuthal rotation) this would ideally be
0\dega.  ``Observation Elevation'' is the constant elevation at which
the telescope was rotated to observe the dipole.  The ``Observed
Amplitude'' is the amplitude of the dipole over the region of the
scan.  This varies slightly over the course of the observation, due to
the rotation of the sky.}

\label{tab:dipole_signal}
\end{center}
\end{table}

	Dipole observations are carried out by rotating the telescope
in azimuth with a azimuthal angular velocity of 20\dega /sec.  The
observed signal from the dipole is a sine wave at 55~mHz.

	The observed signal varies from the canonical dipole amplitude
for two reasons.  First, the dipole fluctuates annually due to the
motion of the Earth around the Sun.  Second, the observation is a
circular pattern at a fixed elevation, and does not span the full
extent of the dipole.  The observed signals are listed
in Table~\ref{tab:dipole_signal}.

\subsubsection{Parasitic Signals}

	The rotating scan pattern of the dipole observation is
sensitive to parasitic signals from Galactic dust and in some cases
the atmosphere.  The Galactic dust signal is modeled from frequency
extrapolations of published maps (\cite{FORECAST}, \cite{SFD_Dust}).
The dust signal is much smaller than the dipole signal at 150~GHz,
except near the Galactic plane.  Data within 5\degs of the Galactic
plane are neglected in data analysis.  Elsewhere, the dust model is
fit to the data along with the dipole model.  Overall normalization is
taken as a free parameter to account for uncertainties in the
frequency extrapolation of the dust signal.  In practice, the dust
model does not affect the dipole calibration due to its low amplitude
and lack of a dipole-like spatial component.

	An additional small signal, believed to be atmospheric, is
observed in the beginning of the \maximai\ dipole calibration.  In
\maximai\ we began the dipole observation near the beginning of the
flight, while the telescope was still ascending from \sima 21.5~km to
the final observing altitude of \sima 38.5~km.  The additional signal
was observed during the first third of the observation (altitude
$<$30~km).  We believe that this signal is atmospheric for four
reasons: 1, it is highly correlated in all the optical bolometers; 2,
it is spectrally consistent with atmospheric emission, being larger
for the higher frequency detectors; 3, it is spatially stable on the
scale of a few minutes, but varies on longer scales; 4, the magnitude
of the signal declines steadily with altitude.

	The atmospheric signal is corrected using data from the
410-GHz bolometers.  These data, which are relatively insensitive to
the dipole and sensitive to the parasitic signal, are used as a
template for the parasitic signal in the 150-GHz and 230-GHz data.  A
correction is applied for the CMB sensitivity of the 410-GHz
detectors, as calibrated by planet observations.  As with Galactic
dust, we find that fitting the believed atmospheric signal does not
affect our final dipole calibration values.  It does slightly
increase calibration uncertainty, because of noise in the 410-GHz data
used to `model' the parasitic signal.

\subsection{Dipole Data Analysis}\label{calib:dipdata}

\begin{figure}[hp]
\centerline{\epsfig{width=5.0in,angle=0,file=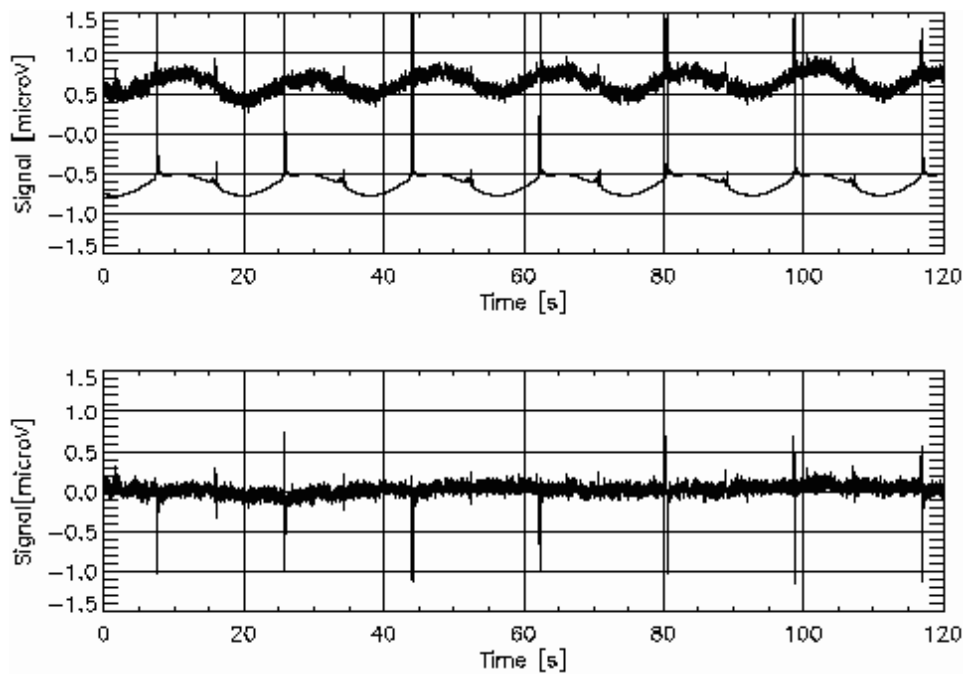}}
\caption[\maximaii\ Dipole Data] {\maximaii\ 150-GHz Dipole Data and
Fit.  {\bf{Top panel:}} The top trace is the data from a 150-GHz
bolometer during observations of the CMB dipole.  An overall gradient
has been removed and the offset is arbitrary.  The sinusoidal signal
is the CMB dipole modulated by the rotation of the telescope (\sima 18
second period).  The large periodic spikes are caused by intense dust
signals near the Galactic plane.  The lower trace is a model curve,
with amplitude fitted to the data, including the CMB dipole and a
Galactic dust map.  {\bf{Bottom panel:}} The difference between the
model and the fit in the top panel are shown.  The model deviates from
the data near the Galactic plane crossing.  These highly localized
signals are not well fit with \sima 10\mins pointing
reconstruction accuracy of the dipole observation.}
\label{fig:dipole}
\end{figure}

	During each flight, the dipole was observed for \sima 30
minutes (100 rotations).  For each detector, the effects of electronic
filters and bolometer time constants are first deconvolved from the
entire data stream.  Data from each rotation are then fit
independently according to the model,
\begin{equation}
T_{detector} = (A * T_{CMB,Model}) + (B * T_{dust}) + (C * N_{Drift}) + D
,\end{equation}
\noindent in which $T_{detector}$ is time stream of detector data in
voltage units, $T_{CMB,Model}$ is the CMB dipole model in units of
temperature contrast, $T_{dust}$ is the Galactic dust model,
$N_{Drift}$ is linear drift, and $A$, $B$, $C$, $D$ are fitting
constants.  $A$ is the calibration of the detector to CMB signals.
Data collected within 5\degs of the Galactic plane are not used for
fitting because the pointing reconstruction of the dipole observation
is not accurate enough for localized features.

\begin{figure}[hp]
\centerline{\epsfig{width=5.0in,angle=0,file=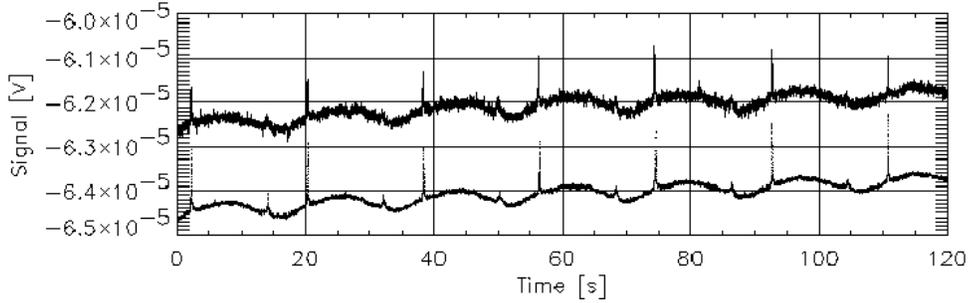}}
\caption[\maximai\ Dipole Data] {\maximai\ 150-GHz Dipole Data and
Fit.  Similar to the \maximaii\ data in the top panel of
Figure~\ref{fig:dipole}.  The top trace is raw data with arbitrary
offset, while the bottom trace is a model including dipole, dust, and
a linear drift.  In this case, the model curve also includes an
additional term based on 410-GHz data to account for the believed
atmospheric signal observed at low altitude (the first \sima 1/3 of
the data).  This is why the model curve is not noiseless.}
\label{fig:dipole2}
\end{figure}

	For \maximai\ the believed atmospheric signal is taken into account by modifying the fit to,
\begin{equation}\label{eqn:dipmodelfull}
T_{detector} = (A * T_{CMB,Model}) + (B * T_{dust}) + (C * N_{Drift}) + D + (E * T_{410})
.\end{equation}
\noindent $T_{410}$ is time stream data from a 410-GHz detector and
E is an additional fitting parameter.  Because the 410-GHz data do
have some very small sensitivity to the CMB, $A$ is no longer an
unbiased calibration.  To account for this, the 410-GHz data are first
calibrated using planet data and the parameter $A$ is
corrected,
\begin{equation}
A' = A - (E * Cal_{410}).\end{equation}
\noindent $Cal_{410}$ is the planet-based calibration of the 410-GHz
data and $A'$ is the true calibration of the low frequency channel.
In practice the 410-GHz term doesn't affect calibration values by more
than 0.5~$\sigma$, though it does increase uncertainties.  The correction
from $A$ to $A'$ has a negligible effect on both the calibration and
the calibration uncertainty due to the small value of the $E$ parameter.

	Each rotation yields a calibration value ($A$ or $A'$ above)
and an associated error range.  These are combined statistically, with
2-$\sigma$ outliers excluded.  Data from \sima 80 rotations are analyzed
from each flight, with 2 to 8 excluded as outliers for each detector.

\subsection{Dipole Calibration Error Sources}\label{calib:diperror}

\subsubsection{Detector Noise}

	Dipole calibration uncertainty (1-4\%) is dominated by
detector noise at the dipole observation frequency of 56~mHz.  The
fitting routine is found numerically to reject noise beyond a
fractional bandwidth of \sima 0.5.  Noise is effectively reduced by a
further factor of $\sqrt{2}$ by the known phase of the dipole model.
Raw detector noise near 56~mHz for a 150-GHz bolometer is typically
150~nV~Hz$^{-0.5}$.\footnote{In fact, detector noise performance at
these frequencies below the 1/f knee (\sima 0.5 Hz) is much less
consistent than at high frequencies, and can vary by a
factor of four or more between detectors.}  Considering bandwidth
and phase constraints, this yields an expected \sima 20-nV noise level
for a single dipole fit.  For a typical 150-GHz detector the amplitude
of the dipole response is \sima 70~nV.  We therefore expect a
statistical uncertainty of \sima 30\% from a single rotation.

	Detector noise is the only source of statistical uncertainty
in the calibration, and can be estimated directly from the scatter of
the individual, single rotation calibrations.  Such analysis yields
single rotation statistical uncertainties of 10\% to 30\% for 150-GHz
detectors in either flight.  These numbers are somewhat lower than
predicted due to imperfect understanding of detector noise at very low
frequencies.

	An integration of 80 to 90 dipole observations per flight
provides a total statistical uncertainty of 1.4\% to 4.2\% for 150-GHz
detectors in \maximai\ and 1.1\% to 2.5\% in \maximaii.  Note that
when combining data from multiple detectors, we use the highest
statistical uncertainty of the combined channels.

	For the beginning of the \maximai\ dipole scan, a parasitic
signal is seen, and is modeled as described above.  This modeling is
based on data from a 410-GHz detector that contributes additional
detector noise to the calibration uncertainty.  Because the
coefficient $E$ in Equation~\ref{eqn:dipmodelfull} is small, the noise
contribution from the 410-GHz detector is suppressed and increases the
final calibration uncertainty by only about 0.5\%.

	Because the CMB responsivity of the 230-GHz detectors is
60-70\% that of the 150-GHz detectors, they have a proportionally
higher statistical error.

\subsubsection{Other Error Sources}

	In addition to statistical uncertainty, there are a number of
known systematic effects, though none have a significant impact on the
calibration.  First, dipole pointing reconstruction
(\S \ref{pdata:dipole}) is accurate to \sima 10\mins and contributes
negligibly (\sima 0.1\%) to the calibration uncertainty.  Second, the
characterization of the high pass filter in the bolometer readout is
accurate to \sima 0.3\%.  Third, the dipole model derived from the
\cobe\ measurement is accurate to 0.68\%.  Finally, the signal from the
dipole is small enough that bolometer saturation is negligible
(Table~\ref{table:linearity}, Appendix~\ref{chap:callinear}).

\begin{table}
\begin{center}
\begin{tabular}{c|ccc}
\\ & Dipole & Planet & Stimulator
\\ & Responsivity & Responsivity & Responsivity
\\ & Change & Change & Change
\\ & \tiny{(\maximai/2)} & \tiny{(\maximai/2)} & \tiny{(\maximai/2)}
\\ \hline\hline
\\ 150-GHz & \small{$<$0.05\%/$<$0.05\%} & \small{0.7-2.5\%/0.2-0.5\%} & \small{0.1-0.5\%/0.1-1.5\%}
\\ 230-GHz & \small{$<$0.05\%/$<$0.05\%} & \small{1.0-6.1\%/0.2-0.4\%} & \small{0.2-4.0\%/0.1-0.5\%}
\\ 410-GHz & \small{$<$0.05\%/$<$0.05\%} & \small{1.9-7.7\%/0.4-1.0\%} & \small{0.5-3.3\%/0.3-2.7\%}
\end{tabular}
\caption[Calibration Linearity]{A summary of detector response
linearity during observations of various calibration sources.
``Responsivity Change'' is the fractional change in detector
responsivity due to the optical load from the calibrator.  For the
dipole calibration there was no responsivity change within noise
limits.  For the planet and
stimulator calibrations, quoted values are derived from the maximum of
the signal.  Stimulator power was reduced in \maximaii\ to
increase linearity.  The impact of these responsivity changes on
calibration accuracy is discussed in the text.

These values are derived from measured changes in bolometer
resistance, as described in Appendix~\ref{chap:callinear}}.
\label{table:linearity}
\end{center}
\end{table}

\section{Planets}\label{calib:planet}

\begin{figure}[hp]
\centerline{\epsfig{width=2.in,angle=90,file=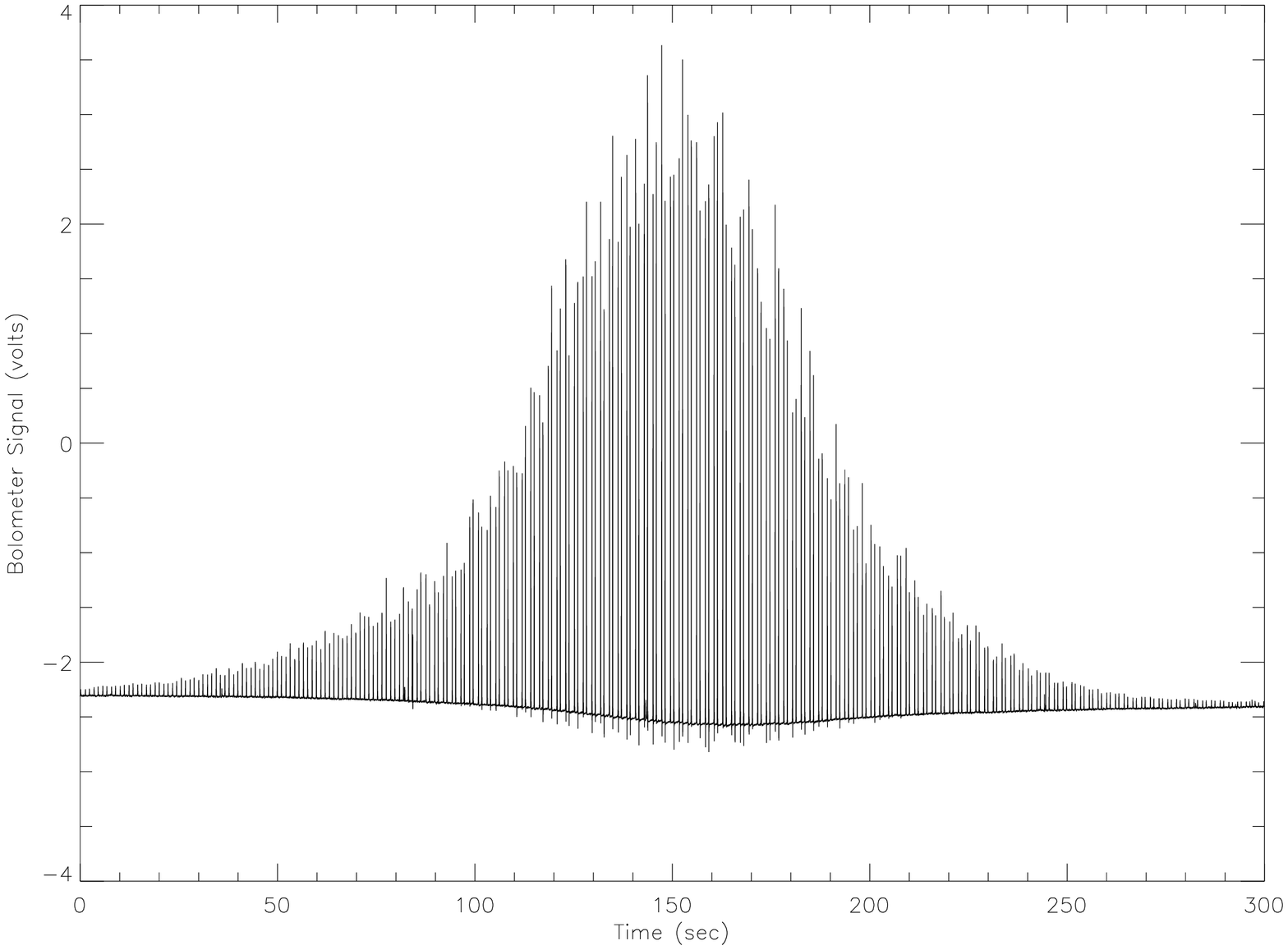}
\epsfig{width=2.in,angle=90,file=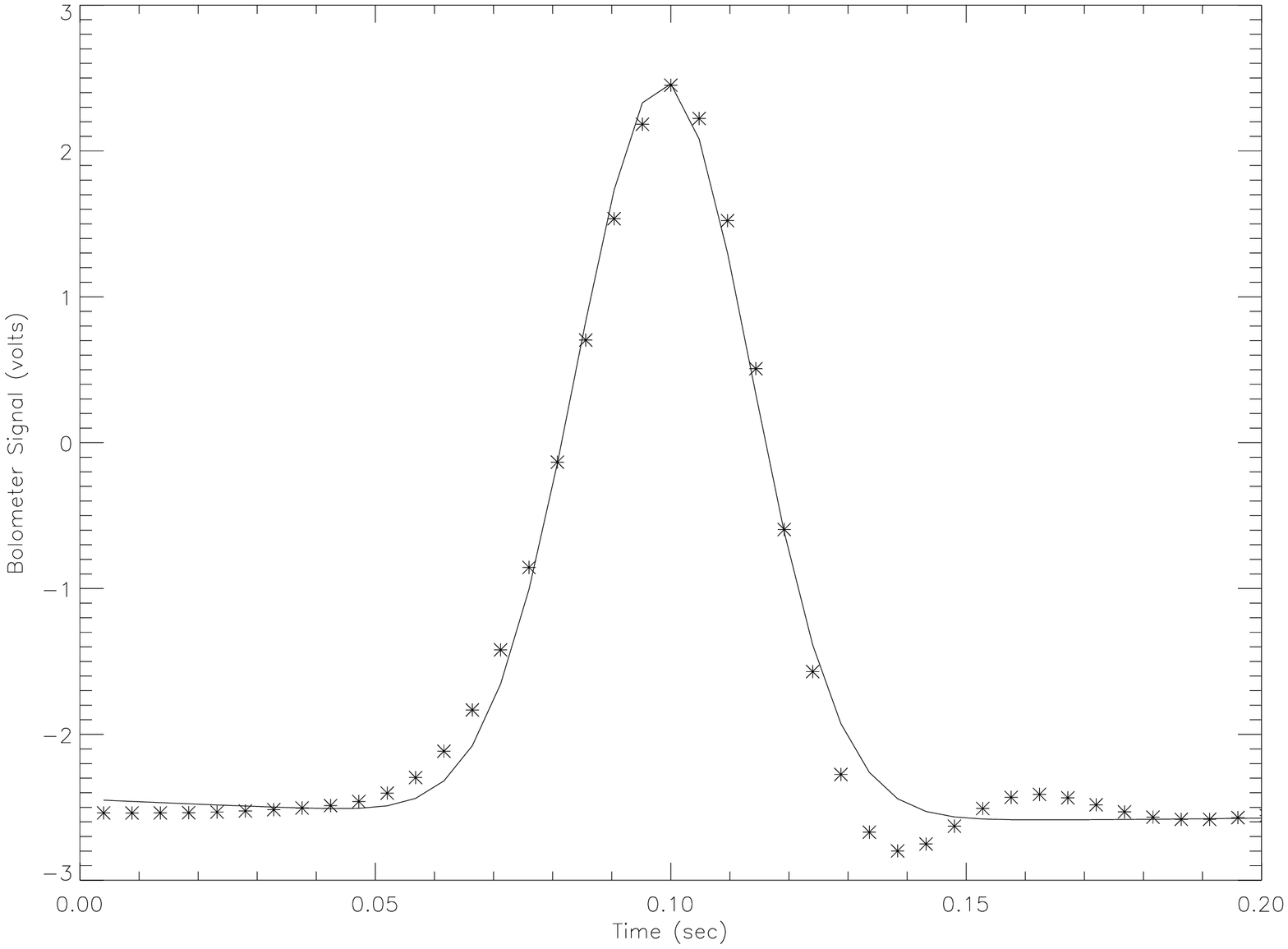}}
\centerline{\epsfig{width=2.in,angle=90,file=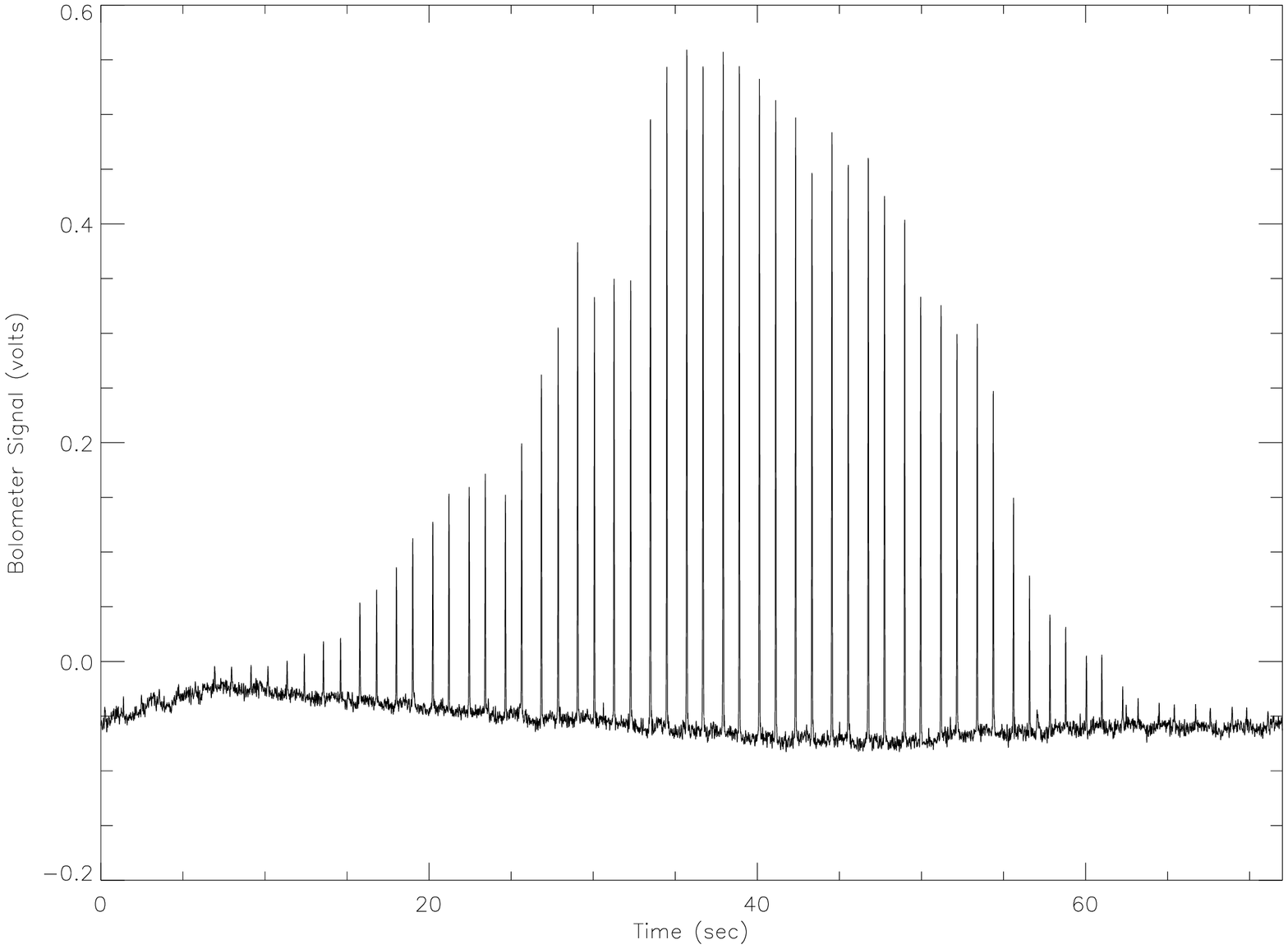}
\epsfig{width=2.in,angle=90,file=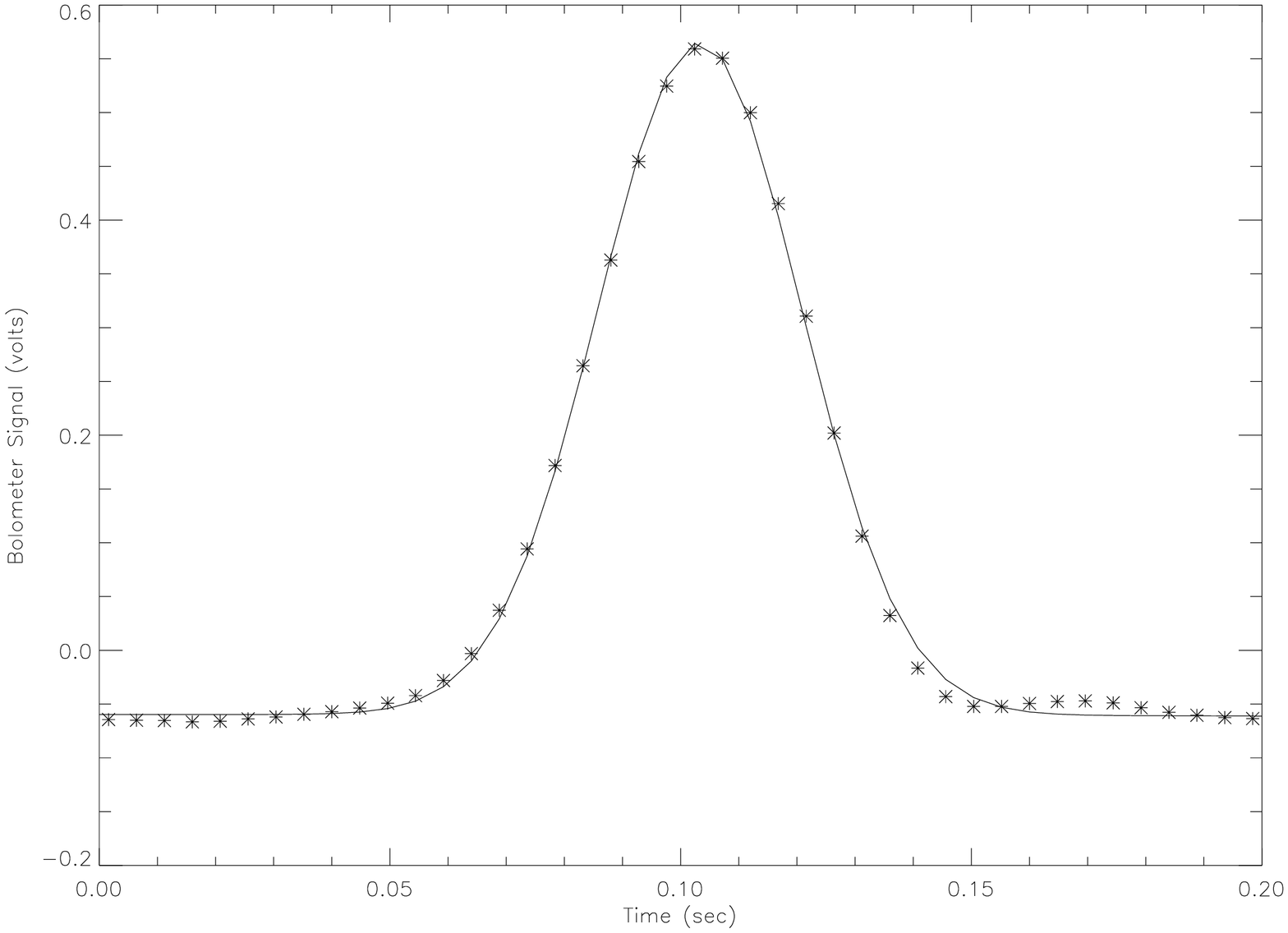}}
\caption[Planet Calibration]{Data from observations of Mars and
Jupiter.  {\bf{Left Top:}} Raw data from the entire \maximai\ Jupiter
observation for a 150-GHz detector.  Each of the closely spaced
vertical lines is a single pass of the planet.  The modulation of the
envelope is caused by the drift of the planet in elevation.  The
apparently non-gaussian shape of the envelope is caused by known
pointing control imperfections.  {\bf{Right Top:}} Raw data from a
single pass of Jupiter for a 150-GHz detector.  These data are an
expanded view of one of the vertical spikes in the plot on the left.
The scan speed is determined by the modulation of the primary mirror.
Signal-to-noise is \sima 1000.  The solid line is a gaussian fit.  The
`bump' on the right side of the plot is caused by bolometer time
constants and electronic filters.  Deconvolution of these effects
removes the bump.  {\bf{Left and Right Bottom:}} As above, for a
\maximaii\ Mars observation.  The signal-to-noise ratio is \sima 150.}
\label{fig:planet_data}
\end{figure}

\begin{figure}[ht]
\centerline{\epsfig{width=4.5in,angle=0,file=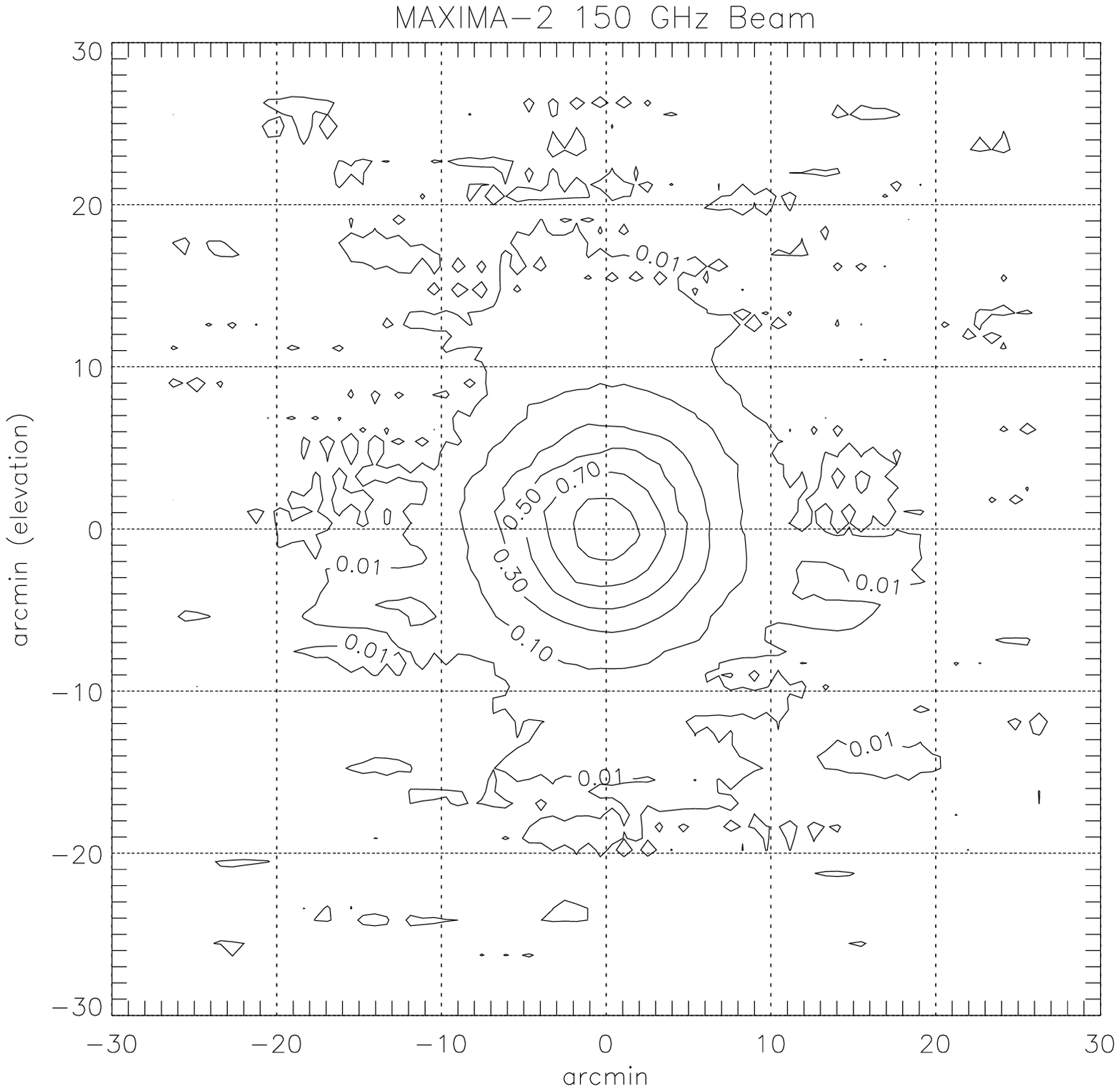}}
\caption[Beam Map] {Beam map of a 150-GHz from \maximaii.  The two
dimensional plot is obtained from the data in
Figure~\ref{fig:planet_data} by the methods detailed in \cite{CDWThesis}.
The overall normalization is used for calibration.  In addition, the
integrated beam size determines the `beam dilution' factor for a given
source size.}
\label{fig:beam_map}
\end{figure}

		In each flight observations are made of a planet:
Jupiter in \maximai\ and Mars in \maximaii.  These data serve several
purposes.  They are used to measure optical parameters of the
telescope, including the size, shape, and absolute position of the
each beam, and to calibrate the primary mirror modulation.  They are
also used to measure the electronic filters and bolometer time
constants.  These measurements are described in \cite{CDWThesis}.

		Here we discuss the use of planet data for
responsivity calibration.  Planet data are the only absolute
calibration source for the 410-GHz detectors.  They are also used to
confirm the dipole calibration of the 150-GHz and 230-GHz detectors.

\subsection{Planet Data}\label{planet:data}

		During planet observations, the telescope tracks the
planet in azimuth while remaining at fixed elevation.  As the planet
drifts through the elevation of the observation, the modulation of the
primary mirror ``slices'' the beams across the planet many times.  As
the planet drifts in elevation the spatial response of each beam is
measured in two dimensions.

		Responsivity calibration is obtained from the maximum
voltage response.  Expected signals are derived from published
measurements and models of planet temperatures and emissivities
(\cite{Jupiter}), combined with the spectral response of the
detectors.

		A correction is applied for beam dilution - the
fraction of the telescope beam filled by the planet.  Jupiter had an
angular diameter of 46.5\asec during \maximai\ and Mars had an angular
diameter 12.7\asec during \maximaii.  The beam dilution is the
integral of the spatial response of the detector over the area of the
planet, normalized by the integral of the entire spatial response.
The dilution factors vary from 3.4$\times 10^{-3}$ to 4.4$\times
10^{-3}$ for \maximai\ and 3.1$\times 10^{-4}$ to 4.4$\times 10^{-4}$
for \maximaii.  This correction is the dominant error source for the
planet calibration in both flights.

		An additional correction of roughly 5\% for \maximai\
and 1\% for \maximaii\ is applied for the reduction in responsivity
caused by the optical load from the planet
(Table~\ref{table:linearity}).  This effect was neglected in the
initial \maximai\ data analysis and caused an apparent small
systematic discrepancy between the dipole and Jupiter calibrations.
This discrepancy was within the error range of the Jupiter calibration
and does not affect the CMB map or power spectrum.

\subsection{Planet Calibration Error Sources}\label{calib:planerror}

		The dominant error term for the planet calibration is
the uncertainty in the beam dilution factor.  The uncertainty in the
integrated beam response is 5\% to 10\%.  In addition, there is a
possibility of small, broad side-lobes that are not measured in the
beam maps.  We assign an uncertainty of 10\% from beam shape errors.
Beam shape error, especially that due to broad side-lobes, is partially
correlated between detectors because of their shared optics.

		Uncertainties in the effective brightness temperature
of the planets contribute 5\% to calibration error.  The brightness
temperature of Mars has been modeled to this accuracy, both by
extrapolations from high frequency observations (\cite{Mars1}) and by
physical modeling (\cite{Mars2}).  The atmospheric properties for
Jupiter make modeling relatively difficult.  Our expected Jupiter
signal is based on published brightness ratios between Jupiter and
Mars (\cite{Jupiter}).  The planet temperature uncertainty is fully
correlated between all the detectors.

		Measurements of the detector spectra contribute 1-2\%
error at 150~GHz, 3-7\% at 230~GHz, and 2-3\% at 410~GHz.
Measurements of the peak planet voltage contribute 1-4\% error; one
detector in \maximaii\ was anomalously noisy, increasing this term to
\sima 10\%.  Uncertainty in the bolometer saturation is negligible
(Table~\ref{table:linearity}).

\section{Time Dependent Calibration}\label{calib:relative}

\begin{figure}[ht]
\centerline{\epsfig{width=3.0in,angle=90,file=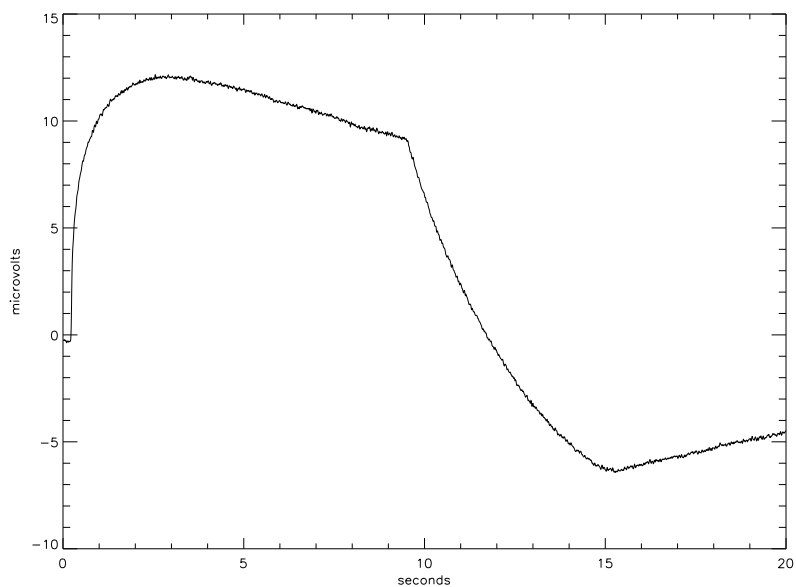}}
\caption[Internal Relative Calibration Signal] {An internal relative
calibration event.  The voltage response of a 150-GHz detector in
\maximaii\ to the stimulator lamp is shown.  Power to the stimulator
is constant for 0.2~sec$<$t$<$ 9.5~sec and is zero elsewhere.
The rounding of the response is caused by the stimulator on/off time
constant, which is \sima 100 times longer than the bolometer
time constant.  The sloping of the signal in the `on' state and the
slow settling of the baseline after the event are caused by a weak
high pass filter in the bolometer readout electronics.}
\label{fig:stim_event}
\end{figure}

		The responsivity of bolometers varies with their
operating temperature.  A temperature increase of 1~mK in the thermal
reservoir leads to a responsivity reduction of 1-2$\%$.  The
temperature varied by \sima 6~mK over the course of data collection in
\maximai\ and by \sima 21~mK in \maximaii.  The larger variation in
\maximaii\ was due partly to the length of the flight, and partly to
technical difficulties with the adiabatic demagnetization
refrigerator.

\begin{figure}[ht]
\centerline{\epsfig{width=3.0in,angle=90,file=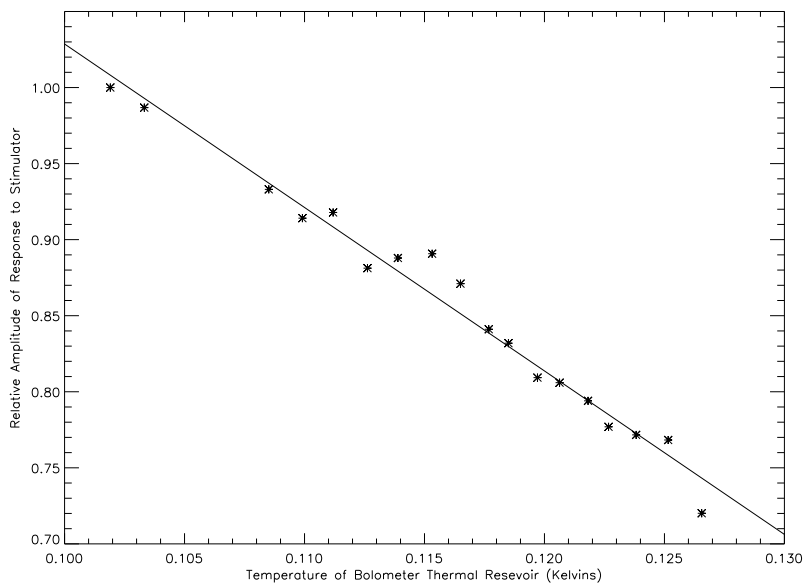}}
\caption[Temperature Dependence of Calibration] {Temperature
dependence of the responsivity of a 150-GHz detector.  These data were
collected during \maximaii.  The large range of
temperature was due to the length of the flight and a partial failure
of the ADR.  Points at 0.105~K and 0.107~K are
omitted due to high noise.  The point near 0.127~K, measured shortly
after sunrise, shows less responsivity than would be expected from the
nighttime data.  During \maximai, the temperature of the thermal
reservoir was more stable, varying from 98-104~mK over the course
of data collection.}
\label{fig:relative_cal}
\end{figure}

		Responsivity variations are monitored using the
internal millimeter wave source (stimulator) described in
Section~\ref{receiv:stim}.  The stimulator is activated for 10-sec
periods, once every 20 minutes during the flight.  The signal in a
150-GHz detector from a stimulator event is shown in
Figure~\ref{fig:stim_event}.  The response to the stimulator signal in
the 230-GHz and 410-GHz detectors is more than twice than that in the
150-GHz detectors.  In addition, the stimulator location is asymmetric
with respect to the detector array, with the four closest detectors
having twice the response of the four farthest ones.

		To obtain relative calibration values from stimulator
events, we begin by subtracting an overall gradient from each event to
remove the effects of detector drift.  We then perform a linear fit
between pairs of stimulator events.  The slope of this fit is the
calibration ratio between the events, while the offset of the fit is
simply an offset in the detector data.

		Once the relative calibration at each stimulator event
is known, we fit the values to a linear function of the temperature of
the bolometer thermal reservoir (Figure~\ref{fig:relative_cal}).  This
fit is combined with the absolute calibration to obtain the overall
calibration as a function of time throughout the flight (\S
\ref{calib:final}).

\subsection{Relative Calibration Error Analysis}\label{calib:stimerror}

		The relative calibration between stimulator events is
affected by random variations (detector noise or stimulator
instability), but is not affected by systematics that are consistent
between stimulator events.  Uncertainties in the spectra of the
detectors and the beam filling of the stimulator signal are purely
stable.  The reduction in bolometer responsivity due to the large
optical load of the stimulator is nearly stable
(Table~\ref{table:linearity}).  Though it does vary with bolometer
temperature, this variation contributes less than 0.1\% error to the
relative calibration.

		Random errors in the comparison of stimulator
events are 1-2\%.  Instabilities in the stimulator current account for
$<$0.5\% of this, while detector noise accounts for the rest.

	We treat the relative calibration as a linear function of
temperature and take the optical load during the flight as
constant.  These assumptions are supported by bolometer models
(\cite{CDWThesis}, \cite{SabrinaBolo}) and contribute negligibly to
calibration error.

\section{Combined Calibration}\label{calib:final}

		The overall calibration for each detector was obtained
by combining an absolute calibrator with the relative calibration.
The absolute calibrator is the CMB dipole for 150-GHz and 230-GHz
detectors and is the planet scan for 410-GHz detectors.  The relative
calibration is based on the temperature of the bolometer thermal
reservoir and the responsivity-temperature relation obtained in
Section~\ref{calib:relative}.  Temperature is monitored continuously.

		The overall calibration error is the combined error
from the absolute calibrator and the relative calibration.  Relative
calibration error varies over the course of the flight.  Quoted values
are based on averages over the CMB observations.  Relative calibration
error is subdominant for \maximaii\ (1\% to 2\%) and negligible for
\maximai\ ($<$0.1\%).

		The published \maximai\ data are conservatively
assigned the highest calibration uncertainty of the detectors used,
4\%.

\section{Pre-flight Responsivity Tests}\label{calib:preflight}

		Rough measurements of the bolometer responsivity were
made before each \maxima\ flight as a diagnostic of receiver
performance.  For these measurements, an optical load at 273 K (0\degs
C) is placed at the entrance of the receiver optics.  A spinning fan
blade coated with millimeter wave emitting material at \sima 300~K
(room temperature) periodically blocks the optics from the colder
load.  This setup gives a \sima 28-K chopped signal against a \sima
300-K background at receiver entrance.  An internal neutral density
filter reduces chopped signal and external loading by a factor of 100.

		In principle, stimulator data taken on the ground can
be used to transfer pre-flight responsivity tests to in-flight
calibration.  In practice, the errors in this procedure are far larger
than those for either the dipole or planet calibrations.
Uncertainties in the emissivity of the load, the temperature contrast,
and the transmittance of the neutral density filter are more than
30$\%$.  While a determined effort might reduce this uncertainty
somewhat, reaching the accuracy of the in-flight calibrations would be
impractical or impossible.

\chapter{Data Processing and Analysis}\label{chap:data}

	This chapter describes the processing and analysis of \maxima\
data to obtain CMB maps and power spectra.  It begins with an overview
(\S \ref{data:intro}), followed by a look at the key steps
(\S \ref{data:bolo} -~\ref{data:ps}).  Pointing reconstruction and
calibration have already been described in Chapters~\ref{chap:point}
and~\ref{chap:calib}.  Results are presented in
Chapter~\ref{chap:results}.  Tests of systematic effects at various
stages are described in Chapter~\ref{chap:systematic}.

	References of particular interest are \cite{StomporData},
which deals with noise estimation and map making using \maximai\ data
as an example, and \cite{JBMadcap}, which describes \madcap, a software
package used in \maxima\ data analysis.

\section{Introduction}\label{data:intro}

	The process of extracting cosmological information from
experimental data consists of two general stages.  First is the
reduction of raw data into time ordered pointing, calibration, and
detector data.  During this data reduction, experimental details are
accounted for, low quality data are removed, calibration is
determined, telemetry glitches and cosmic ray hits are identified, and
a pointing solution is found.  Data with glitches or poorly determined
pointing are flagged.  These tasks are described in
Chapters~\ref{chap:point} (Pointing Reconstruction)
and~\ref{chap:calib} (Calibration) and in Section~\ref{data:bolo} in
this chapter (Detector Data Preparation).

	The second general stage of data treatment, converting
time-ordered data into CMB maps and power spectra, is less dependent
on details of the experiment and is more numerically intensive.
First, the detector noise spectrum is estimated, calibration is
applied, and glitches in the data are replaced with unbiased noise
(\S \ref{data:noise}). Then, the time-ordered data along with the time
domain noise estimate are used to produce a CMB map and spatial noise
correlation matrix (\S \ref{data:map}).  Finally, the map and noise
correlations are used to estimate the power spectrum of the CMB
(\S \ref{data:ps}).  More detailed discussions, and treatments of
concerns beyond the scope of this chapter, can be found in
\cite{StomporData}, \cite{JBMadcap}, and \cite{BJK98}.

\section{Bolometer Data Preparation}\label{data:bolo}

\begin{table}
\begin{center}
\begin{tabular}{cc|cc}
\\Flight & CMB Scan & Total & Time Stream
\\ & & Measurements & Segments
\\ \hline\hline
\\ \maximai & 1 & 1,060,050 & 11
\\ \maximai & 2 &   917,900 & 10
\\ \maximaii & 1 & 1,477,146 & 6
\\ \maximaii & 2 & 1,412,874 & 6
\end{tabular}
\caption[Structure of \maxima\ data sets]{The total number of
measurements and the number of time stream subdivisions for each CMB
observation.}
\label{tab:chunks}
\end{center}
\end{table}

	The CMB data consist of approximately $2\times 10^{6}$
measurements per detector for \maximai\ and $3\times 10^{6}$
measurements per detector for \maximaii.  The raw output of the
experiment was approximately $4\times 10^{6}$ measurements for
\maximai\ and $9\times 10^{6}$ measurements for \maximaii, including
CMB data, calibration, and test data.

	The first stage of data preparation is to isolate the detector
time streams corresponding to the CMB observations and to divide them
into shorter segments (Table~\ref{tab:chunks}).  Internal relative
calibration events cause a very large signal in the optical channels,
and are a natural break-point between time stream segments.  These events
occur at intervals of roughly 250,000 measurements (20 minutes); this
is the maximum length of any time stream segment.  The time streams
are also divided when the scan pattern of the telescope is changed.
Further subdivisions are added with the requirement that noise
properties be approximately stationary within each segment.  The
shortest segments are roughly $3\times 10^4$ elements in \maximai\ and
$1\times 10^5$ elements in \maximaii.  Gaps between segments are at
least $2\times 10^4$ elements (\sima 100 sec).  This gap length,
combined with an electronic 15-mHz highpass filter, eliminates
significant noise correlations between segments.

	Within each time stream segment, overall offset, gradient, and
quadratic components are subtracted.  Electronic filters and bolometer
time constants will be deconvolved in a later stage of data analysis.

	Measurements compromised by `glitches' - short transients such
as cosmic ray hits and telemetry drop outs - are flagged and excluded
from data analysis.  Typically, \sima 2\% of data are flagged as
glitches.  The gaps left in the time stream are short (typically \sima
10 measurements) and frequent; the noise on either side cannot be
considered uncorrelated.  Treatment of these gaps is discussed in
Section~\ref{data:noise}.

\section{Composition of Bolometer Data}\label{data:TOD}

	The time ordered detector data are modeled as a sum of sky
signal, noise, and a parasitic signal synchronous with the modulation
of the primary mirror.  Sky signals are filtered by the detector time
constants and electronics.  Various noise components are subject to
some or all of these filters.\footnote{Johnson noise, for example, is
subject to electronic filters, but not time constants, while photon
noise is subject to all the filters.  In practice these difference
are not important to data analysis.}  Mirror scan synchronous noise is
assumed to be subject to all filters.  Denoting the raw time stream
for a single detector for a single segment of a flight as $d_F$, the
$i^{th}$ time sample $d_F(i)$ within these data is
\begin{equation}\label{eqn:dat_struc_1}
d_F(i) = \sum_{j} F(i,j)[t_{sky}(\gamma(j)) + x(\alpha(j))] + n_t(i)
.\end{equation}
\noindent $F(i,j)$ is the combined electronic and time constant
filter, $t_{sky}$ is the temperature fluctuation of the CMB (and
foregrounds) as a function of position, $\gamma$ is the pointing
solution as a function time, $x$ is the mirror scan synchronous signal
as a function of mirror orientation, $\alpha$ is the orientation of
the mirror as a function of time, and $n_t(i)$ is the total
time stream noise including applicable filters.

\section{Noise Estimation, Gap Filling, and Filter Deconvolution}\label{data:noise}

	Once the deglitched time ordered data are available, the
frequency power spectrum of the time domain noise is estimated.  This
estimate is used for map making (\S \ref{data:map}), and also to
restore time stream continuity over the short gaps left by
deglitching.  After noise estimation, the effects of instrumental
filters are removed from the data and the calibration is applied.

\subsection{Time Ordered Data Power Spectrum}

	The noise power spectrum $P_{n}(f)$ is defined as
$|\widetilde{n_t}|^2$.  $P_{n}(f)$ for \maxima\ consists of
approximately white noise above \sima 0.5~Hz.  At lower
frequencies the noise power spectrum is described by a power law,
$1/{f^n}$, with 1.0$<$n$<$2.5.  Above 20~Hz, the noise power spectrum
drops rapidly due to the effects of an electronic filter.

	The power spectrum $P(f)$ of the full data (signal + noise) is
similar to that of the noise.  The sky signal is very
small compared to the time domain noise, but in some \maximai\ channels and most
\maximaii\ channels, the mirror scan synchronous signal is
significant, causing peaks in the power spectrum at the frequency of
the scan and its first harmonic (0.45~Hz and 0.90~Hz).

\subsection{Estimation of Signal Free Noise}\label{estim1}

	Noise estimation is simplest under the assumption of a highly
noise dominated time stream, free from both CMB and parasitic signals.
In this case, noise estimation consists of five steps.

	The first step is prewhitening, which reduces the noise
correlation length.  The data are convolved with a filter selected to
yield constant $P(f)$ for $f \rightarrow 0$.  Gaps in the convolved
data are widened to account for the width of the filter.  The second
step is the estimation of the power spectrum for continuous blocks of
data, using standard methods (\cite{NumRecipe}).  Next, time stream
gaps are filled with a constrained realization of the estimated noise
based on neighboring valid data (\cite{Filling}).  Data within the
gaps are still flagged as invalid for purposes of map making.  In the
fourth step, the now continuous time stream is used to re-estimate the
noise power spectrum.  A deviation from the previously estimated power
spectrum usually indicates a poorly selected prewhitening filter.  In
this case the process is repeated with a new filter.  Finally, all
filters (instrumental and prewhitening) are deconvolved and
time stream gaps are widened accordingly.

\subsection{Treatment of Data with Signal and Noise}

	In the case of significant signal, as with the mirror scan
synchronous signal, the above noise estimation procedure is performed
iteratively.  On each iteration, the CMB sky and scan synchronous
signal are estimated by map making (\S \ref{data:map}).  The expected
signals are then subtracted from the time stream yielding an estimated
noise-only time stream.  The signal estimation is then repeated and
residual signal is again subtracted from the time stream, yielding
increasingly pure noise.

	When this process converges and an accurate noise estimate is
obtained, we return to the raw time stream of signal and noise.
Instrumental filters are deconvolved from this full time stream and
the gaps are correspondingly widened.  The gaps are filled using the
corresponding samples in the best estimate of the noise-only
time stream.  The calibration obtained in Chapter~\ref{chap:calib} is
then applied.  The full time stream with gaps filled and filters
deconvolved is an input of the map making procedure ($d_t$ in
Section~\ref{data:map}).

\subsection{Time Domain Noise Correlations}

	After an estimate of the noise power spectrum is found, a
time-time noise correlation matrix is calculated.  This matrix will be
used in map making.  Under the assumption of stationary, gaussian
noise, noise correlation matrix elements are given by the Fourier
transform of the noise power spectrum as
\begin{equation}
N_t(i,j) \equiv \tilde{P}(t=(i-j)\Delta)
,\end{equation}
\noindent where $\Delta$ is the sample interval.  In practice, this
calculation is prone to numerical errors.  These can be reduced by
smoothing the noise power spectrum, $P(f)$, before deconvolving the
prewhitening filter.

	We also make the approximation that $N_t(i,j)$ is zero for
$|i-j| \geq \lambda_c$, where $\lambda_c$ is the time stream correlation
length.

\section{Map Making}\label{data:map}

	Map making is the next stage of data analysis.  The time
stream data and the time-time noise correlations are combined with the
pointing solution (\S \ref{point:data}).  The products of map making
are a vector of sky temperature contrast for each pixel, $m_p$, and
the pixel-pixel noise correlation matrix, $N_p$.

	The removal of instrumental filters reduces
Equation~\ref{eqn:dat_struc_1} to,
\begin{equation}\label{eqn:dat_struc_2}
d_t = A m_p + n_t + B x_q
,\end{equation}
\noindent where $d_t$ is the deconvolved time stream of signal and
noise, $m_p$ is a vector of sky temperature contrast for each pixel,
assigning each time sample to exactly one pixel, and $n_t$ is the
noise-only time stream with correlations given by $N_t$, and $x_q$ is
a vector of the effective temperature contrast of the mirror scan
synchronous signal.  $A$ is a pointing matrix, and $B$ is a pseudo
pointing matrix assigning each time sample to a mirror orientation.

	Neglecting the $B x_q$ term, a closed form solution for sky
signal $m_p$ and the associated pixel-pixel noise correlation matrix
$N_p$ is given by:
\begin{eqnarray}
m_p & = & (AMA^T)^{-1}AMd_t,\label{eqn:map_sol_1}\\
N_p & = & (AMA^T)^{-1}(AMN_tMA^T)(AMA^T)^{-1}.\label{eqn:map_sol_2}
\end{eqnarray}
\noindent A minimum variance solution is obtained if $M\equiv{N_t}^{-1}$.
Though this matrix inversion is not impractical for \maxima\ data
sections (up to 250,000 samples), various approximations to
${N_t}^{-1}$ (\cite{StomporData}) have been used and found to be
consistent with the exact approach.

\subsection{Removal of Unwanted Signals}\label{data:ssynch}

	The $B x_q$ term in Equation~\ref{eqn:dat_struc_2} represents
the mirror scan synchronous signal.  However, the treatment of this
signal applies equally well to other unwanted signals, such as overall
temperature offsets.

	Mirror positions can be treated as extra pixels, observed
simultaneously with the real (sky) pixels.  Concatenating $B$ and
$x_q$ with the corresponding matrixes for the sky data yields
\begin{eqnarray}
A^\prime & = & [A,B],\\
m_p^\prime & = & \left[ \begin {array}{c} m_p \\ x_p \end {array} \right],
\end{eqnarray}
\noindent where $A^\prime$ and $m_p^\prime$ are effective pointing and
signal matrixes including both sky and scan synchronous signals.
Unlike the original $A$ matrix, $A^\prime$ assigns each time sample to
two pixels: one for the sky coordinate and one for the mirror
orientation.  The solution for $m_p^\prime$ and the corresponding
$N_p^\prime$ are still given by Equations~\ref{eqn:map_sol_1}
and~\ref{eqn:map_sol_2}.  Once the map of sky pixels and extra pixels
is calculated, the signal at the extra pixels is marginalized to
obtain the sky map.\footnote{The extra pixels approach is not the only
way to account for unwanted signals.  Other methods involve
marginalization earlier in map making.}  A further refinement is
required because the scan synchronous signal is not stable over the
entire flight, or even over every data segment.  To account for this,
a different set of template pixels is used for every several minutes
of data.  The stability of the signal over these time scales is tested
directly.

	This procedure relies on the orthogonality of the scan
synchronous signal and the sky stationary signal from the CMB.  Due to
the double modulation of the telescope in azimuth, the sky signal
varies between primary mirror scans; subtraction of mirror synchronous
signals does not remove sky stationary signals.  This is confirmed by
a power spectrum analysis of maps made with and without treatment of
the scan synchronous signal.  The power spectra are consistent, though
the maps without the scan synchronous signal treatment appear
excessively noisy.

	The same approach can be used for a variety of applications.
For example, overall offsets of data segments are assigned to
fictitious pixels before combining them.  Similarly, we use the $A$
matrix to assign all corrupted (glitch) data to a single fictitious
pixel.

\subsection{Combined Map}\label{combine}

	The final CMB map and pixel-pixel noise correlation matrix
includes data from multiple photometers and multiple independent data
sections from each photometer.  Assuming that these data are
uncorrelated, the equations in this section can be applied directly to
the combined data set.  The time stream data and pointing matrixes are
concatenated, and the noise correlation matrixes are combined into a
single block diagonal matrix.  The numerical cost is not increased
substantially given the known sparsity of the resulting noise
correlation matrix.  Alternatively, it may be desirable to analyze
blocks of data separately, and add (or subtract) them later.  This
process involves the noise-weighted addition of the CMB maps, and
the summing of the inverses of the noise correlation matrixes.

	Regardless of the method used, it is difficult to combine
data with little spatial overlap.  We rely upon the common structure
in the overlap region to constrain the unknown relative offset between
data sections.  Insufficient overlap will cause spurious shifts
between sections of the map, or, in extreme cases, will lead to a
singularity and cause the map making (or map combination) to fail.

	Combining data from different detectors, each with some
calibration error, can make overall error estimation difficult.  Due to the high accuracy
of the CMB dipole calibration, we neglect this effect and assign the
combined data the highest calibration uncertainty of the individual
channels.

\section{Angular Power Spectrum Estimation}\label{data:ps}

	The final stage of data analysis is estimation of the angular
power spectrum of the CMB, based on the temperature contrast map and
pixel-pixel noise correlation matrix.  Because there is no closed form
for the most likely angular power spectrum, iterative estimation is
necessary.  For this process we use \madcap, a parallel (supercomputer)
software package for CMB data analysis.  \madcap 's CMB power spectrum
estimation is an implementation of Newton-Raphson iteration, an
approach discussed in \cite{BJK98}.\footnote{The \madcap\ package also
includes an implementation of the map making algorithm - one of
several used in \maxima\ data analysis.}

	Power spectrum estimation requires a pixel-pixel correlation
matrix for the CMB map, including both signal and noise.  Because
the CMB signal and the instrumental noise are assumed to be
realizations of independent random Gaussian processes, the total
pixel-pixel correlation matrix, $D_p$, is the sum of the signal
correlation matrix, $S_p$, and the noise correlation matrix $N_p$.
The latter has already been found.  A signal correlation matrix
element, for an assumed CMB power spectrum, is given by
\begin{equation}
S_{pp^\prime} = \sum_{l} \frac{2 l + 1}{4 \pi} {B_l}^2 C_l P_l (X_{pp^\prime})
.\end{equation}
\noindent $C_l$ is the CMB power at multipole $l$, $P_l$ is the
$l^{th}$ Legendre polynomial, and $X_{pp^\prime}$ is the angle between
the pixels.  $B_l$ is spherical harmonic decomposition of the
telescope beam shape - the angular window function.  We assume an
effective circularly symmetric beam profile for any given combination
of photometers as described in \cite{WuBeam}.  If necessary, the
angular window function for the map pixel size may be similarly
included.

		Due to the limited sky coverage of the experiment, each
multipole is not treated as an independent variable.  Instead they are
grouped into bins (typically 8 to 12 bins of width 75 to 150
multipoles).  Bins may or may not be weighted and/or overlapping.  In
practice we use either top hat shaped bins for simplicity, or use
overlapping bins weighted to eliminate residual correlations between
bins.  If bin width is not too narrow, the difference is very small.

	Calculating $S_p$, and therefore $D_p$, for a given multipole
binning and assumed power level within each bin, allows us to find the
probability distribution for the map, $d_p$, using,
\begin{equation}\label{eqn:quadestim}
P(d_t|C) \propto \exp \left\{ - \frac{1}{2}\left( {d_t}^T {D_p}^{-1} d_t + Tr[\ln D_p] \right) \right\}
,\end{equation}
\noindent where $C$ denotes the power level in each bin, as used in the calculation of $D_p$.

	Newton-Raphson iteration (\cite{BJK98}) makes the assumption
that the logarithm of this function is quadratic and determines the
deviation of the assumed power at each bin from the maximum of the
quadratic.  Any analytic function is increasingly quadratic near a
peak, so calculating this correction iteratively will converge upon a
peak.  The curvature of the probability distribution around the peak
is used as a direct measure of statistical uncertainty in each power
spectrum bin.  Issues of non-convergence and the possibility of
convergence on local maxima are explored in the references.

	Power spectrum estimation is the most numerically intensive
step in the data analysis process.  \maxima\ power spectra have been
calculated using supercomputers at NERSC\footnote{National Energy
Research Scientific Computing Center at the Lawrence Berkeley National
Laboratory} and at the University of Minnesota.  A power spectrum from
the 3\mins pixelization \maximai\ map requires about 20 hours on 256
processor nodes on the NERSC T3E supercomputer.

\subsection{Power Spectrum Uncertainties}\label{ps:errors}

	Uncertainties in the angular power spectrum are derived from
instrumental noise, limited sky coverage (\ie~sample variance), and
uncertainties in calibration, instrumental filters, pointing
reconstruction, and beam shape measurements.  Most of these effects
vary strongly with \ella .  Foreground signals are a
different kind of uncertainty, discussed in Section~\ref{sys:foreg}.

	The quadratic estimator directly assesses the effects of
instrumental noise and sample variance.  Noise is most important at
high multipoles where there are few observations per mode.  It is
quantified by the pixel-pixel noise correlation matrix $N_p$.  Sample
variance is most important at low multipoles where the fewest modes
are observed; poorly sampled modes do not strongly affect the
probability of a given measured map in Equation~\ref{eqn:quadestim}.
The combination of these two random effects is the dominant error
source at all multipoles.  The inherent asymmetry of these errors is
modeled using the offset log-normal distribution of \cite{BJK2K}.

	Calibration uncertainty (Chapter~\ref{chap:calib}) is
independent of the multipole bin.  As such it does not affect the
shape of the power spectrum.  Calibration errors are most important
for combining data sets.  Given the high accuracy of the \maxima\
calibration from the CMB dipole ($\leq $3-4\%), we treat the calibration
as perfect during data analysis, and assign the highest calibration
uncertainty of a set of channels to their combination.

	Beam shape errors, including errors in the approximation of
circular symmetry and variations between channels in a combined data
set, affect the spatial window function $B_l$.  These effects are
discussed in great detail in \cite{WuBeam}.  Beam shape errors are
significant only at high multipole bins, reaching \sima 15\% at
$l=1000$.

	Neither instrumental filters nor pointing reconstruction
contribute significantly to angular power spectrum errors.
Uncertainties in instrumental filters are small at frequencies
contributing to CMB observations.  Pointing error
(Chapter~\ref{chap:point}) tends to systematically decrease power at
high multipoles, effectively blurring out the small scale features in
the CMB.  This effect becomes \sima 10\% at $\ell = 1000$, but is less
important than beam shape error at all angular scales.

\chapter{Results}\label{chap:results}

In this chapter we present the results of the \maxima\ experiment.
Data were analyzed using the methods described in
Chapter~\ref{chap:data}.  This chapter starts with an overview of the
\maxima\ data products (\S \ref{results:intro}), presents the CMB map
and angular power spectra (\S \ref{results:map}, \S \ref{results:ps}),
and concludes with cosmological interpretations (\S\ref{results:cosm}).
Chapter~\ref{chap:systematic} presents the results of systematic 
error tests, including difference maps and power spectra.

\begin{figure}[t]
\centerline{\epsfig{width=5.5in,angle=0,file=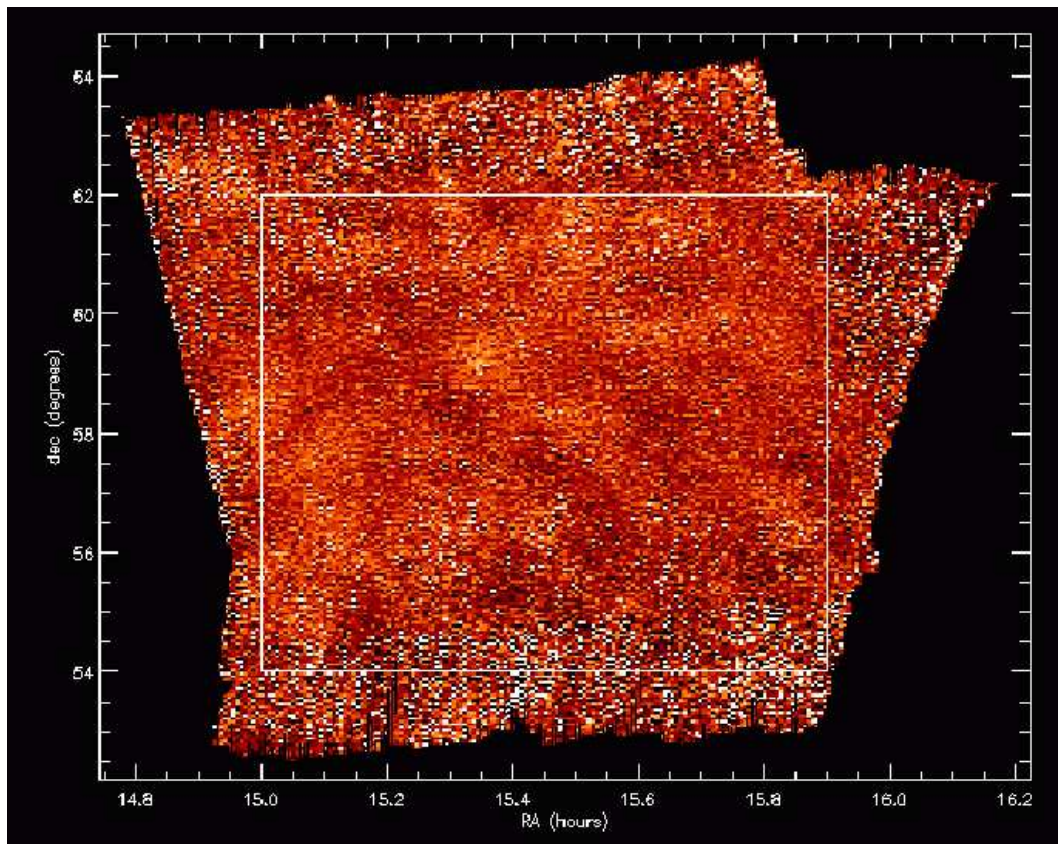}}
\caption[CMB Temperature Map]{The 3\mins resolution \maximai\ map
derived from three 150-GHz detectors.  The indicated central region
has the highest signal-to-noise and best cross-linking and was used in
the \cite{LeeResults} analysis.  The color scale covers a temperature
contrast of -750~\micro K (black) to +750~\micro K (white).}
\label{fig:map0}
\end{figure}

\section{Introduction}\label{results:intro}

Science data have been derived from the subset of \maximai\ detectors
which show the highest sensitivity and which pass all consistency and
systematic tests.  CMB data are obtained from three 150-GHz detectors,
designated as `B25,' `B34,' and `B45.'  Data from a 230-GHz detector,
`B33,' were initially included, but later failed consistency tests at
\ella $>$785 and were omitted from higher resolution analyses.  Data
from a 410-GHz detector, `B22,' are used to monitor dust and
atmospheric signals (Chapter~\ref{chap:systematic}).

The original analysis at 5\mins resolution was published in
\cite{HananyResults}.  A companion paper, \cite{BalbiResults}, used
these data for cosmological parameter estimation.  Results of a 3\mins
resolution analysis have been published in \cite{LeeResults}, with
cosmological parameters estimated in \cite{StomporResults} and
\cite{AbroeResults}.  Results from the \maximaii\ data are not
available at this time.

Publicly available \maxima\ data can be obtained by request or at:

\noindent http://cosmology.berkeley.edu/maxima/data\_release/

A number of papers have been published comparing \maximai\ results
with those of other CMB experiments (e.g. \cite{WangCompare},
\cite{JaffeMaxiboom}).  Power spectra derived from \maximai,
\boomerang, \cbi, \dasi, and \vsa\ are generally consistent with each other
and with the weaker constraints of previous generation experiments.
\cite{JaffeMaxiboom} is a combined analysis of the 5\mins \maximai\
data and the \boomerang\ data, including all noise correlations.

\section{The 3-arcminute CMB Map}\label{results:map}

The current best map from \maximai\ uses the data from three 150-GHz
detectors, analyzed with 3\mins square pixels.  For \sima 10\mins
instrumental resolution, the 3\mins pixel window
function has little effect and can be ignored.  The
map consists of \sima 40,000 pixels; a central rectangular region
consists of \sima 23,000 pixels with uniform sampling, good
cross-linking, and a signal-to-noise of \sima 5 per 10\mins beam-size.
(Figure~\ref{fig:map0})

	Foregrounds are negligible over the entire area
(\S \ref{sys:foreg}).  Observed signals on angular scales from
10\mins to 5\degs are consistent with Gaussianity under a variety of
tests including the method of moments, cumulants, the Kolmogorov test,
the $\chi^2$ test, and Minkowski functionals tests (\cite{WuGauss}).

\section{Angular Power Spectra}\label{results:ps}

\begin{figure}[ht]
\centerline{\epsfig{width=4.0in,angle=0,file=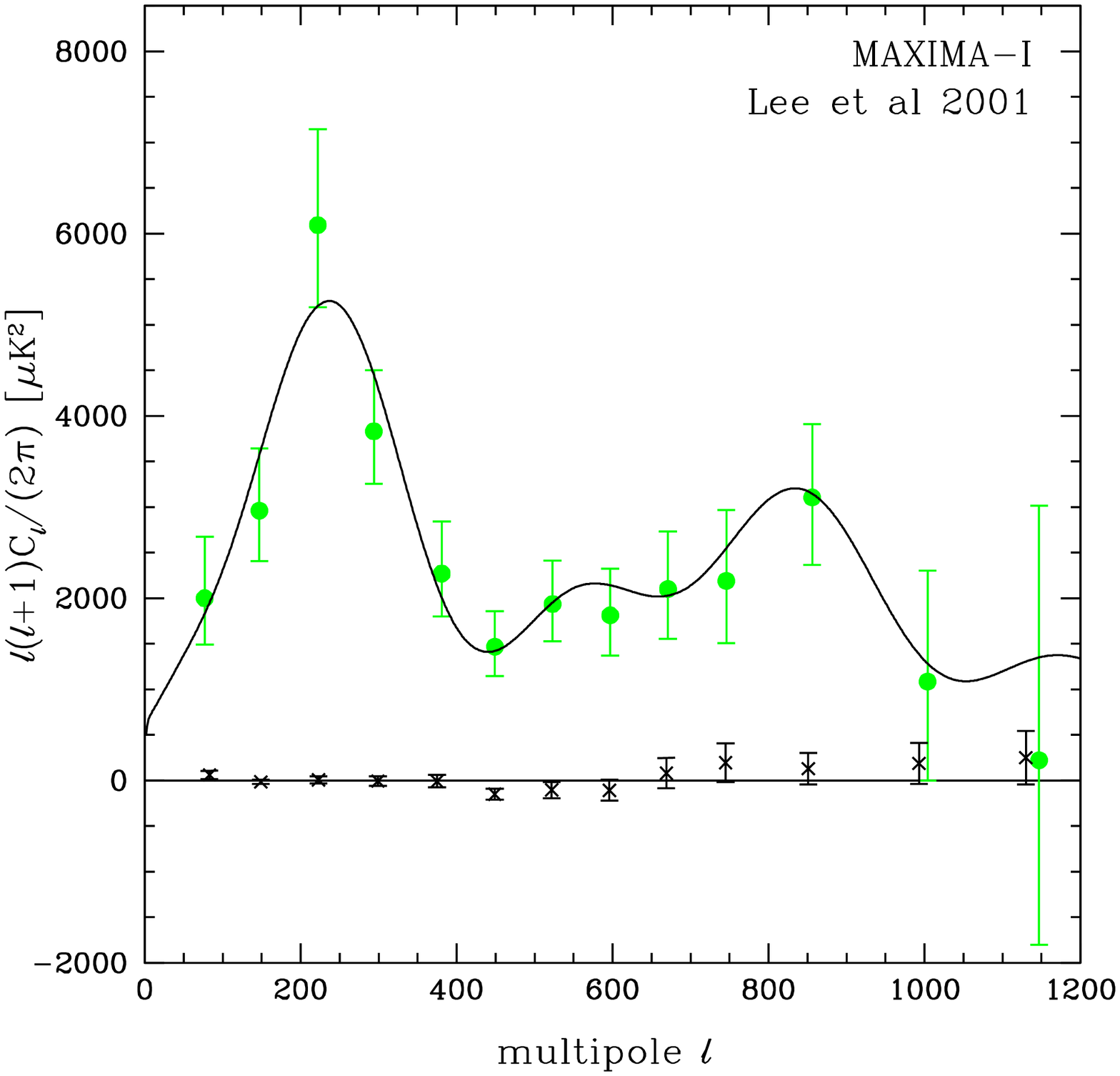}}
\caption[\maximai\ Angular Power Spectrum]{The power spectrum of the
CMB using a hybrid analysis of 5\mins resolution (up to \ells =
335) and 3\mins resolution (over \ells = 335) maps.  Error bars
show the statistical uncertainties from
Table~\ref{tab:lee_cl_estimates}.  The solid curve is the power spectrum
of the best fit model from \cite{BalbiResults} with $\Omega_b = 0.1$,
$\Omega_{cdm} = 0.6$, $\Omega_{\Lambda} = 0.3$, $n = 1.08$, and $h =
0.53$.  The crosses are the power spectrum of the difference between
one detector and the combination of the other two.  (\cite{LeeResults})}

\label{fig:hybrid}
\end{figure}

The \cite{LeeResults} analysis uses only the central \sima 23,000
pixel region of the 3\mins map.  This greatly reduces the
computation time needed for consistency tests requiring a large number
of power spectrum calculations (\S \ref{sys:consistency}).  A
power spectrum computed from the full map agrees with that of the
central region, but was not published in \cite{LeeResults} due to the lack
of systematic error testing at the time.

\begin{table*}
\begin{center}
\begin{tabular}{cclccl} \hline \hline
$\ell_{eff}$ & $[\ell_{min},\ell_{max}]$ & $\ell(\ell+1)C_{\ell} /2\pi$ & Beam Error & Pointing Error &$\Delta T$ \\
   &         & ($\mu K^{2}$) & (\%) & (\%) & ($\mu K$) \\ \hline 

77 & [ 36, 110]& $1999_{-506}^{+675}$ & $\pm 0$ & $\pm 0$  &          $45_{-6}^{+7}$ \\
147 & [ 111, 185]& $2960_{-554}^{+682}$ &  $\pm 0.6$ &  $\pm 0.2$ &        $54_{-5}^{+6}$ \\
222 & [ 186, 260]& $6092_{-901}^{+1052}$ &  $\pm 1.5$ &  $\pm 0.4$ &       $78_{-6}^{+6}$ \\
294 & [ 261, 335]& $3830_{-577}^{+670}$ &  $\pm 2.5$ &  $\pm 0.8$ &        $62_{-5}^{+5}$ \\
381 & [ 336, 410]& $2270_{-471}^{+569}$ & $\pm 3.5$ &  $\pm 1.2$ &        $48_{-5}^{+6}$ \\
449 & [ 411, 485]& $1468_{-325}^{+387}$ & $^{+5}_{-4.5}$  &  $\pm 1.7$ &        $38_{-4}^{+5}$ \\
523 & [ 486, 560]& $1935_{-408}^{+475}$ & $^{+6.5}_{-6}$  & $\pm 2.3$ &        $44_{-5}^{+5}$ \\
597 & [ 561, 635]& $1811_{-441}^{+511}$ & $^{+8}_{-7} $&  $\pm 3.0$  &        $43_{-6}^{+6}$ \\
671 & [ 636, 710]& $2100_{-546}^{+629}$ &  $^{+9.5}_{-8.5}$ &  $\pm 3.7$ &        $46_{-6}^{+6}$ \\
746 & [ 711, 785]& $2189_{-680}^{+777}$ &  $^{+11}_{-10}$ &  $\pm 4.6$ &        $47_{-8}^{+8}$ \\
856 & [ 786, 935]& $3104_{-738}^{+805}$ &  $^{+14}_{-12}$&  $\pm 5.6$ &        $56_{-7}^{+7}$ \\
1004 & [ 936, 1085]& $1084_{-1085}^{+1219}$ & $^{+18}_{-15}$ &  $\pm 7.7$  &    $33_{-22}^{+13}$ \\
1147 & [ 1086, 1235]& $223_{-2025}^{+2791}$ &  $^{+25}_{-18}$ &  $\pm 10.2$  &    $15_{-15}^{+29}$
\\
\hline
\end{tabular}
\end{center}
\caption[Power Spectrum Data]{The uncorrelated angular power spectrum
from \maximai\ (\cite{LeeResults}).  The first four \ells bins are derived from the 5\mins
resolution map, while the last nine are derived from the 3\mins
map.  $\ell_{min}$ and $\ell_{max}$ give the dominant range for each
of the overlapping bins.  Statistical errors are 68\% confidence offset-log normal
probability distributions with a constant prior (\cite{BJK2K}), and are
purely uncorrelated.  Beam errors for the effective combined beam of
the detectors (\cite{WuBeam}) are highly correlated between
bins.  Pointing uncertainty is an upper limit.

Complete results are available at:
http://cosmology.berkeley.edu/maxima/data\_release/ }
\label{tab:lee_cl_estimates}
\end{table*}

At high \ella , essentially all information is contained in the high
signal-to-noise region and errors are not substantially increased by
restricting the map area.  At low \ella , the restricted region
significantly increases sample variance errors.  To avoid a loss of
precision, the power spectrum up to \ells = 335 was calculated from
the full scan region using the well tested 5\mins resolution map of
\cite{HananyResults}, while the higher \ells data were calculated from
the restricted region of the 3\mins map.  The combined power spectrum
benefits from large sample area at low \ella, and high resolution at
high \ells with relatively low computational cost.

This combined power spectrum is presented in
Table~\ref{tab:lee_cl_estimates}.  A constant 8\% error due to
calibration uncertainty is not listed in the table.

Since publication of the \cite{LeeResults}, systematic testing
of the full 3\mins map and associated power spectrum has been
completed.  This power spectrum (Figure~\ref{figure:3arcminps}) is in
strong agreement with that of \cite{LeeResults}.

\begin{figure}[ht]
\centerline{\epsfig{width=4.0in,angle=270,file=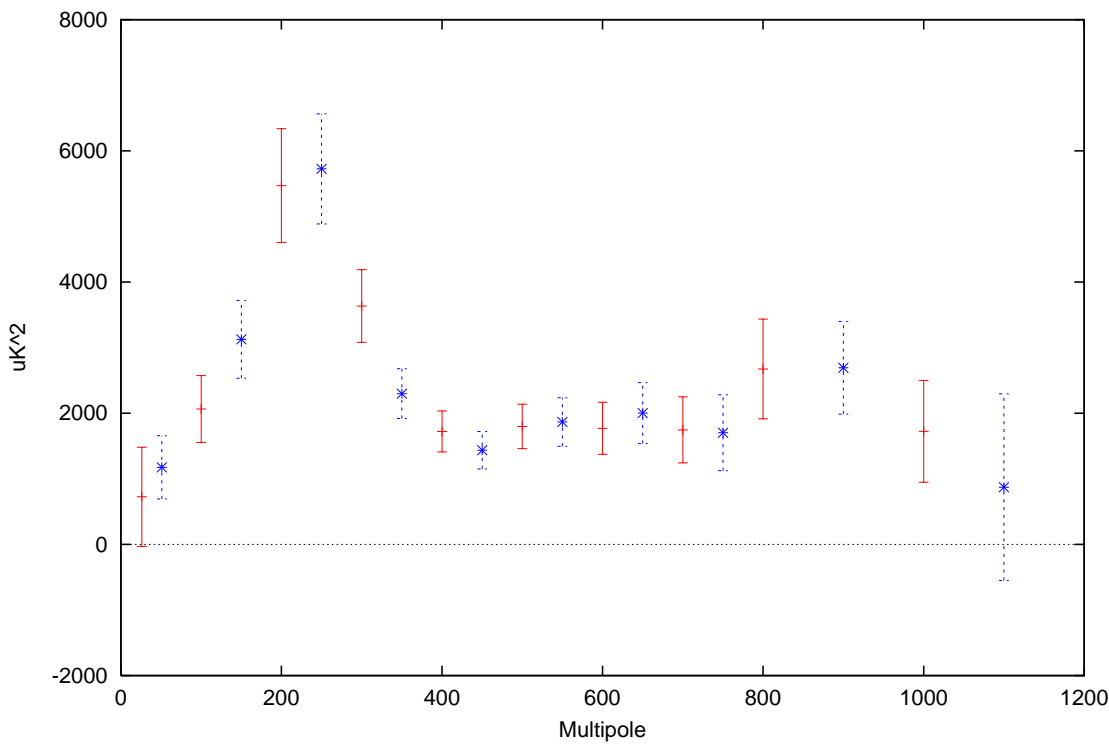}}
\caption[3-arcminute Full Map Power Spectrum] {Angular power spectra
from the full 3\mins CMB map.  Dashed (red) and starred (blue)
data points represent separate interleaved analyses.  Either of these
sets alone represents the statistical weight of the experiment.  These
analyses show no substantial deviations from the \cite{LeeResults}
data.}
\label{figure:3arcminps}
\end{figure}

\section{Cosmological Implications}\label{results:cosm}

The most obvious feature of the angular power spectrum is a clear
peak at $\ell \simeq 220$ followed by relatively low power and the
suggestion of additional peaks at higher \ella .  The presence of a sharp
peak is consistent with an inflationary Big Bang and rules out the
majority of cosmological defect models.  The spatial Gaussianity of
the CMB map provides further evidence for inflationary models.

Within standard inflationary models, cosmological parameters have been
determined by Baysian (\cite{BalbiResults}, \cite{StomporResults}) and
frequentist methods (\cite{AbroeResults}).  All of these are found to
yield consistent best fit models and error ranges.  The results quoted
here are from the \cite{StomporResults} analysis based on the
\cite{LeeResults} power spectrum unless otherwise noted.

Seven independent parameters were varied: $C_{10}$, the amplitude of
fluctuations at \ells = 10; $\Omega_b h^2$, the physical density of
baryons; $\Omega_{cdm} h^2$, the physical density of cold dark matter;
$\Omega_{\Lambda}$, the cosmological constant; $\Omega_{tot}$, the total
energy density; $n_s$ the spectral index of primordial fluctuations;
and $\tau_c$, the optical depth of reionization.  Parameters are
sampled over the following ranges:
\begin{eqnarray}
C_{10} && \mbox{is continuous} \nonumber\\
\Omega_b h^2 & = & 0.00325, 0.00625, 0.01, 0.015, 0.02, 0.0225, ..., 0.04, 0.045, 0.05, 0.075, 0.1\nonumber\\
\Omega_{cdm} h^2 & = & 0.03, 0.06, 0.12, 0.17, 0.22, 0.27, 0.33, 0.4, 0.55, 0.8\nonumber\\
\Omega_{\Lambda} & = & 0.0, 0.1, 0.2, ..., 1.0\nonumber\\
\Omega_{tot} & = & 0.3, 0.5, 0.6, 0.7, 0.75, ..., 1.2, 1.3, 1.5\nonumber\\
n_s & = & 0.6, 0.65, 0.7, 0.75, 0.8, 0.85, 0.875, ...1.2, 1.25, ..., 1.5\nonumber\\
\tau_c & = & 0, 0.025, 0.05, 0.075, 0.1, 0.15, 0.2, 0.3, 0.5\nonumber
\end{eqnarray}

Likelihoods are interpolated between grid points for additional
resolution.  The likelihood at each grid point is calculated using an
offset log normal approximation (\cite{BJK2K}), including statistical
uncertainties and systematic error due to calibration and beam
functions.  The subdominant systematic effects of pointing uncertainty
are neglected.  Top hat priors are applied to the Hubble parameter
($0.4 <h <0.9$), the age of the universe ($t >10~Gyr$), and the matter
density ($\Omega_m >0.1$).  Additional low \ells constraints are
provided by the \cobe\ results of \cite{Gorski}.

\begin{figure}[ht]
\centerline{\epsfig{width=4.0in,angle=0,file=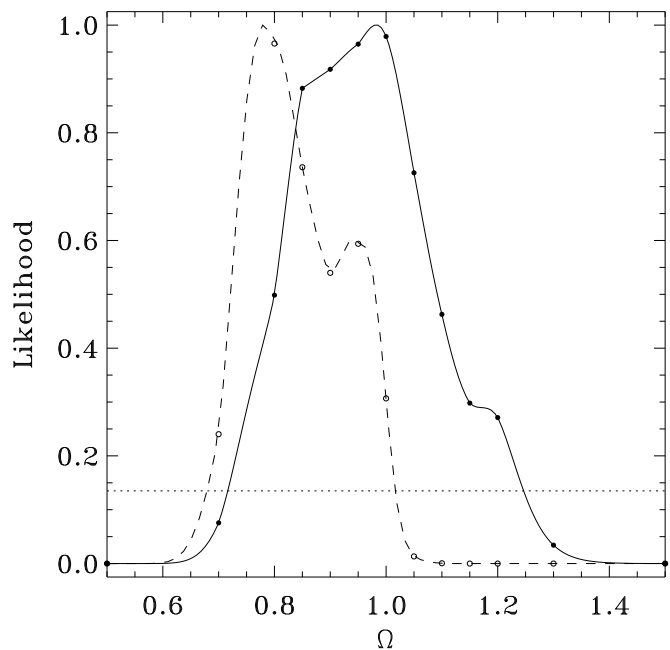}}
\caption[Constraints on Total Energy Density] {Likelihood function of
$\Omega_{tot}$.  The solid line is obtained by maximizing over other
parameters while the dashed line adds the constraints that $\Omega_b
h^2 = 0.0190 \pm 0.0024$ and $h = 0.65 \pm 0.07$.  The horizontal line
represent the 95\% confidence limits.  This plot, based on the 5\mins
analysis is virtually identical to that derived from the 3\mins
analysis.  (\cite{BalbiResults})}
\label{fig:omegatot}
\end{figure}

Constraints on individual parameters and combinations are found by
explicit marginalization of all other parameters over the range of
our sample grid.  We obtain 95\% confidence limits on total density
$\Omega_{tot}=0.9_{-0.16}^{+0.18}$, baryon density $\Omega_b h^2 = 0.033 \pm
0.013$, and power spectrum normalization $C_{10} = 690_{-125}^{+200} \mu
K^2$.  A constraint on dark matter density $\Omega_{cdm}h^2 =
0.17_{-0.07}^{+0.16}$ is largely based on priors (\cite{JaffeMaxiboom}).  

The optical depth to reionization, $\tau_c$, and the index of
primordial scalar fluctuations, $n_s$, are degenerate and obey the
relationship $n_s = (0.99 \pm 0.14) + 0.46 \tau_c$ at 95\% confidence
for $\tau_c < 0.5$.  Setting $n_s$ to 1.0 gives an upper
limit of $\tau_c < 0.26$, while setting $\tau_c$ to 0.0 gives $n_s
= 0.99 \pm 0.14$ (both at 95\% confidence).  Regardless of $\tau_c$,
we find $n_s > 0.8$ at 99\% confidence.

\begin{figure}[ht]
\centerline{\epsfig{width=4.0in,angle=0,file=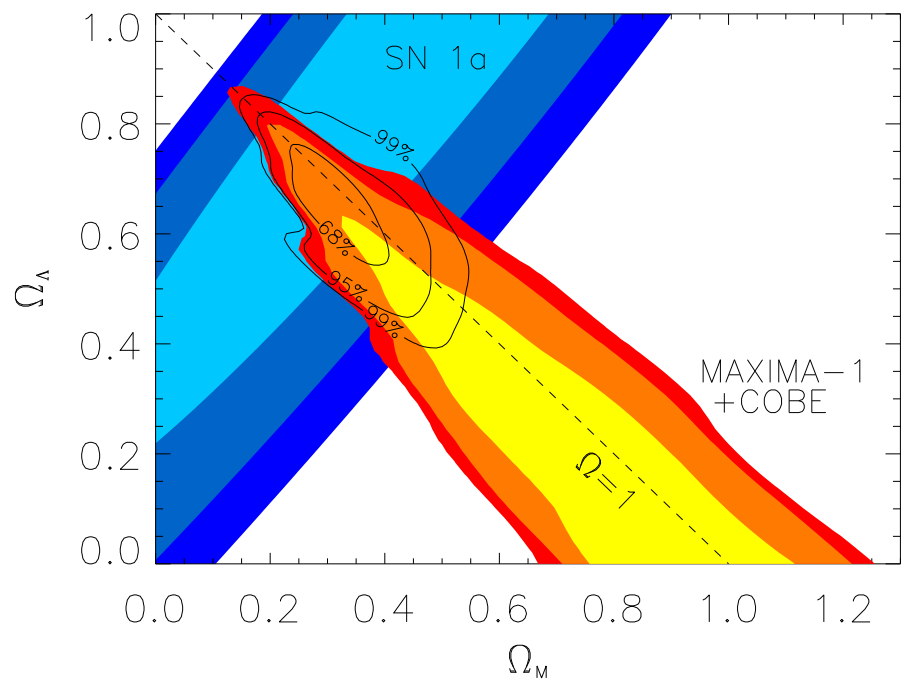}}
\caption[\maximai/\cobe/Supernovae Constraints on Matter/Vacuum Energy]
{Constraints on $\Omega_{m}$ and $\Omega_\Lambda$ from the combined
\maximai\ and \cobe\ \dmr\ data sets as well as those from high redshift
supernovae data (\cite{Perlmutter}, \cite{Reiss}).  Likelihood
contours at 68\%, 95\%, and 99\% confidence are shaded for each.
Outlines around $\Omega_m = 0.32$ and $\Omega_\Lambda = 0.65$ are the
joint likelihood contours.  (\cite{StomporResults})}
\label{fig:matterlambda}
\end{figure}

The overall best fit model parameters (\cite{StomporResults}) are
\begin{equation}(\Omega_{b},\Omega_{cdm},\Omega_\Lambda,\tau_{c},n_{s},h) = (0.07,0.68,0.1,0.0,1.025,0.63)
,\end{equation}
\noindent with $\chi^2 = 30$ for the 41 \maximai\ + \cobe\ power spectrum
points and $\chi^2 = 4$ for the 13 \maximai\ points.  The low estimate
of vacuum energy is consistent with recent supernovae results
because of the strong degeneracy between matter and vacuum energy.
Combining the \maximai\ + \cobe\ results with supernovae data
(\cite{Perlmutter}, \cite{Reiss}) yields a combined best fit on
$(\Omega_{m}, \Omega_{\Lambda})$ of $(0.32{+0.14\atop -0.11},
0.65{+0.15\atop -0.16})$ at 95\% confidence
(Figure~\ref{fig:matterlambda}).

\chapter{Foregrounds and Systematics}\label{chap:systematic}

	CMB temperature fluctuations are \sima $10^{-5}$~K.  Such a
small signal can be obscured by systematic effects including celestial
foregrounds, far side-lobe contamination, atmospheric emission, and
instrumental instability.  Section~\ref{sys:foreg} deals with
astronomical foregrounds.  Section~\ref{sys:concerns} summarizes other
potential systematic problems.  Section~\ref{sys:consistency}
describes general consistency tests.

\section{Foregrounds}\label{sys:foreg}

	Foregrounds are the best known source of systematic error for
CMB observations and have been widely discussed in the literature
(e.g. \cite{Bouchet_fore}).  \cite{JaffeDust} is a detailed treatment of
diffuse Galactic foregrounds in the \maximai\ scan region.

	\maxima\ deals with foregrounds in three ways.  Scan regions
are chosen for low foreground emission.  Foreground maps and point
source catalogues are used to model expected signals.  Spectral
discrimination provides empirical limits on foreground signals.
Subtraction of modeled and spectrally identified foregrounds is
viable, but has not been necessary for \maxima.

	Figure~\ref{fig:foreg_general} is a schematic of the relative
importance of foregrounds to different types of CMB experiments.  The
main foregrounds for \maxima\ are Galactic dust (\S \ref{foreg:dust}) and
point sources (\S \ref{foreg:point}).  Synchrotron and free-free emission
and zodiacal dust are secondary concerns (\S \ref{foreg:other}).

\begin{figure}[ht]
\centerline{\epsfig{width=4.5in,angle=0,file=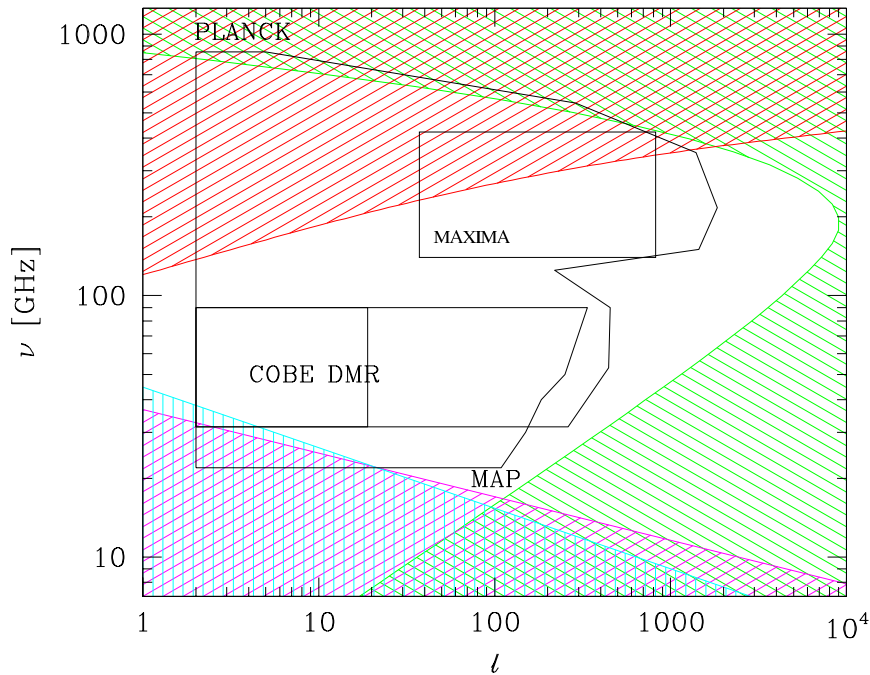}}
\caption[Foregrounds] {A schematic of the four main celestial
foregrounds for CMB observations.  The vertical axis is the optical
frequency of the observation, while the horizontal axis is the \ells
mode observed.  The shaded regions indicate that foreground anisotropy
is at least comparable to CMB anisotropy; proper selection of scan
regions can greatly reduce the impact of foregrounds.  The upper
region (red) indicates Galactic dust.  The v-shaped region on the
right (green) indicates point sources - IR at high frequency and radio
at low frequency.  The bottom regions represent synchrotron emission
(steeper, blue) and free-free emission (shallower, magenta).  The
black outlines indicate the observed frequencies and angular scales of
\maxima\ and several satellite experiments. (Figure by Martin White)}
\label{fig:foreg_general}
\end{figure}

\subsection{Galactic Dust}\label{foreg:dust}

	Galactic dust is the primary foreground contaminant for the
spectral bands and angular range of \maxima.  Dust emission has a
characteristic temperature of 17~K to 21~K and a spectral opacity
index of 1.5 to 2.7, with peak emission at 100~\micro\ to 200~\micro\
(3~THz to 1.5~THz).  Averaging over the entire sky, dust emission is
comparable to CMB anisotropy at 150~GHz and much larger at 230~GHz and
410~GHz.  The spatial anisotropy of Galactic dust decreases at smaller
angular scales as $\ell^{-3}$ (\cite{Dustspec1},\cite{Dustspec2}).

\begin{figure}[ht]
\centerline{\epsfig{width=4.0in,angle=0,file=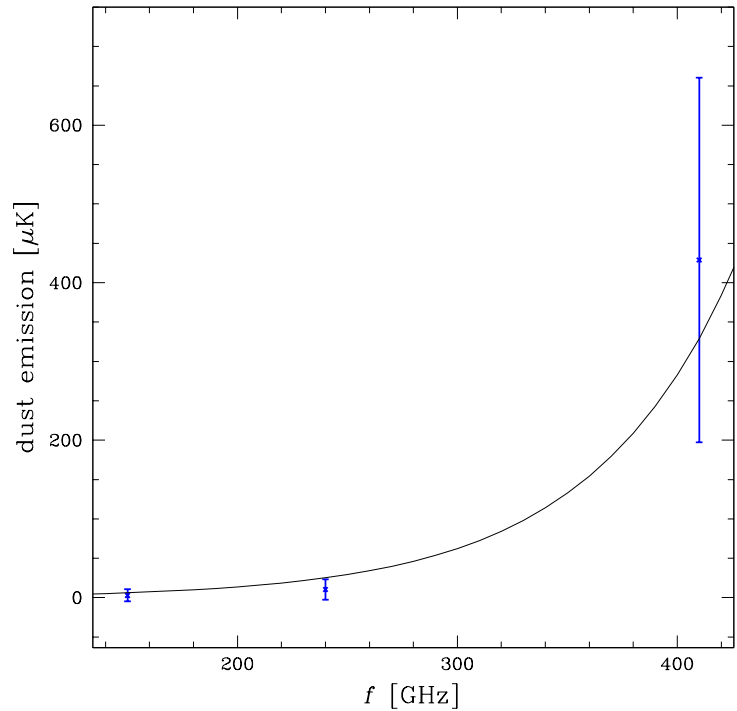}}
\caption[Observed Dust Emission] {Thermodynamic temperature of dust
observed in each of the three \maximai\ bands (\cite{JaffeDust}).
Values are derived from cross-correlation of \maximai\ and
\cite{SFD_Dust}.  Dust is not strongly detected at 150~GHz or 230~GHz.
The curve is the average emission from the \cite{FDS_Dust} ``Model 8''
prediction in the scan region.}
\label{fig:dust_spec}
\end{figure}

	Dust models based on \cobe/\dirbe, \iras/\issa, and \cobe/\firas\
data (\cite{SFD_Dust}, \cite{FDS_Dust}) have been used, both for sky
selection and for modeling of expected signals.  Dust data and models
are manipulated via the FORECAST software package (\cite{FORECAST}).
We have restricted observations to regions of the sky with low dust
contrast, $>$50\degs from the Galactic plane.  The dust in the scan region
observed in \maximai\ has a predicted in-band equivalent temperature of
10.0~\micro K (34.5~\micro K) at 150~GHz (230~GHz) with rms
fluctuations of 2.5~\micro K (8.8~\micro K) at 150~GHz (230~GHz).  For
\maximaii, the average equivalent temperature is 9.5~\micro K (32.8~\micro
K) at 150~GHz (230~GHz) with rms fluctuations of 2.5~\micro K
(8.5~\micro K) at 150~GHz (230~GHz).  These values are normalized to the CMB thermal spectrum.  The expected dust signal is
\sima 50 times higher at 410~GHz than 150~GHz.

	Correlation with the \cite{SFD_Dust} dust map is used to
directly quantify the effects of dust on \maximai\ (\cite{JaffeDust}).
The measured dust levels are consistent with zero and are 
\sima 1~$\sigma$ lower than the FORECAST projections at 150~GHz and 
230~GHz (Figure~\ref{fig:dust_spec}).

\subsection{Point Sources}\label{foreg:point}

	Infrared and radio point sources are most important at small
angular scales.  A random distribution of unresolved point sources
will cause apparent excess power which increases as $\ell^2$.

	Catalogued point sources are relatively easy to investigate
and can be removed from the map if needed (\cite{SGS_point},
\cite{GawiserSmoot_point}).  The \maximai\ observing region was
selected to be free of known bright sources.  The \maximaii\ observing
region is well away from the Galactic plane, but no extra precautions
were taken to avoid known point sources.  In neither case are known
point sources expected to measurably affect the angular power spectra.
Power spectra calculated omitting the regions around the brightest
known sources are consistent with that of the full map.

	Dimmer, uncatalogued point sources are more difficult to
handle.  For some sources, spectral arguments based on the measured
power spectrum at 410~GHz can be used to rule out significant
contributions at lower frequencies.  However, it is possible to
postulate a large number of faint point sources with exotic spectra,
the effects of which are highly model dependent.  Reasonable estimates
(\cite{GJS_point}) indicate that uncatalogued point sources are
unlikely to be significant for our observations.

\subsection{Other Foregrounds}\label{foreg:other}

	Synchrotron and free-free emission are diffuse Galactic
foregrounds which peak at radio frequencies and are fairly weak at
150~GHz and higher.  \maxima\ scans, which are well away from the
Galactic plane, are subject to very little synchrotron and free-free
emission.  Estimates based on \cite{Bouchet_fore} yield expected
contributions of less than 1~\micro K in all of our optical bands.
This agrees with upper limits from a correlation analysis of
synchrotron and \maximai\ maps (\cite{JaffeDust}).

	Zodiacal dust is a diffuse foreground concentrated on the
ecliptic plane of the solar system.  Zodiacal dust is not a major
foreground for CMB anisotropy due to its low emissivity and smooth
spatial distribution.  \maxima\ scans are conducted approximately
70\degs from the ecliptic plane, where the column density and
anisotropy of zodiacal dust are negligible.

\section{Other Systematic Concerns}\label{sys:concerns}

	Though foreground contamination is the most universal
systematic error for CMB anisotropy measurement, a variety of other
effects must also be considered.  Their origins may be optical
(e.g. far side-lobe pickup, atmospheric emission) or non-optical
(e.g. radio frequency pickup, instrumental noise instability).  No
list of systematic concerns can be complete; it is possible to
postulate any number of instabilities in detector noise and operating
temperature, unexpected atmospheric phenomena, or artifacts of readout
and data acquisition.  Many of the tests described later in this
chapter are of a generic nature, sensitive to broad classes of
problems.

\subsection{Far Side-Lobe Contamination}\label{sys:sidelobe}

\begin{figure}[hp]
\centerline{\epsfig{width=2.5in,angle=90,file=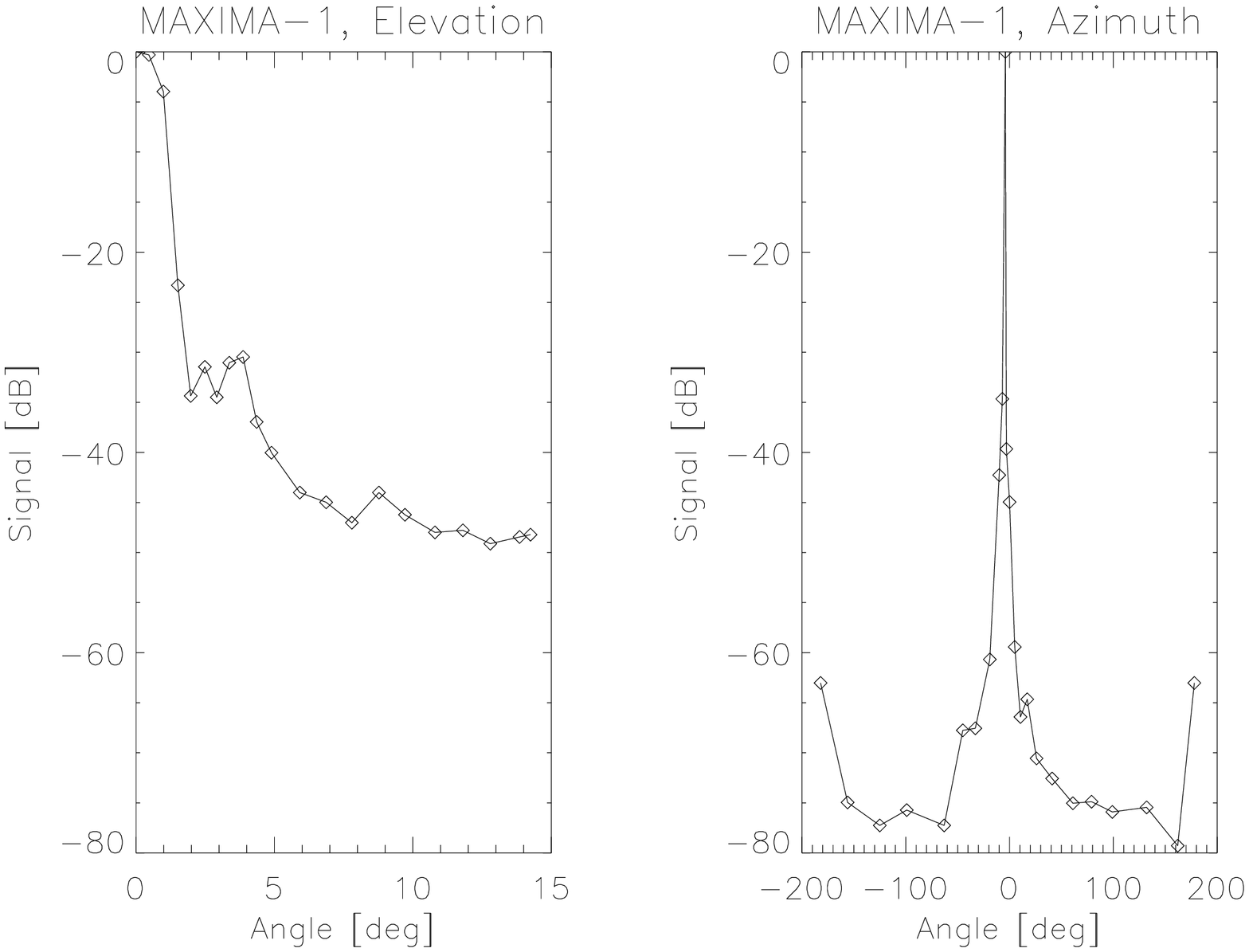}}
\centerline{ }
\centerline{\epsfig{width=2.5in,angle=90,file=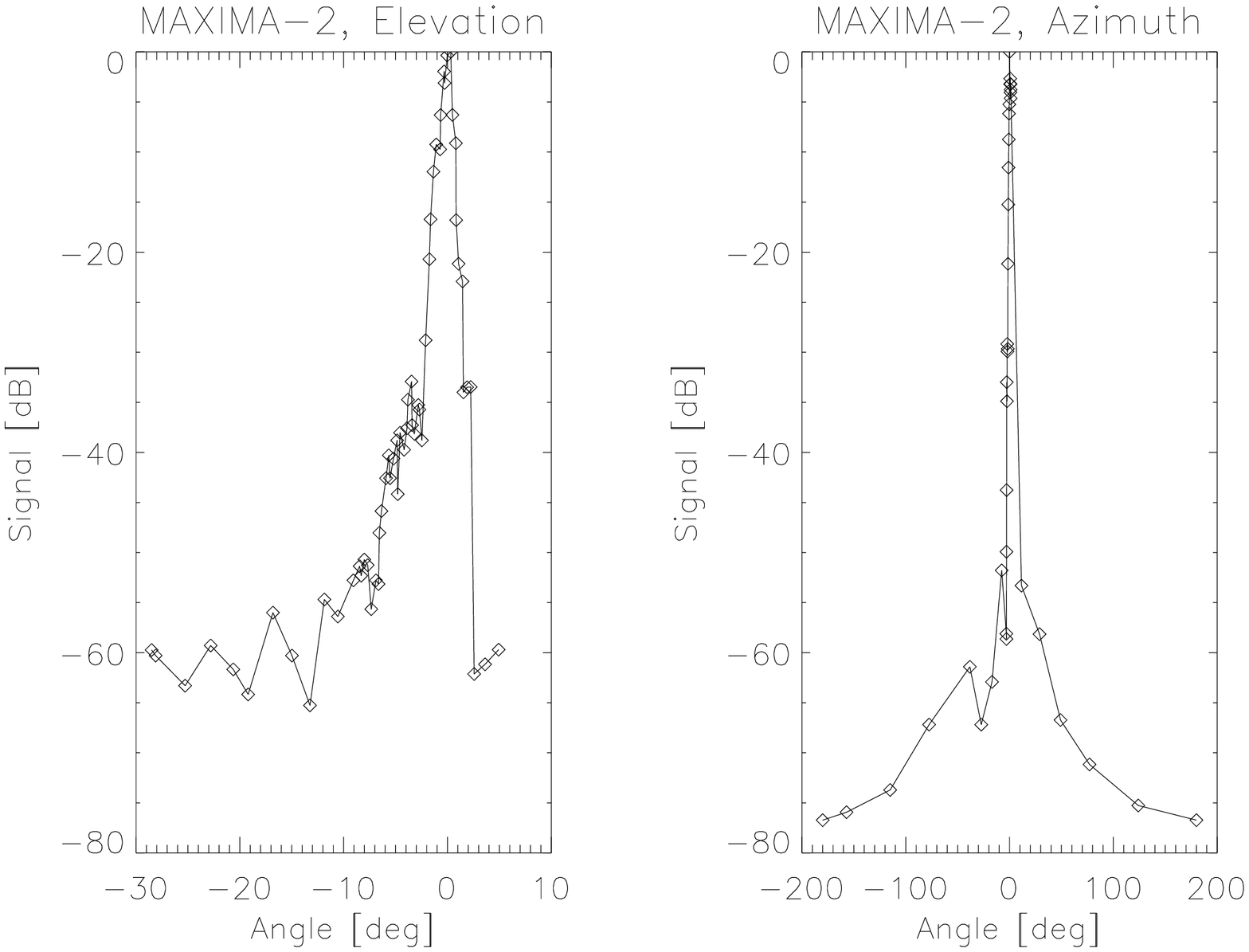}}
\caption[Side-Lobes]{Data from pre-flight side-lobe tests.  The source
was roughly 30~m from the telescope.  {\bf{Left Top:}} Test data in
the elevation direction for \maximai.  The angle is that of the
telescope above the test source.  The source is held at fixed
elevation while the telescope is aimed at different elevations.
{\bf{Right Top:}} Test data in the azimuth direction for \maximai .
The telescope beam is at fixed elevation (\sima 30\dega).  The source
is moved around the telescope at the same elevation.  The flat data at
roughly -78~dB represent the noise limit of the measurement.  The
apparent back-lobe is suspected to be an artifact of the measurement
technique.  If real, such a back-lobe would not affect the CMB data.
{\bf{Left Bottom:}} As above, for \maximaii.  Most of these data were
collected with the source at higher elevation than the telescope beam
(negative angles on the x-axis).  The measured response drops more
quickly than for \maximai, probably because of improvements in the test setup.
{\bf{Right Bottom:}} As above, for \maximaii.  No apparent back-lobe
is observed.  This is also likely because of improved measurement techniques.}
\label{fig:sidelobe}
\end{figure}

	Side-lobe contamination refers to spurious signals from bright
sources outside the main lobe of the telescope beams.  The temperature
contrasts of the balloon, the earth, the Sun, and the Moon are up to
ten orders of magnitude greater than that of the CMB.

	Pre-flight side-lobe measurements were made using a directional
Gunn Oscillator with variable attenuation as a 150-GHz test source
(Figure~\ref{fig:sidelobe}).  Measurement imperfections, such as
reflections of the source from the ground, affect these tests; the
actual side-lobe sensitivity may be lower than the measurements
indicate.  Side-lobe response is most overestimated in the lower
elevation direction.  The measured attenuation at 15\degs below the
beam is sufficient to prevent earth-based side-lobe contamination.

	The possibility of side-lobe contamination is minimized by
collecting data at night, with the Sun well below the horizon and with
the Moon far from the scan region (Chapter~\ref{chap:flights}).

	Likely side-lobe contaminants (the Sun, Moon, and features on
the Earth) are not oriented constantly with respect to the scan.
Differences maps of the two CMB scans for each flight
(\S \ref{sys:scans}) are therefore sensitive to side-lobe signals.

\subsection{Scan Synchronous Signals}

	\maxima\ data show a small spurious signal synchronous with
the modulation of the primary mirror, with a CMB temperature amplitude
of up to 200~\micro K at 150~GHz.  It is possible that the scan
synchronous signals in \maximai\ and \maximaii\ have different
sources.

	These signals might be radio frequency pickup (\S
\ref{receiv:rfi}); the radio signal from the on-board telemetry
transmitters is partially blocked by the primary mirror mechanism.
The modulation of the mirror could vary the degree of pickup by the
receiver, leading to a scan synchronous signal.  This model is
plausible for \maximai, during which the receiver was relatively
susceptible to radio interference.  It is less likely for \maximaii\
because the signals are more prevalent despite additional radio
frequency filtering.

	A second possibility is variation in atmospheric optical
loading as a function of mirror position.  This could be explained by
a tilt in the mirror motion, causing the telescope beams to change
elevation.  Under this model, we would expect the synchronous signal
to display a strong spectral signature.  Atmosphere accounts for less
than 1\% of the optical load at 150~GHz but about half of the optical
load at 410~GHz.  The observed scan synchronous signals are similar at
all three optical frequencies.  In \maximaii\ the signals vary greatly
between detectors, but show no clear correlation with the spectral
bands.  We conclude that atmospheric loading variations are unlikely
to account for the synchronous signals.

	Scan synchronous signals may also be caused by electrical or
mechanical pickup from the mirror drive.  It is difficult to disprove
these effects or to estimate their magnitude.  However, subsequent
tests of the integrated system in the laboratory have failed to
reproduce the observed in-flight scan synchronous signals.

	Regardless of the source, the map making procedure in
Chapter~\ref{chap:data} includes a treatment of parameterized
parasitic signals (\S \ref{data:ssynch}).  The assumption that the
spurious signal is stable over periods of several minutes has been
confirmed in data analysis.

	In addition to primary mirror scan synchronous pickup, a
signal synchronous to the azimuth modulation of the entire telescope
has also been measured.  It is relatively small (up to \sima
50~\micro\-K) and much slower than the primary modulation (40 to 70
seconds).  This signal is strongly rejected by the faster modulation
of the primary mirror; treating it explicitly has no effect on maps
and power spectra.  It is ignored in the final data analysis.

\subsection{Telescope Pendulum Motion}

	Pendulum motion is a well known danger for balloon-borne
telescopes.  Pendulum motion changes the telescope elevation angle and
therefore the observed atmospheric load.  The two main pendulum modes
of the \maxima\ telescope have frequencies of roughly 0.6~Hz and
0.05~Hz.  Of these the mode at 0.6~Hz is relevant to CMB
observations.  The 0.6-Hz mode is suppressed by a factor of \sima 10
using passive pendulation dampers.\footnote{The pendulation dampers,
built by Geneva Observatory, are highly damped harmonic oscillators
consisting of weighted spheres rolling in oil filled spherical
cavities.}

	Pendulation modes may be excited by the attitude control
motors.  The telescope is driven in azimuth using a reaction wheel
(\S \ref{point:control}) that is symmetric about the rotation axis.
However, the telescope is not symmetric and its moment of inertia
tensor may have non-zero off-diagonal terms, coupling the azimuth
drive to the pendulum modes.  

	Pendulation amplitude is $<$10\secs based on pre-flight tests.
No pendulum motion is observed during flight on time scales of \sima 1
minute or less within the 1\mins accuracy of the pointing
reconstruction.  Detector data, including the 410-GHz data most
sensitive to atmospheric emission, show no signal at 0.6~Hz or
0.05~Hz.

	An irregular variation of up to 20\mins is observed on time
scales of \sima 20 minutes.  This slow pendulation shows a strong
correlation with the altitude of the telescope and is believed to
result from the coupling of the pendulum modes to vertical motion of
the balloon.  Regardless of the cause, oscillations on such long time
scales do not affect CMB observations.

\subsection{Secondary Data Errors}

	Errors in calibration, pointing reconstruction, beam
measurement, optical filters, and electronics can affect CMB maps and
power spectra.  Errors in these data contribute to the error estimates
in Chapter~\ref{chap:results}.

\begin{figure}[ht]
\centerline{\epsfig{width=4.8in,angle=0,file=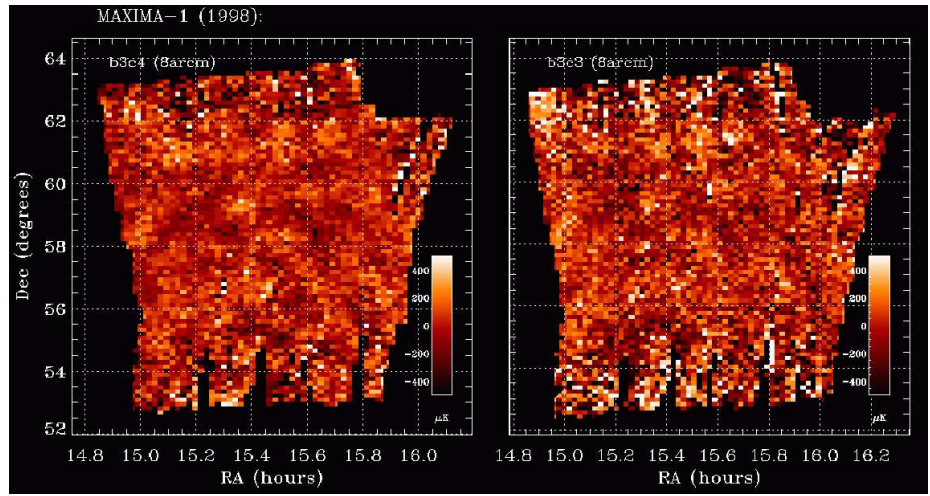}}
\caption[Single Detector Maps]{5-arcminute resolution \maximai\ maps
computed from two single detectors.  Neither map is Weiner filtered.
{\bf{Left:}} The map from a 150-GHz detector (B34), the least noisy
detector from \maximai.  {\bf{Right:}} The map from a 230-GHz detector
(B33), the noisiest detector used in published results.}
\label{fig:map_b34b33}
\end{figure}

\begin{figure}[ht]
\centerline{\epsfig{width=3.0in,angle=0,file=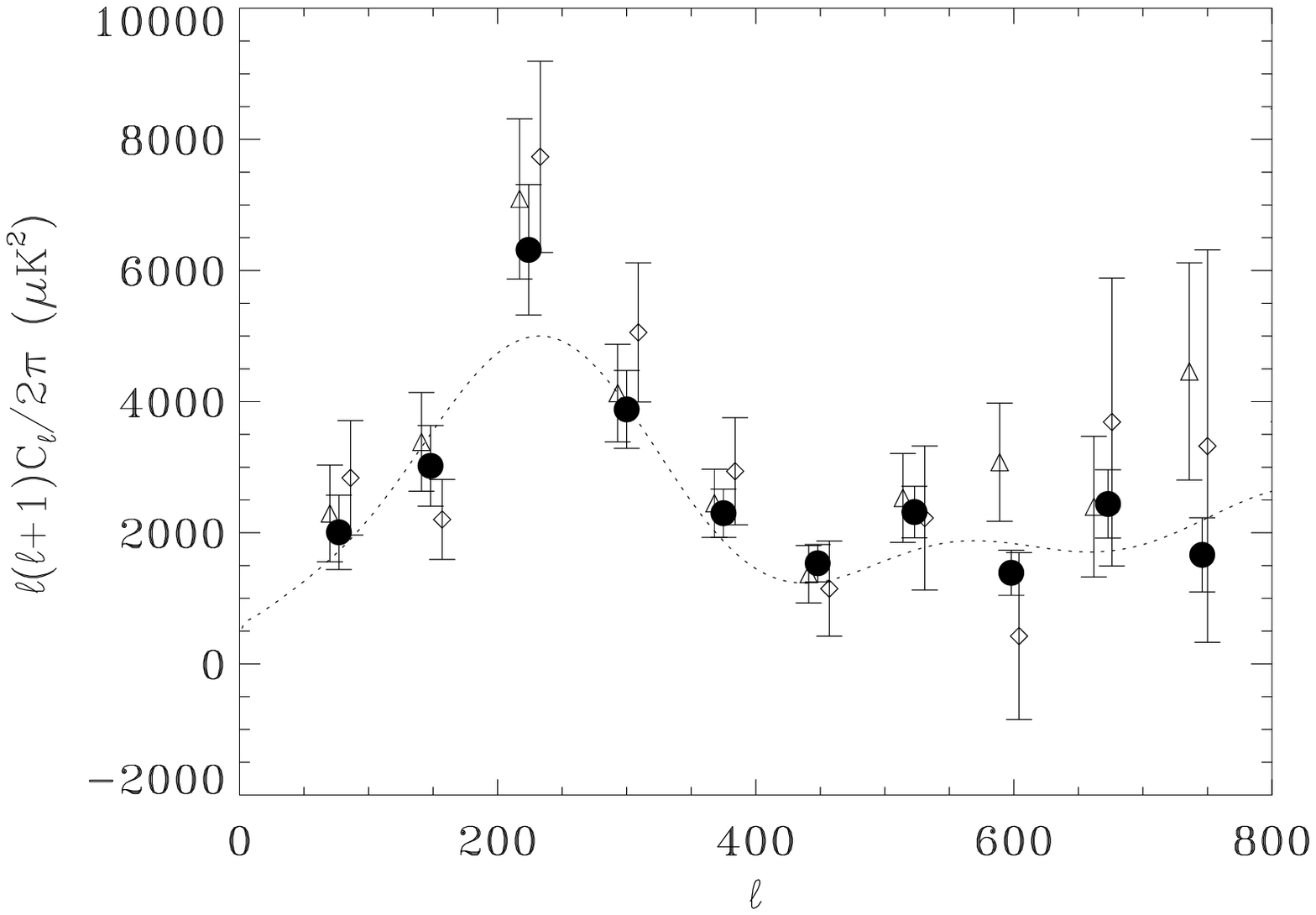}
\epsfig{width=3.0in,angle=0,file=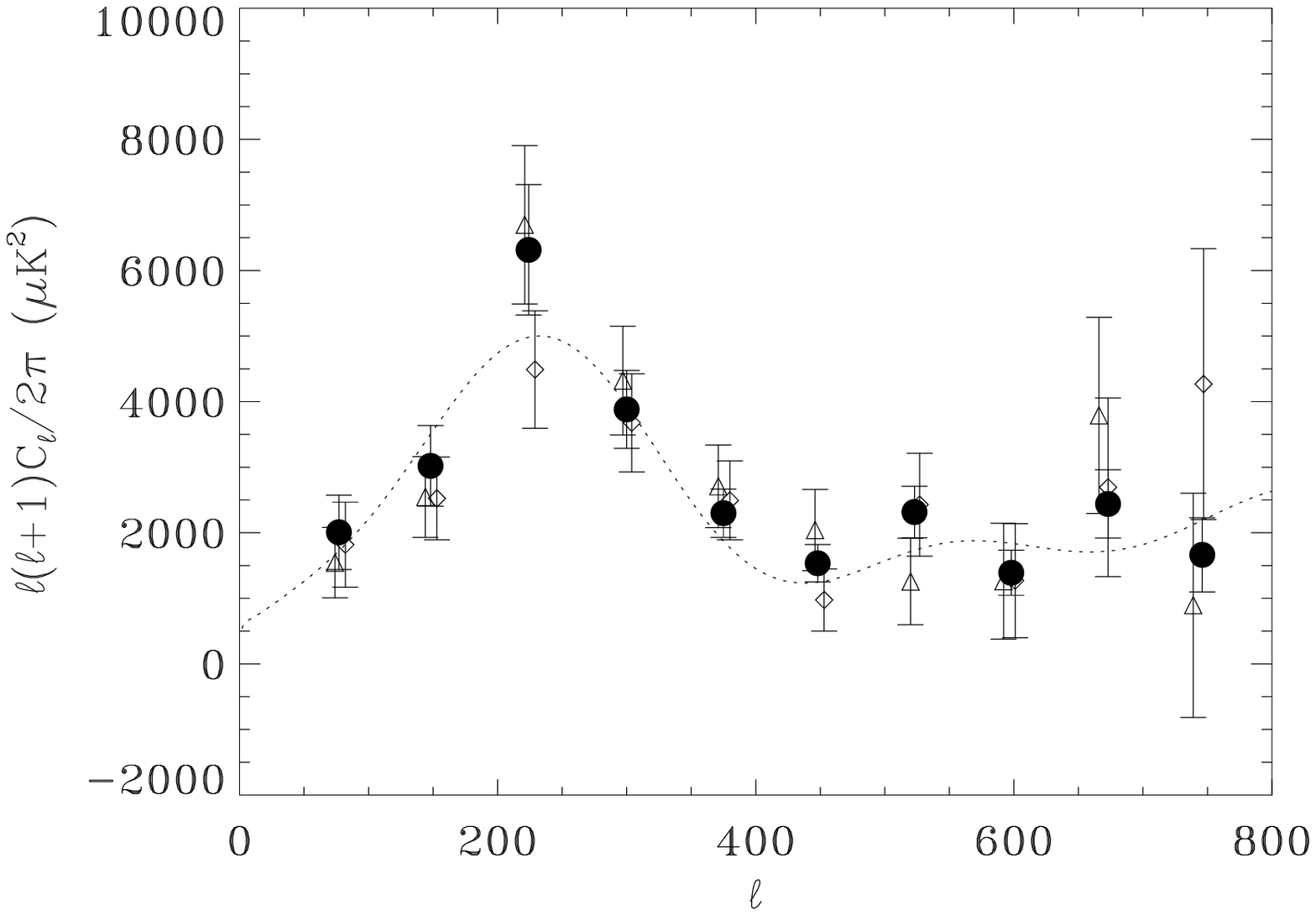}}
\caption[Single Detector Power Spectra]{The angular power spectra
derived from 5\mins resolution \maximai\ maps.  Each panel shows
the published (\cite{HananyResults}) 4-channel combined data (filled
circles), as well as two single channel power spectra.  In each bin, the
single channel power spectra have been horizontally offset for
readability.  {\bf{Left:}} Triangles (B45) are 150-GHz data and
diamonds (B33) are 230-GHz data.  {\bf{Right:}} Triangles (B34) and
diamonds (B25) are 150-GHz data.}
\label{fig:ps_full_vs_individual}
\end{figure}

\section{Consistency Tests}\label{sys:consistency}

	Systematic errors are examined using general consistency
checks that are sensitive to broad classes of effects.

	Many systematic tests involve the use of difference maps.
Difference maps are generated with the same techniques used for
summing maps (\S \ref{combine}).  Assuming that the individual maps
include only well calibrated sky-stationary signals, difference maps
should be consistent with combined detector noise.  These maps are
subject to a variety of statistical tests, including power spectrum
and $\chi^2$ analyses.

	The data shown in this chapter are a representative subset of
the tests conducted on published \maximai\ data.  Corresponding
systematic tests of \maximaii\ data are in progress.  Further
discussions of these and other systematic error checks are found in
\cite{StomporSystem}.

\subsection{Cross-channel consistency}\label{sys:channel}

\begin{figure}[ht]
\centerline{\epsfig{width=4.8in,angle=0,file=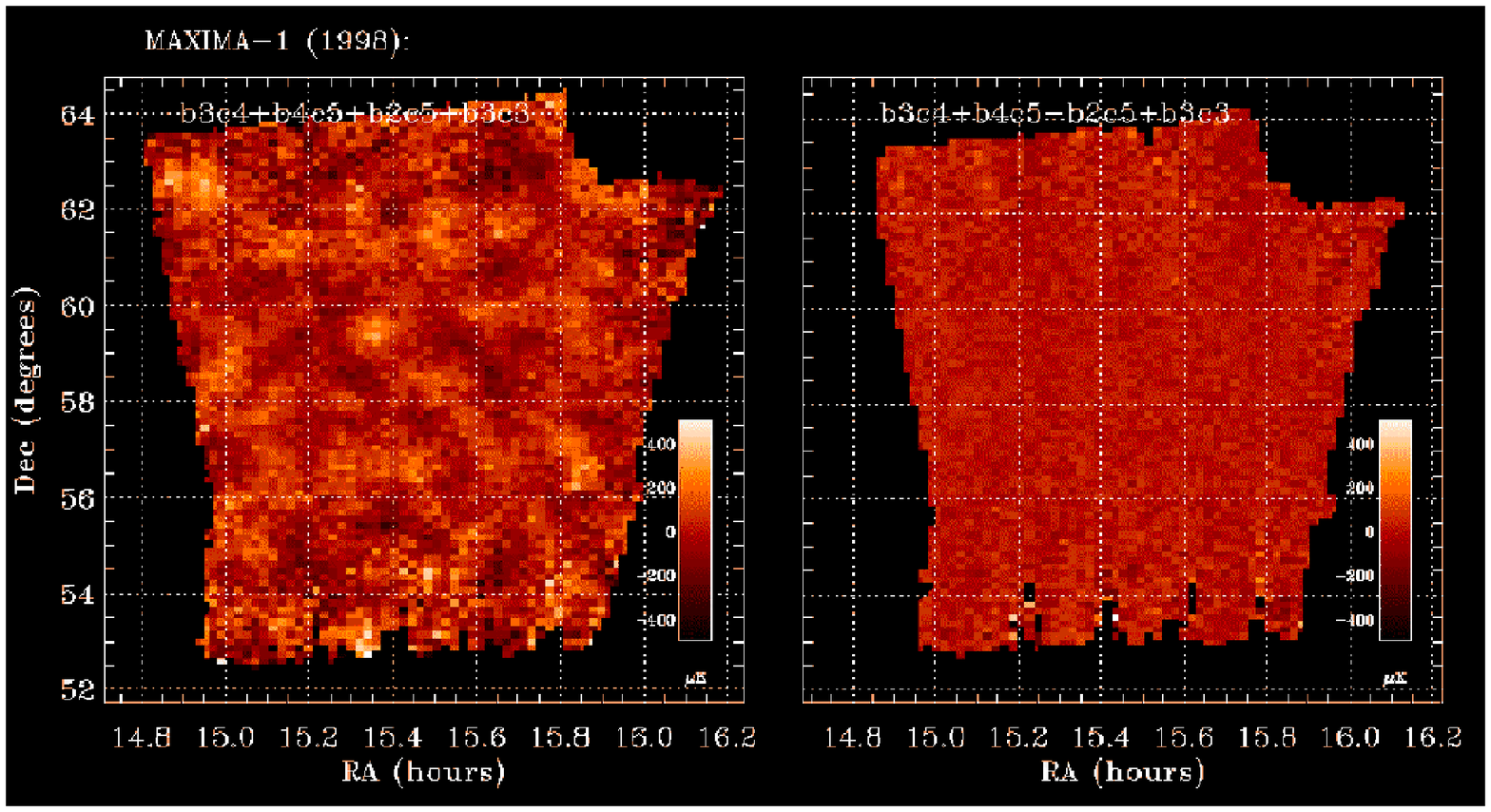}}
\caption[4-Channel Sum and Difference CMB Maps]{5\mins resolution
\maximai\ maps.  {\bf{Left:}} The combined 4-detector map of
\cite{HananyResults}.  This map has not been Weiner filtered.
{\bf{Right:}} A difference map from the same four detectors.  The
detectors are summed pairwise, and these pairs are differenced to
produce this map.  The summed pairs are B34 (150~GHz) + B45 (150~GHz)
and B25 (150~GHz) + B33 (230~GHz).  The difference map is consistent
with combined detector noise.}
\label{fig:map_double_difference}
\end{figure}

\begin{figure}[ht]
\centerline{\epsfig{width=4.0in,angle=0,file=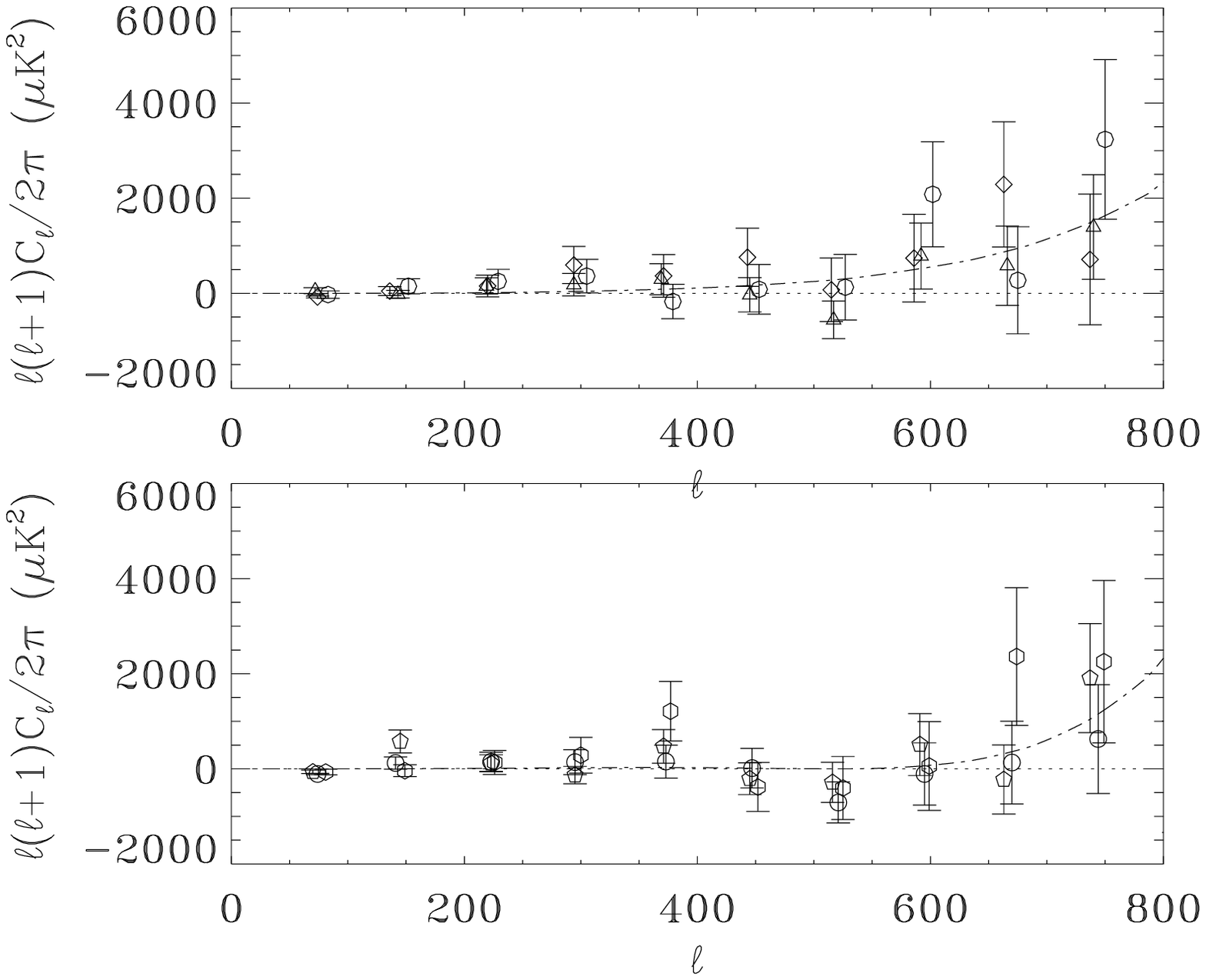}}
\caption[Difference Angular Power Spectra]{Angular power spectra of
difference maps from pairs of single detectors.  The four detectors
used in \cite{HananyResults} are differenced in every possible
pairing.  Each panel shows three of the six combinations.  A small
amount of residual power is seen at high \ells for some pairings,
caused by differences in the beam patterns of the detectors
(\cite{WuBeam}).}
\label{fig:ps_difference}
\end{figure}

	Data from multiple detectors are analyzed separately and in a
variety of combinations to test consistency between channels and with
the combined map.  These tests are sensitive to certain instrumental
problems, to spectral consistency, to the relative spatial offsets
of the detectors, and to calibration consistency.

	Channels and their combinations are compared in several ways.
First, independent maps are compared directly.
Figure~\ref{fig:map_b34b33} shows the maps from a 150-GHz and a 230-GHz
detector.  Cross-correlation is used to determine the relative
amplitude and position of the measured signals.  In all cases, maps
are found to be aligned to much better than 1\mina , with power
levels deviating by less than the calibration uncertainty.

	Second, angular power spectra from independent maps are
compared.  Figure~\ref{fig:ps_full_vs_individual} show the power
spectra from four channels, compared to that of the combination of those
channels.

	Finally, difference maps and angular power spectra are
generated.  Figure~\ref{fig:map_double_difference} shows the
difference and sum of a pair of two-channel combinations.
Figure~\ref{fig:ps_difference} shows angular power spectra from single
channel difference maps.

\subsection{Cross-scan consistency}\label{sys:scans}

\begin{figure}[ht]
\centerline{\epsfig{width=4.8in,angle=0,file=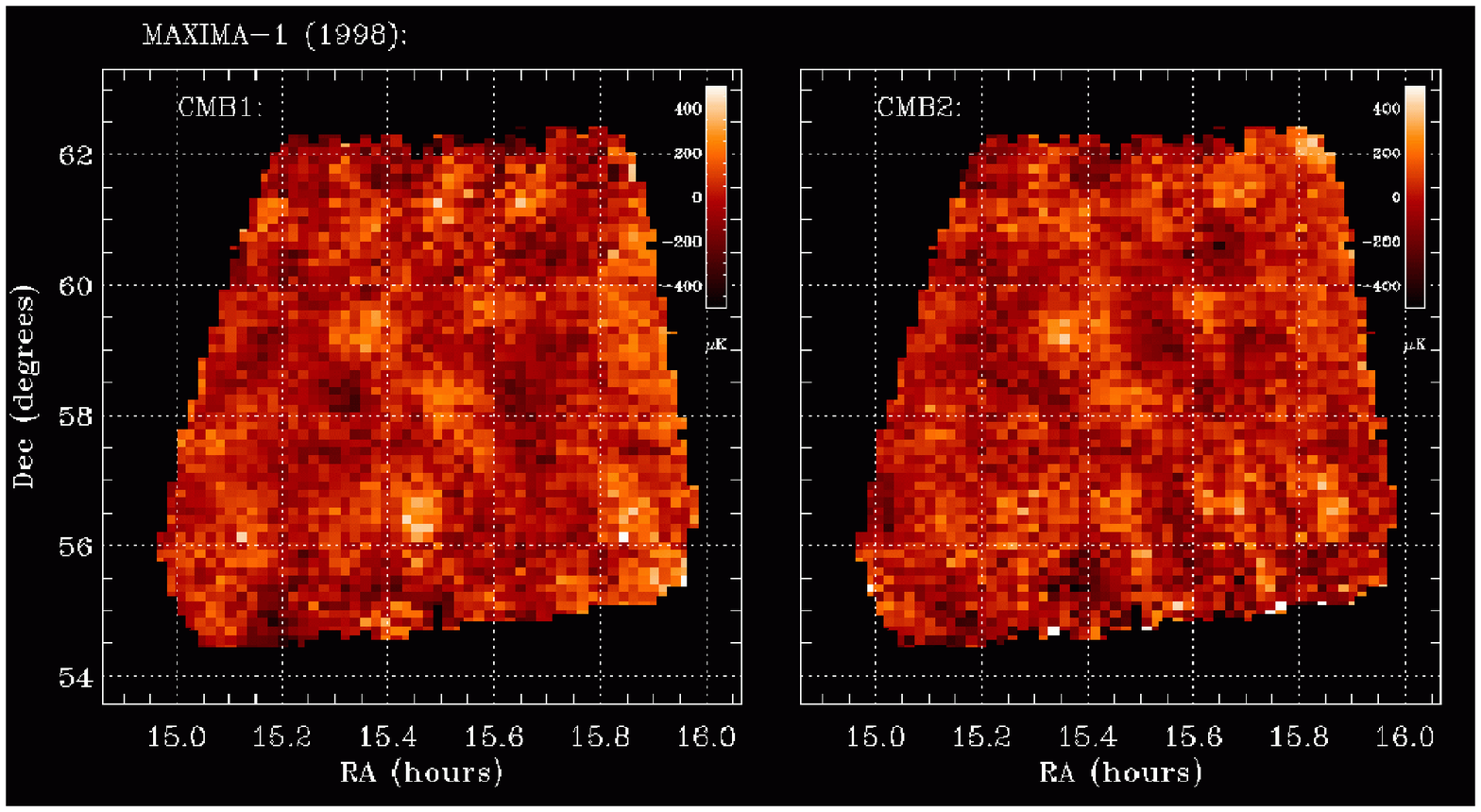}}
\caption[Repeated Observations]{5\mins resolution 4-channel maps from
the two CMB observations of \maximai.  The first scan is shown on the
left, and the second scan is shown on the right.  Only the overlapping
region of the two scans is shown.  Each of these maps has a much
lower pixel sensitivity than the combined map, due to the lower number
of measurements and the lack of cross-linking.}
\label{fig:map_cmb12}
\end{figure}

	Each \maxima\ flight includes two largely overlapping CMB
scans.  The two scans have different scan geometry, telescope
orientation, and uncorrelated detector noise.

	Cross-scan consistency tests are sensitive to far side-lobe
contamination, to atmospheric pickup, to the time dependence of the
responsivity calibration, to time varying instrumental effects, and to
pointing reconstruction.  Figure~\ref{fig:map_cmb12} shows the maps
and for the two CMB observations of \maximai\ and
Figure~\ref{fig:ps_cmb12} shows the power spectrum of the their
difference.
	
\begin{figure}[ht]
\centerline{\epsfig{width=5.0in,angle=0,file=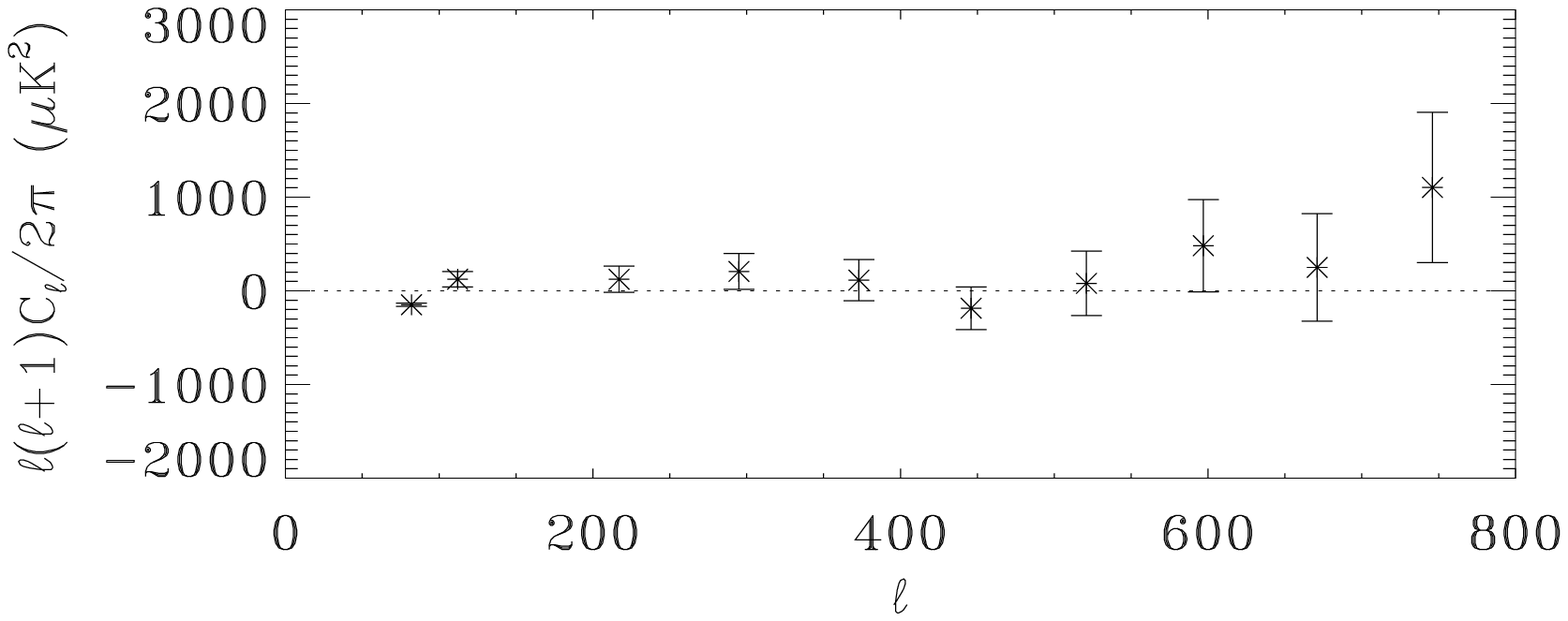}}
\caption[Repeated Observations Power Spectrum]{An angular power
spectrum derived from the difference of the two \maximai\ CMB
observations.  Based on the \cite{HananyResults} 5\mins analysis.}
\label{fig:ps_cmb12}
\end{figure}

\subsection{Dark Channel Maps and Power Spectra}\label{sys:dark}

	The receiver includes three non-optical `detectors'.  Data
from these devices are processed by the same electronics as the
optical data, and are sensitive to different subsets of potential
instrumental problems.  The first is a ``dark'' bolometer, not coupled
to an optical feedhorn.  The second is a bolometer-style NTD
thermistor, thermally coupled to the ADR.  The third is a temperature
independent resistor, with impedance similar to that of the
bolometers.  Four additional sets of dark data are obtained from
``bias monitors'', \ie~the locked-in signal from the bolometer AC-bias
generators, not connected to the bolometers or cryogenic preamplifiers.
Figure~\ref{fig:ps_dark} shows a power spectrum of the dark bolometer in
\maximai.

\begin{figure}[ht]
\centerline{\epsfig{width=4.0in,angle=0,file=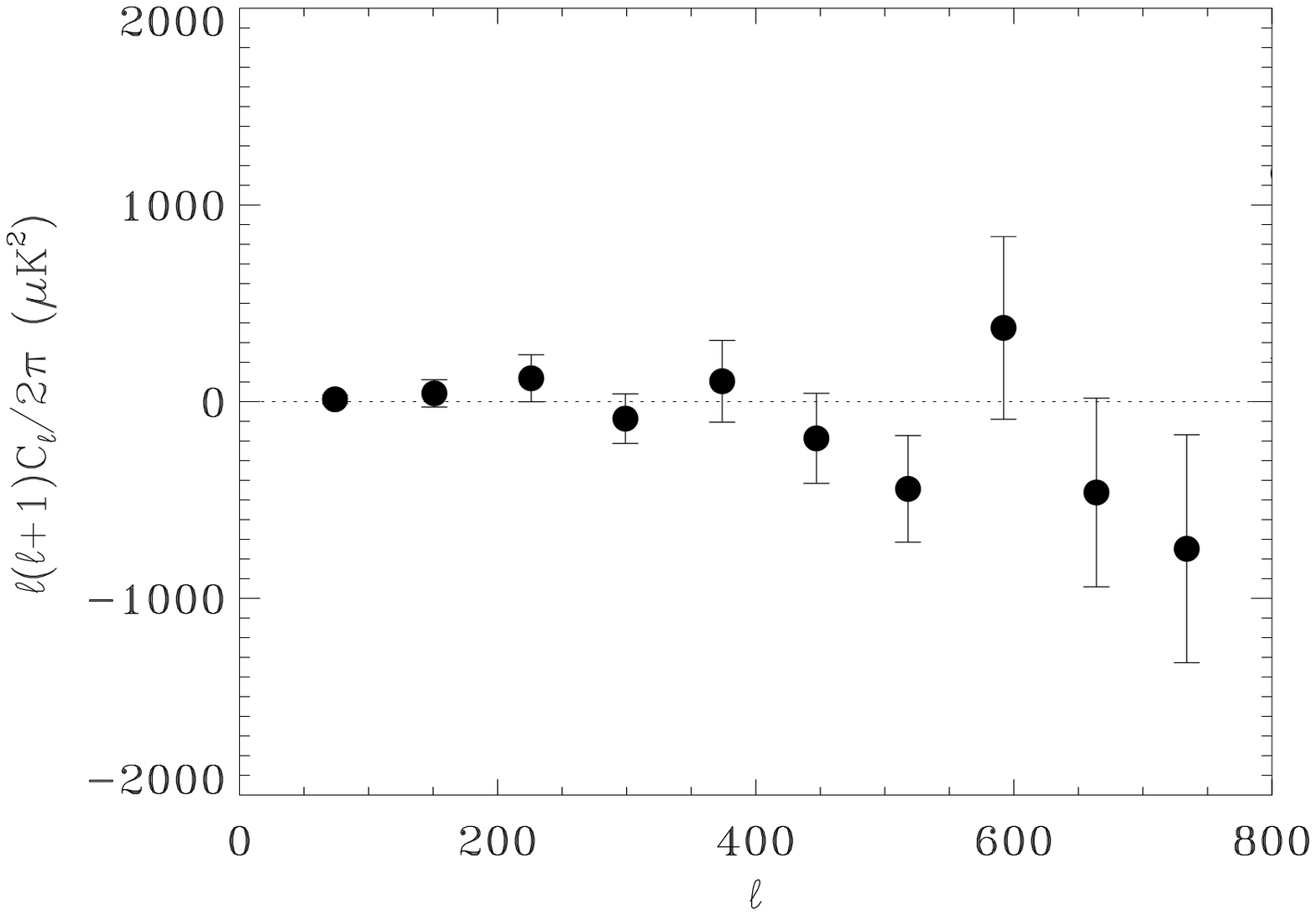}}
\caption[Non-optical Bolometer Data]{An angular power spectrum from an
optically insensitive detector in \maximai.  Because this ``dark''
data has no real mapping onto the sky, the pointing solution from an
optical bolometer (B34) is used.  Similarly, the CMB temperature
units on this plot are based on the calibration of an optical
bolometer (B34) for purely comparative purposes.  The measured power
is consistent with detector noise.}
\label{fig:ps_dark}
\end{figure}

\subsection{Other Consistency Tests}

	Description of all systematic and consistency tests is beyond
the scope of this document; more details are found in
\cite{StomporSystem}.  In addition to those already discussed,
systematic tests have included: selective omission of various subsets
of time domain data; variations in acceptance criteria for pointing
and detector data; separate analyses of different sections of the CMB
maps; variations in map pixelization and resolution; and variations
in data analysis.

\chapter{Future Work: Polarization}\label{chap:future}

	This chapter discusses CMB polarization anisotropy and the
motivation for measuring it (\S \ref{future:pol}).
Section~\ref{maxipol} describes \maxipol, the polarization sensitive
follow-up to \maxima.

\section{Polarization Anisotropy}\label{future:pol}

	CMB polarization anisotropy is a fundamental prediction of inflationary Big Bang models.
Thomson scattering of CMB photons from free electrons at
the surface of last scattering causes a net linear polarization where
there is a non-zero local quadrupole (\cite{Kosow_pol}).  The degree
of linear polarization is a measure of anisotropy at the time of last
scattering;  polarization probes the epoch of last
scattering directly, unlike temperature fluctuations that can arise
in part after last scattering.  This ``localization in
time'' makes polarization a strong constraint on the origin of
anisotropies, complementary to temperature (\cite{Primer}).

\subsection{$E$-modes and $B$-modes}

	Linear polarization of photons is described by two orthogonal
components.  
CMB polarization is naturally discussed in terms of the
components $E$ and $B$ (\cite{EBDecomp}).  This form has two
advantages over the Stokes parameters $Q$ and $U$.  First, the $E$ and $B$
components are independent of coordinate rotation.  Second, each has a
distinct symmetry under parity inversion.  $E$-mode polarization is
symmetric under parity, while $B$-mode polarization is anti-symmetric
under parity.  The $E$ component is named for its likeness to an
electric field (\ie~a pure gradient vector field).  The $B$ component
is likened to a magnetic field (\ie~a pure curl vector
field).\footnote{The $E$ and $B$ components are often called $G$ and
$C$ for the same reason.}

	Because Thomson scattering turns local quadrupoles - which
are parity symmetric - into polarization, it can only produce
$E$-modes.  However, spatial modulation of the perturbations over
the surface of last scattering can convert $E$-modes into $B$-modes
(\cite{Primer}).  Whether such a conversion actually occurs depends
on the source of the quadrupole.  Density (scalar) fluctuations do not
create $B$-modes, but gravitational waves are tensor fluctuations
that do create $B$-modes (\cite{Kamionkowski}).  $B$-modes are also
created by vortical flows, but this mechanism is not expected to
contribute significantly.  Primordial $B$-mode polarization is
expected to be two or more orders of magnitude smaller than $E$-mode
polarization.

	$E$-modes are also converted into $B$-modes by gravitational
lensing (\cite{B-lensing}).  This effect is itself of interest, but
for $B$-mode CMB measurement it is significant foreground.  It can be
subtracted, though only partially, from $B$-mode maps to help search
for gravity wave signatures (\cite{B-cleaning1}, \cite{B-cleaning2}).

	Angular power spectra of $E$-mode and $B$-mode polarization
anisotropy are denoted as $C_l^{EE}$ and $C_l^{BB}$ respectively (in
this context the temperature power spectrum is generally denoted as
$C_l^{TT}$).  In addition, correlations between $E$-mode and
temperature anisotropy lead to a finite cross power spectrum,
$C_l^{TE}$.  Due to parity invariance, $B$-mode CMB polarization
anisotropy is uncorrelated with $E$-mode and temperature anisotropies;
$C_l^{TB}$ and $C_l^{EB}$ are zero in the absence of foreground
effects.

\subsection{Cosmology with CMB Polarization}

	CMB polarization provides cosmological information that cannot
be obtained from temperature anisotropy alone. The partial
polarization of the CMB is a basic prediction of our model of
structure formation.  If the present structure in the universe grew
through gravitational instability from primordial fluctuations, their
existence at the time of last scattering would ensure CMB
polarization.

	Polarization allows discrimination between types of primordial
fluctuations.  Parameter estimation from temperature anisotropy relies
on the assumption of purely adiabatic (scalar) primordial
fluctuations.  Temperature anisotropy alone has ruled out pure
isocurvature models, but more general models involving a mix of
adiabatic and isocurvature fluctuations are still possible
(\cite{BMT1}).  While different combinations of perturbations may lead
to equivalent temperature anisotropy, they can be distinguished in the
polarization anisotropy (\cite{BMT2}).

	In addition, polarization measurements constrain cosmological
parameters and can break degeneracies in temperature data.  A great
example of this is the reionization of the universe.  There is a
strong degeneracy in temperature anisotropy between the optical depth
of reionization, $\tau_c$, and the overall amplitude of primordial
density fluctuations, $A_s^2$.  In the $E$-mode power spectrum,
reionization causes a distinct peak at $\ell\approx 20$, which breaks
the degeneracy (\cite{Reionization}).

\begin{figure}[ht]
\centerline{\epsfig{width=4.0in,angle=0,file=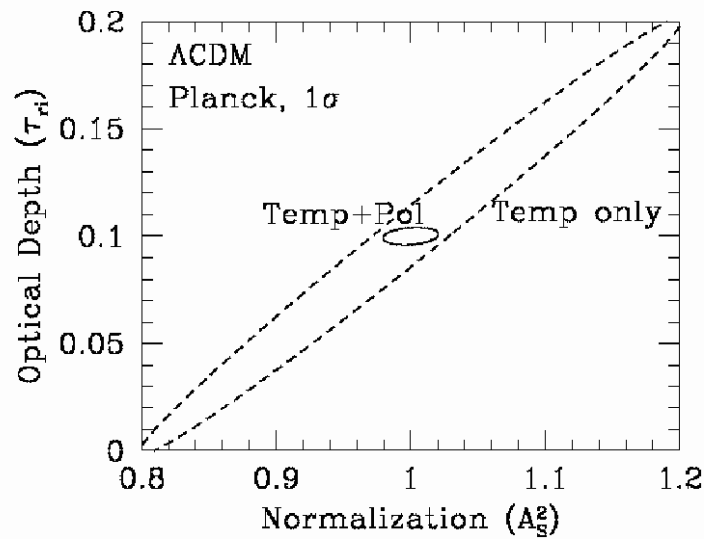}}
\caption[Reionization Degeneracy]{Measurement of reionization using
temperature anisotropy (dashed) and both temperature and polarization
anisotropy (solid), based on the projected sensitivity of the Planck
Surveyor.  Temperature anisotropy alone cannot break the degeneracy
between the optical depth of reionization, $\tau_c$, and the amplitude
of density fluctuations, $A_s^2$.  (Figure by M. White)}
\end{figure}

	$B$-mode polarization, though smaller than $E$-mode
polarization, is especially interesting.  $B$-modes are a measure of
the long wavelength gravitational waves predicted by inflation.  The
amplitude of these waves is proportional to the energy scale of
inflation.  Thus, the detection of $B$-modes would provide
overwhelming evidence for inflation and new information about
inflationary physics (\cite{InfGravWaves}).

\subsection{Detection of CMB Polarization}

	The low level of CMB polarization, and the poorly understood
foregrounds and systematic error sources, make measuring polarization
an experimental challenge.  If models currently favored by CMB
temperature anisotropy are accurate, the $E$-mode polarization
anisotropy is expected to be about an order of magnitude smaller than
the temperature anisotropy.  Several experiments (\cite{PIQUE},
\cite{POLAR}) have already set upper limits near the predicted level,
and very recently the DASI experiment is believed to have made a
detection (\cite{DASIpol}).

	$B$-mode polarization is probably not detectable by current
generation experiments.  Large bolometric arrays currently under
development promise a large increase in sensitivity.  However, even
with arbitrary sensitivity, primordial gravity wave signatures may be
obscured by the effects of weak lensing and other foregrounds.
	
\section{MAXIPOL}\label{maxipol}

	\maxipol\ is the polarization sensitive follow-up to \maxima.
The primary goal is a robust detection of $E$-mode polarization near
the spectral peak of the CMB.  This includes detections of power in
both the $C_l^{TE}$ and $C_l^{EE}$ power spectra.  \maxipol\ is
designed primarily for detection and is not expected to measure the
shape of the power spectra to high accuracy.  Two or three \maxipol\
flights are planned.

	\maxipol\ shares the main advantages of \maxima\ - high
sensitivity, well optimized scan strategy, excellent calibration, and
precise pointing reconstruction - but differs as follows: the
bolometric receiver is retrofitted for polarization sensitivity; the
scan strategy is modified and the scan region is much smaller (\sima
10~deg$^2$ over 2-3 scan regions per flight); and the flight time is
longer and data are taken during both day and night (\sima 30 hours
per flight, with 15-20 hours of CMB observation).  Projected power
spectrum sensitivity is shown in Figure~\ref{fig:maxipolps}.

\subsection{Polarimetry}

\begin{figure}[ht]
\centerline{
\epsfig{width=2.3in,angle=270,file=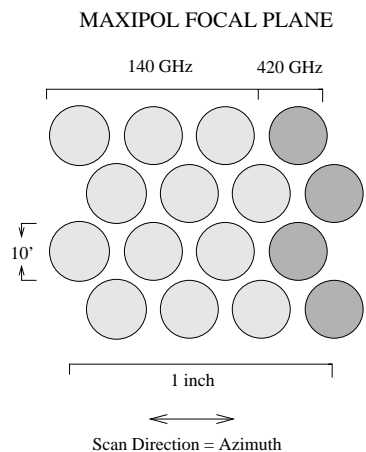}}
\caption[Polarized Focal Plane]{The 16 element, two color \maxipol\
focal plane array.  The 140-GHz detectors observe near the peak of the
CMB signal, while the 420-GHz detectors monitor foregrounds and
atmospheric emission.  All photometers share a single linear
polarization, while the polarization of incident light is rotated by a
spinning halfwave plate at the telescope aperture stop.  All beams
have 10\mins FWHM.}
\end{figure}

	The polarimeter uses a combination of a rotating halfwave
plate (HWP) and a stationary polarizing wire grid.  This results in
modulated sensitivity to both linear polarizations in each detector.
This technique is common in infrared and millimeter astronomy, though
it is new in CMB science.

	The HWP rotates continuously at a frequency of 2~Hz, turning
the incident polarization vector at 8~Hz.  The linearly polarizing
wire grid is between the HWP and the focal plane.  Each detector
independently produces maps of CMB temperature, and the Stokes
parameters $Q$ and $U$ which are converted to $E$ and $B$ for power
spectrum estimation.  Because each detector is used to measure all
three components, cross-calibration does not bias the polarimeter.

	As in \maxima, 16 single-color photometers are used, but in
\maxipol, 12 operate at optical bands around 140~GHz with a bandwidth
of 45~GHz and four operate around 420~GHz with a bandwidth of 35~GHz.
These bands cover the first and third orders of the HWP, respectively.
The actual detector spectra are the same as those of \maxima\ at
150~GHz and 410~GHz.\footnote{The change in nominal center reflects the HWP
bands.  Given the bandwidth and variation between channels, either
convention is reasonable.}  Broad spectral coverage is essential for
monitoring atmospheric emission and foreground contributions from
Galactic dust.  All photometers have a beam size of 10\mins FWHM.

	The fastest optical modulation in \maxipol\ is the 8-Hz
rotation of the polarization.  This modulation strongly rejects noise
at lower frequencies, including scan synchronous signals.  The
polarization rotation frequency is the fourth harmonic of the physical
rotation frequency of the HWP; the phase sensitive detection
effectively discriminates against spurious signals at lower or higher
harmonics.  Though an `offset' signal is observed at the polarization
rotation frequency, laboratory measurements have found that it is
stable to at least the detector noise level.

	The 0.45-Hz primary mirror modulation from \maxima\ is not
used in \maxipol.

\subsection{Scan Strategy}

\begin{figure}[ht]
\centerline{
\epsfig{width=5.0in,angle=0,file=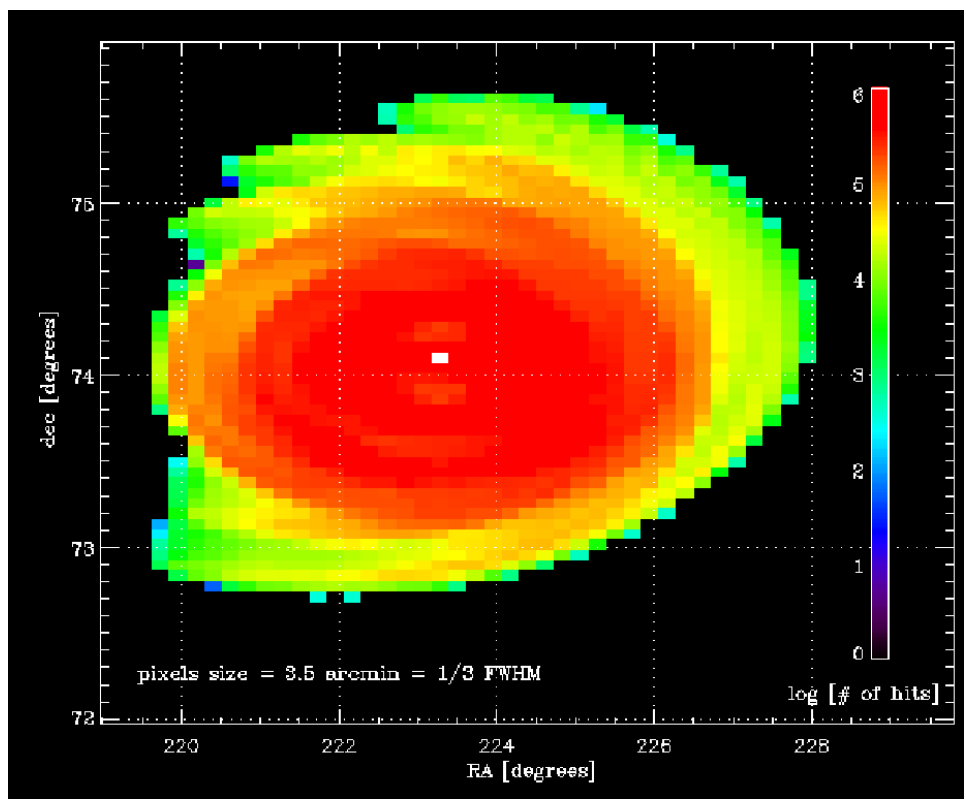}}
\caption[Scan Strategy]{A pointing simulation of a 13-hour \maxipol\
observation.  An area of about 10~deg$^2$ is covered by the 12 140-GHz
photometers.  The color code corresponds to the log of the number of
detector samples in each 3.5\mins square pixel.  This scan is
realizable in a Spring \maxipol\ flight; it consists of \sima 10 hours
of nighttime data and \sima 3 hours of daytime data.}
\label{fig:polscan}
\end{figure}

	The observing strategy for \maxipol\ provides a very deep
integration.  For the first flight, we will observe two or three scan
regions of 4 to 10 square degrees, for a total of about 400 beam-size
pixels with an expected noise of 1.4~\micro K in each of $Q$ and $U$
and 0.35~\micro K in temperature (at 140~GHz).

	In order to focus on such a small region, the scan pattern is
modified from that of \maxima.  The gondola is modulated in azimuth
with an amplitude of \sima 2\degs peak-to-peak and a period of \sima 15
seconds.  Unlike \maxima, the center of the modulation is fixed in
right ascension and declination and follows the rotation of the sky in
both azimuth and elevation.  The rotation of the sky sweeps out a
`bowtie' pattern about the center of the scans (Figure~\ref{fig:polscan}).

\subsection{Flights}

	Balloon flights of up to 36 hours can be achieved during brief
launch windows in Spring and Fall from the National Scientific Balloon
Facility in Fort Sumner, New Mexico.  Using our scan strategy, a
region with very low Galactic dust can be observed continuously for up
to 13 hours in the Spring and for up to six hours in the Fall.

	Two modifications were required to make CMB observations
during daylight hours.  A great deal of additional side-lobe shielding
was required to prevent solar radiation from heating surfaces near the
optical path.  Similar shielding was used successfully for the
\boomerang\ experiment during daylight observations in 1998.  In
addition, one of the CCD camera star sensors used for pointing
reconstruction was modified to observe stars in the daytime.
Development of this sensor has been one of my main contributions to
\maxipol.

	Camera tests during daylight hours at altitudes 20~km to 40~km
have confirmed that we reliably detect stars of visible magnitude 2.5
and higher, and detect certain stars of visible magnitude 3.0.  These
limits are more than adequate for the \maxipol\ scan strategy.

\subsection{Beyond MAXIPOL}

	Current polarization experiments have reached the
sensitivities required to measure $E$-modes.  The next wave of data
from experiments such as \maxipol, MAP and DASI will begin to
characterize the $C_l^{EE}$ and $C_l^{TE}$ power spectra.  After this,
more experiments devoted to CMB polarization will be required.  Of
particular promise are large bolometric arrays for use in ground-based
and balloon-borne polarimeters.

	The Planck satellite will provide a great deal of information
about $E$-mode polarization and has the benefit of full sky coverage.
However, sub-orbital experiments with deeper integration have the
potential to discover more about the CMB, especially low amplitude
$B$-mode fluctuations.

	$B$-mode polarization may not be measurable due to confusion
caused by gravitational lensing.  As more is learned about CMB
polarimetry, the challenge of measuring $B$-modes must be weighed
against this possibility.  The additional science goal of studying
gravitational lensing provides another motivation to observe CMB
polarization beyond the level of $E$-modes.

\begin{figure}[ht]
\centerline{\epsfig{width=5.5in,angle=0,file=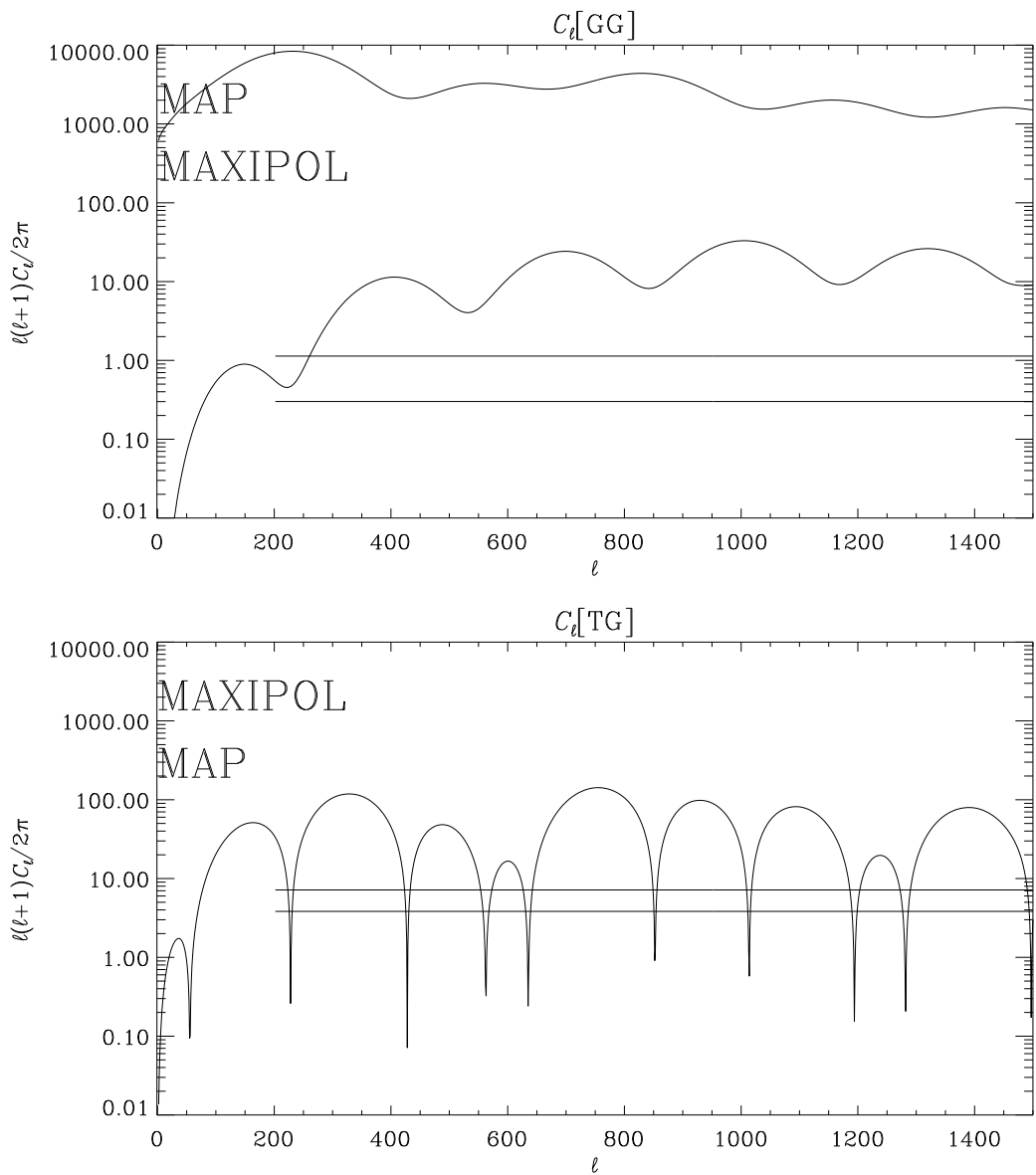}}
\caption[Power Spectrum Estimation]{Projected performance of \maxipol\
and MAP in measuring the $E$-mode polarization.  Model curves
are based on \maximai\ best fit parameters.  {\bf{Top:}} The top curve
shows the temperature power spectrum, $C_l^{TT}$, while the lower curve
shows the $E$-mode power spectrum, $C_l^{EE}$.  The horizontal lines
represent the error on $C_l^{EE}$ for MAP (above) and \maxipol\
(below) assuming a single band power estimate for $200<\ell <3000$.
The very deep integration of \maxipol\ makes it comparable in power to
MAP.  {\bf{Bottom:}} The model curve shows the $E$-mode and
temperature cross power spectrum, $C_l^{TE}$.  Horizontal lines are
the single band error projections for \maxipol\ (above) and MAP
(below).  Sample variance affects the $C_l^{TE}$ and $C_l^{EE}$ power
spectra differently, so the relative power of the experiments varies
for the two measurements.}
\label{fig:maxipolps}
\end{figure}

\clearpage
\addcontentsline{toc}{part}{Bibliography}

{\ssp
\bibliographystyle{apj}
\bibliography{thesis}
}

\appendix
\clearpage
\addcontentsline{toc}{part}{Appendices}
\part*{Appendices}

\chapter{The MAXIMA Collaboration}\label{chap:collab}

The following collaborators have been authors on one or more \maxima\
result papers:

\bigskip

{\ssp 
\small{
\begin{tabbing}
Matt~E.~Abroe (1)\hspace{2in}\=Adrian~T.~Lee (4,13)\\
Peter~Ade (2)\>Phil~D.~Mauskopf (2)\\
Amedeo~Balbi (3)\>C.~Barth~Netterfield (14)\\
Domingos~Barbosa (4,5)\>Sang~Oh (13)\\
James~Bock (6,7)\>Enzo~Pascale (9)\\
Julian~Borrill (8)\>Bahman~Rabii (4,13,15)\\
Andrea~Boscaleri (9)\>Paul~L.~Richards (13)\\
Paolo~de~Bernardis (10)\>George~F.~Smoot (4,13,15)\\
Pedro~G.~Ferreira (11)\>Radek~Stompor (8,16)\\
Shaul~Hanany (1)\>Andrew~E.~Lange (7)\\
Viktor~Hristov (7)\>Celeste~D.~Winant (13,15)\\
Andrew~H.~Jaffe (12)\>Jiun-Huei~Proty~Wu (17)\\\end{tabbing}}

\noindent \line(1,0){350}

\tiny{
\noindent 1 School of Physics and Astronomy, University of Minnesota/Twin Cities, Minneapolis, MN, USA

\noindent 2 Department of Physics and Astronomy, University of Wales, Cardiff

\noindent 3 Dipartimento di Fisica, Universit\`a Tor Vergata, Roma, Italy

\noindent 4 Division of Physics, Lawrence Berkeley National Laboratory, Berkeley, CA, USA

\noindent 5 CENTRA, Insitituto Superior Tecnico, Lisboa, Portugal

\noindent 6 Jet Propulsion Laboratory, Pasadena, CA, USA

\noindent 7 California Institute of Technology, Pasadena, CA, USA

\noindent 8 National Energy Research Scientific Computing Center, Lawrence Berkeley National Laboratory, Berkeley, CA, USA

\noindent 9 IROE-CNR, Firenze, Italy

\noindent 10 Dipartimento di Fisica, Universit\`a La Sapienza, Roma, Italy

\noindent 11 Astrophysics, University of Oxford, UK

\noindent 12 Imperial College, UK

\noindent 13 Department of Physics, University of California, Berkeley, CA, USA

\noindent 14 Department of Physics and Astronomy, University of Toronto, Canada

\noindent 15 Space Sciences Laboratory, University of California, Berkeley, CA, USA

\noindent 16 Department of Astronomy, University of California, Berkeley, CA, USA

\noindent 17 National Taiwanese University, Taipei, Taiwan

}}

\chapter{Calibration Linearity}\label{chap:callinear}

To calculate the linearity of the detectors during calibrations, we use
a combination of theoretical models, measured bolometer properties, and
in-flight measurements of bolometer resistance.

The responsivity of a current biased NTD bolometer
(\cite{SabrinaBolo}) is

\begin{equation}\label{eqn:responsivity_1}
S = \frac{\alpha IR}{G_{d} - \alpha I^{2} R}
,\end{equation}

\noindent where $S$ is the responsivity, $I$ is bias current, $G_{d}$
is the differential thermal conductance which varies as $T^\beta $
with $\beta $ between 1.0 and 3.0 depending on the heat sinking of the
bolometer.  In the case of \maxima\ $\beta = 1.0$ is a good
approximation.  $\alpha$ is

\begin{equation}
\alpha = \frac{1}{R}\frac{\delta R}{\delta T}
.\end{equation}

\noindent $R$ and $T$ are the NTD resistance and temperature, related
by

\begin{equation}
R(V,T) = R_{o}exp\left[\left(\frac{A}{T}\right)^{n} - \frac{eVL}{dk_{b}T}\right]
,\end{equation}

\noindent where $V$ is the voltage across the bolometer, $d$ is NTD
thickness, and $L$ is the characteristic spacing between NTD
impurities.  $A$, $R_{o}$, and $n$ are constants measured before
flight.  In \maxima, $n$ is 0.5 and $eVL \ll dk_{b}T$, so $R$ reduces
to

\begin{equation}\label{eqn:roft}
R(T) = R_{o}e^{\sqrt{\frac{A}{T}}}
.\end{equation}

Equation~\ref{eqn:responsivity_1}, given the relations above for
$\alpha$ and $R(T)$, reduces to,

\begin{equation}\label{eqn:linbig1}
S = -\frac{1}{2T}\sqrt{\frac{A}{T}}\frac{RI}{G_{d} + \frac{1}{2T}\sqrt{\frac{A}{T}} RI^{2}}
. \end{equation}

\noindent We differentiate this with respect to NTD temperature,
taking into account the dependencies of $R$ and $G_{d}$ on
temperature, with bias current $I$ held fixed\footnote{Strictly
speaking, the bias current changes with the NTD resistance.  However,
this is an extremely small effect with $\frac{\Delta I}{I} \approx
0.01 \%$ during calibrations.},

\begin{equation}\label{eqn:linbig2}
\frac{dS}{dT} = -\frac{1}{4}\frac{IR}{T}\sqrt{\frac{A}{T}}\frac{5\frac{G_d}{T}+\frac{G_d}{T}\sqrt{\frac{A}{T}}}{\left(G_{d}+\frac{1}{2T}\sqrt{\frac{A}{T}} RI^{2}\right)^{2}}
. \end{equation}

Finally, we obtain the fractional first order responsivity change,
$\frac{\Delta S}{S}$, corresponding to a measured temperature change,

\begin{eqnarray}\label{eqn:linfinal}
\frac{\Delta S}{S} & = & \frac{1}{S} \frac{dS}{dT} \Delta T \nonumber\\
& = & -\frac{\Delta T}{2T}\frac{5 + \sqrt{\frac{A}{T}}}{1 + \frac{1}{2T}\sqrt{\frac{A}{T}}\frac{RI^2}{G_d}}
,\end{eqnarray}

\noindent where $\Delta T$ is the change in bolometer temperature.

	The bolometer responsivity variations in
Table~\ref{table:linearity} are calculated using
Equation~\ref{eqn:linfinal}.  $T$ and $\Delta T$ are calculated using
Equation~\ref{eqn:roft} from $R$ and $\Delta R$ as measured in flight.
$A$, $R_{o}$ are constants and $G_{d}$ at the bolometer operating
temperature is measured before flight.

\chapter{Detector Sensitivities}\label{chap:NET}
\begin{table}[hb]
\begin{center}
\begin{small}
\begin{tabular}{c|cc|cc}
\hline
&\multicolumn{2}{c|}{\maximai}&\multicolumn{2}{c}{\maximaii}
\\ \footnotesize Channel & \footnotesize CMB NET & \footnotesize R-J NET & \footnotesize CMB NET & \footnotesize R-J NET
\\ \footnotesize [Freq (GHz)] & \footnotesize \micro K $\sqrt{sec}$ & \footnotesize \micro K $\sqrt{sec}$ & \footnotesize \micro K $\sqrt{sec}$ & \footnotesize \micro K $\sqrt{sec}$
\\ \hline\hline
   B14 [150]    &  142.3 &  84.7 &  257.2 & 145.3 
\\ B15 [150]    &   87.6 &  52.9 &   62.8 &  36.3
\\ B24 [150]    &  177.0 & 106.0 &   94.6 &  54.1
\\ B25 [150]    &   78.4 &  45.2 &  101.5 &  57.6
\\ B34 [150]    &   88.6 &  46.1 &  118.7 &  66.7
\\ B35 [150]    &  185.6 & 108.0 &   90.8 &  51.6
\\ B44 [150]    &  271.1 & 128.0 &  321.7 & 179.7
\\ B45 [150]    &   92.1 &  52.6 &   73.2 &  41.1
\\ B13 [230]    &  316.6 & 107.2 & *      & *
\\ B23 [230]    &  142.3 &  46.4 &  155.7 &  45.3
\\ B33 [230]    &  123.1 &  37.6 &  263.0 &  76.2
\\ B43 [230]    &  270.0 &  89.0 & *      & *
\\ B12 [410]    & 1701.7 &  66.7 & 2656.1 & 104.2
\\ B22 [410]    & 2049.6 &  80.3 & 2080.9 &  81.5
\\ B32 [410]    & 3013.1 & 116.9 & 1699.6 &  65.9
\\ B42 [410]    & 4416.2 & 173.3 & 4102.0 & 160.9
\end{tabular}
\end{small}
\caption[Noise Equivalent Temperatures]{``CMB NET'' is the detector
sensitivity (\S \ref{exper:bolos}) to CMB temperature
fluctuations. Values are derived from calibrated responsivity
(Appendix~\ref{chap:calparms}) and detector noise over a range of
0.1~Hz to 16~Hz.  In the case of \maximaii, for which responsivity
varies significantly over the flight, average values are used.  ``R-J
NET'' is the sensitivity to temperature fluctuations for sources in
the Raleigh-Jeans limit (\ie~at least \sima 15~K at 410~GHz).  R-J NET
and CMB NET are related by the measured detector spectra.
\newline * Dead channel.}  
\end{center} 
\end{table}

\chapter{Calibration Parameters}\label{chap:calparms}

\begin{table}[hb]
\begin{center}
\begin{small}
\begin{tabular}{c|cc|cc}
\hline
&\multicolumn{2}{c|}{\maximai}&\multicolumn{2}{c}{\maximaii}
\\ & \footnotesize Calibration & \footnotesize Calibration & \footnotesize Calibration & \footnotesize Calibration 
\\ \footnotesize Channel & \footnotesize Dipole & \footnotesize Jupiter & \footnotesize Dipole & \footnotesize Mars
\\ \footnotesize [Freq (GHz)] & \footnotesize (10$^{-5}$ V/K)& \footnotesize (10$^{-5}$ V/K)& \footnotesize (10$^{-5}$ V/K)& \footnotesize (10$^{-5}$ V/K)
\\ \hline\hline 
B14 [150]    & 10.09\err 0.26 &  9.20\err 1.16 & 5.13\err 0.242 & 5.73 \err 1.17
\\ B15 [150]    & 11.56\err 1.16 & 10.49\err 1.27 & 9.21\err 0.103 & 12.29\err 1.25
\\ B24 [150]    &  8.90\err 0.19 &  8.17\err 1.01 & 6.20\err 0.145 & 6.96 \err 0.93
\\ B25 [150]    & 10.92\err 0.16 & 10.34\err 1.23 & 7.88\err 0.198 & 9.60 \err 0.98
\\ B34 [150]    &  9.02\err 0.16 &  9.68\err 1.13 & 5.68\err 0.153 & 7.39 \err 0.75
\\ B35 [150]    &  9.04\err 0.30 &  8.41\err 1.04 & 6.04\err 0.078 & 9.21 \err 0.94
\\ B44 [150]    &  8.17\err 0.34 &  9.65\err 1.16 & 5.30\err 0.174 & 7.52 \err 0.81
\\ B45 [150]    & 10.10\err 0.23 &  9.78\err 1.15 & 8.20\err 0.128 & 9.39 \err 0.59
\\ B13 [230]    &  3.46\err 0.28 &  2.87\err 0.36 & *               & *
\\ B23 [230]    &  6.92\err 0.21 &  5.95\err 0.91 & 4.06\err 0.129 & 5.38 \err 0.76
\\ B33 [230]    &  6.11\err 0.19 &  5.65\err 0.81 & 4.61\err 0.262 & 2.80 \err 0.41
\\ B43 [230]    &  6.30\err 0.24 &  5.51\err 0.78 & *               & *
\\ B12 [410]    &  **             &  0.89\err 0.10 & **              & 0.81 \err 0.091
\\ B22 [410]    &  **             &  0.76\err 0.09 & **              & 0.62 \err 0.067
\\ B32 [410]    &  **             &  0.66\err 0.08 & **              & 0.80 \err 0.122
\\ B42 [410]    &  **             &  0.54\err 0.06 & **              & 0.44 \err 0.046
\end{tabular}
\end{small}
\caption[Absolute Calibration Summary]{Absolute calibration parameters
for both \maxima\ flights.  ``Calibration Dipole'' is the responsivity
as measured by the dipole observation.  ``Calibration Jupiter'' or
``Calibration Mars'' is the responsivity as measured by the planet
observation.
\newline * Dead channel.
\newline ** 410 GHz detectors are not calibrated from the CMB dipole.}
\end{center}
\end{table}

\begin{table}[hb]
\begin{center}
\begin{small}
\begin{tabular}{c|cc|cc}
\hline
&\multicolumn{2}{c|}{\maximai}&\multicolumn{2}{c}{\maximaii}
\\ & \footnotesize Absolute & \footnotesize Ratio & \footnotesize Absolute & \footnotesize Ratio 
\\ \footnotesize Channel & \footnotesize Calibration & \footnotesize from & \footnotesize Calibration & \footnotesize from 
\\ \footnotesize [Freq (GHz)] & \footnotesize Ratio & \footnotesize Stimulator & \footnotesize Ratio & \footnotesize Stimulator 
\\ \hline\hline
B14 [150]    &  0.95\err 0.12 & 1.04\err 0.02       & 1.12\err 0.23 & 1.48\err 0.17
\\ B15 [150]    &  0.96\err 0.11 & 1.05\err 0.02       & 1.33\err 0.14 & 1.43\err 0.04
\\ B24 [150]    &  0.95\err 0.11 & 1.04\err 0.02       & 1.23\err 0.17 & 1.39\err 0.10
\\ B25 [150]    &  0.99\err 0.13 & 1.03\err 0.01       & 1.22\err 0.13 & 1.29\err 0.03
\\ B34 [150]    &  1.08\err 0.13 & 1.01\err $<$0.01    & 1.30\err 0.14 & 1.45\err 0.03
\\ B35 [150]    &  0.98\err 0.12 & 1.05\err 0.01       & 1.52\err 0.16 & 1.35\err 0.02
\\ B44 [150]    &  1.18\err 0.15 & 1.00\err $<$0.01    & 1.42\err 0.16 & 1.37\err 0.02
\\ B45 [150]    &  0.98\err 0.12 & 1.02\err $<$0.01    & 1.14\err 0.12 & 1.30\err 0.01
\\ B13 [230]    &  0.86\err 0.12 & 1.03\err 0.02       & *              & *
\\ B23 [230]    &  0.89\err 0.13 & 1.03\err $<$0.01    & 1.33\err 0.19 & 1.49\err 0.06
\\ B33 [230]    &  0.95\err 0.14 & 1.03\err $<$0.01    & 0.61\err 0.10 & 0.75\err $<$0.01
\\ B43 [230]    &  0.89\err 0.13 & 1.02\err $<$0.01    & *              & *
\\ B12 [410]    &  **             & 1.02\err 0.02       & **             & 1.68\err 0.37
\\ B22 [410]    &  **             & 1.06\err 0.01       & **             & 1.74\err 0.46
\\ B32 [410]    &  **             & 1.01\err $<$0.01    & **             & 1.99\err 0.22
\\ B42 [410]    &  **             & 1.02\err 0.04       & **             & 1.58\err 0.01
\end{tabular}
\end{small}
\caption[Time Dependent Calibration Summary]{Time dependent
calibration.  ``Absolute Calibration Ratio'' is the ratio of the
responsivity measured during the planet calibration to that measured
during the dipole calibration.  ``Ratio from Stimulator'' is the ratio
for these two points in the flight, as determined from the internal
relative calibrator.
\newline * Dead channel.
\newline ** 410 GHz detectors are not calibrated from the CMB dipole.}
\end{center}
\end{table}

\end{document}